\documentclass[a4paper,11pt]{article}
\pdfoutput=1 

\usepackage{jcappub} 
\usepackage{aas_macros}

\usepackage[T1]{fontenc} 
\usepackage{comment}
\usepackage{hyperref}

\usepackage{rotating}
\usepackage{adjustbox}
\usepackage{pdflscape}

\title{\boldmath The impact of the IGM thermal state on the Ly$\alpha$ flux 3D power spectrum from linear to highly non-linear scales}

\author[a,b,c,d,1]{Tomáš Šoltinský,\note{Corresponding author.}}
\author[b,c,d]{Gabriele Autieri,}
\author[e]{Vid Iršič,}
\author[a,b,c,d]{Matteo Viel}

\affiliation[a]{INAF - Osservatorio Astronomico di Trieste, Via G.B. Tiepolo, 11, I-34143 Trieste, Italy}
\affiliation[b]{INFN, Sezione di Trieste, Via Valerio 2, I-34127 Trieste, Italy}
\affiliation[c]{IFPU, Institute for Fundamental Physics of the Universe, Via Beirut 2, I-34151 Trieste, Italy}
\affiliation[d]{SISSA - International School for Advanced Studies, Via Bonomea 265, I-34136 Trieste, Italy}
\affiliation[e]{Centre for Astrophysics Research, Department of Physics, Astronomy and Mathematics, University of Hertfordshire, Hatfield, AL109AB, UK}

\emailAdd{tomas.soltinsky@inaf.it}

\newcommand\HI{\hbox{H$\,\rm \scriptstyle I$}~}
\newcommand\HeII{\hbox{He$\,\rm \scriptstyle II$}~}
\newcommand\HeIII{\hbox{He$\,\rm \scriptstyle III$}~}
\usepackage{xcolor}

\def\vid#1{\textcolor{orange}{VI: #1}}

\abstract{
Recently initiated and upcoming spectroscopic surveys, such as DESI and WST, will provide more than $10^6$ high-$z$ quasar spectra, enabling dense sky coverage by the Ly$\alpha$ forest. This is expected to establish the three-dimensional (3D) Ly$\alpha$ forest power spectrum, $P_{\rm 3D,\alpha}$, as a probe of the thermal and ionization history of the intergalactic medium (IGM). To exploit this opportunity, we quantify the imprints of reionization on the post-reionization IGM using high-fidelity numerical models. We use the Sherwood and Sherwood--Relics cosmological hydrodynamical simulations with box sizes up to $160\,h^{-1}\,\rm cMpc$ to investigate the impact of box size, mass resolution, and extracted grid resolution on $P_{\rm 3D,\alpha}$ over $2.4\leq z\leq4.8$. After applying a Zel'dovich control variate correction, simulation volume has a modest impact over most scales and orientations, whereas degrading the mass resolution produces differences of up to $\sim13\%$. Insufficient resolution of the grid used for the optical-depth calculation can artificially enhance small-scale power by up to $\sim35\%$. The timing of \HI reionization leaves only a percent-level imprint on $P_{\rm 3D,\alpha}$ at $z=2.4$, whereas varying the photoheating rate by a factor of two changes the large-scale power by $\sim4$--$8\%$. Our results demonstrate that numerical effects can be comparable to, or exceed, the relic astrophysical signatures encoded in $P_{\rm 3D,\alpha}$, making numerical convergence essential for interpreting precise Ly$\alpha$ forest measurements. The strongest astrophysical imprint arises from spatially inhomogeneous \HI reionization, which enhances the large-scale power by up to $\sim70\%$ at $z=4.2$. This highlights the potential of post-reionization Ly$\alpha$ forest measurements as a probe of the thermal history and spatial morphology of cosmic reionization.
}

\begin{document}
\maketitle
\flushbottom

\section{Introduction}

The Ly$\alpha$ forest, arising from absorption by neutral hydrogen in the intergalactic medium (IGM) along the lines of sight (LOS) to distant quasars, provides a powerful probe of cosmology and astrophysics at high redshifts ($z \sim 2$--5). Large spectroscopic surveys such as the Baryon Oscillation Spectroscopic Survey \citep[BOSS][]{Dawson_2013_BOSS} and its extension eBOSS \citep{Dawson_2016_eBOSS} have enabled precise measurements of Ly$\alpha$ forest statistics over wide areas of the sky, dramatically increasing the number of available quasar sightlines and allowing for three-dimensional mapping of the IGM. These surveys have led to the detection of large-scale correlations and baryon acoustic oscillations (BAO) in the Ly$\alpha$ forest at redshifts $z \sim 2$--3 \citep{Slosar_2011,Busca_2013,Delubac_2015,Bautista_2017,duMasdesBourboux_2020}. Ongoing and future surveys, such as the Dark Energy Spectroscopic Instrument (DESI) \citep{DESI_2016} and the Wide Spectroscopic Telescope (WST) \citep{Mainieri_2024_WST}, will further increase the statistical power of Ly$\alpha$ forest measurements by providing denser sampling of quasar sightlines and extending the redshift coverage. Specifically, DESI has already provided first data \citep{DESI_2024} and is expected to increase the high-$z$ quasar sample from $\sim 3\times10^5$ in eBOSS to $\sim 1.2\times10^6$.

Traditionally, Ly$\alpha$ forest analyses have focused on two primary statistical observables: the one-dimensional (1D) flux power spectrum along individual sightlines, and the three-dimensional (3D) correlation function reconstructed from multiple sightlines \citep{Lepori_2020,Cuceu_2025}. The 1D power spectrum probes small-scale clustering along the line of sight and has been used extensively to constrain cosmological parameters, including the amplitude and shape of the matter power spectrum \citep{Croft_1998,McDonald_2000,Palanque-Delabrouille_2013}, while the 3D correlation function is particularly sensitive to large-scale features such as the BAO peak \citep{Gordon_2023}. These observables provide unique insight into both cosmology and the astrophysics of the IGM. In particular, the Ly$\alpha$ forest is sensitive to the nature of dark matter through its impact on small-scale structure, allowing constraints on models such as warm \citep{Viel_2005,Viel_2013_WDM,Baur_2017,Murgia_2018,Villasenor_2023,Irsic_2024,Garcia-Gallego_2025}, fuzzy \citep{Irsic_2017,Armengaud_2017,Nori_2019,Rogers_2021}, and interacting dark matter \citep{Dvorkin_2014,Xu_2018_Lya,Garny_2018,Mosbech_2026}. Similarly, it can be used to constrain the mass of neutrinos \citep{Palanque-Delabrouille_2015,Palanque-Delabrouille_2015_BOSS,Palanque-Delabrouille_2020,Seljak_2005,Yeche_2017} and primordial magnetic fields \citep{Pavicevic_2025}. It also probes the thermal and ionization state of the IGM, providing information about AGN feedback \citep{Tillman_2025,Pirecki_2025} and the history of reionization and its impact at lower $z$ \citep{Boera_2019,Walther_2019,Montero-Camacho_2019,Molaro_2022,Molaro_2023,Zheng_2026}. The above-mentioned surveys (e.g. eBOSS and DESI) have already provided data for these studies \citep{Chabanier_2019,Karacayli_2024,Parashari_2026}. As a result, the Ly$\alpha$ forest has become a key tool for studying the interplay between cosmology and baryonic physics at intermediate redshifts.

More recently, there has been increasing interest in measuring the full 3D flux power spectrum\footnote{Another novel Ly$\alpha$ forest statistic that has recently been measured is the 1D bispectrum \citep{delaCruz_2025}.}, $P_{\mathrm{3D}, \alpha}$, which provides a unified description of both small- and large-scale structure and enables anisotropic analyses through its dependence on the angle with respect to the line of sight. Besides the modelling efforts \citep{McDonald_2003,Arinyo-i-Prats_2015,Font-Ribera_2018,Rogers_2018,Chabanier_2024,Karacayli_2025}, there are already first measurements of $P_{\mathrm{3D}, \alpha}$ at $z\sim2.3$ based on eBOSS observations \citep{deBelsunce_2024,Karim_2024}.

Similarly to the 1D power spectrum, $P_{\mathrm{3D}, \alpha}$ is sensitive to the thermal state of the IGM and hence contains information on both hydrogen \citep{Montero-Camacho_2019} and helium reionization \citep{McQuinn_2011,Greig_2015} as well as X-ray preheating \citep{Montero-Camacho_2024}. In this study, we further explore the potential of post-reionization Ly$\alpha$ forest observations to constrain reionization by employing simulations with various thermal histories, reionization timing and morphology models from Sherwood \citep{Bolton_2017} and Sherwood--Relics suites \citep{Puchwein_2023}.

Interpreting Ly$\alpha$ forest measurements requires accurate modelling of the flux power spectrum across a wide range of scales. On large scales, the forest can be described as a biased tracer of the underlying matter distribution with redshift-space distortions, leading to the well-known Kaiser-like form \citep{Kaiser_1987,McDonald_2003,Font-Ribera_2013}. Spatial fluctuations in the ionizing background can further modify this large-scale clustering and introduce a scale dependence in the effective Ly$\alpha$ forest bias \citep{GontchoAGontcho_2014}. On smaller scales, non-linear structure formation, thermal broadening, and peculiar velocities introduce significant deviations from linear theory. Several approaches have been developed to model these effects, including perturbative effective field theory (EFT) descriptions \citep{Ivanov_2024,Chadaykin_2025,deBelsunce_2025}\footnote{EFT has also been applied to the Ly$\alpha$ forest 1D power spectrum \citep{Ivanov_2025_PBH,He_2025,Karacayli_2026}.} and phenomenological parametrizations calibrated on simulations. In particular, the model introduced by \citep{Arinyo-i-Prats_2015} provides a flexible framework in which the flux power spectrum is expressed as a product of a linear-theory term and a non-linear correction factor. This formalism, or closely related variants, has been widely used in both simulation-based and observational analyses (e.g. \citep{duMasdesBourboux_2020,deSainteAgathe_2019,Chabanier_2024}).

Recent work has primarily focused on large-scale measurements of the Ly$\alpha$ forest power spectrum and correlation function, where the signal is less affected by non-linearities and systematic uncertainties. For example, \cite{Chabanier_2024} presented modelling of the 3D Ly$\alpha$ forest power spectrum on large scales, enabling precise cosmological constraints. In this work, we instead investigate the Ly$\alpha$ forest 3D power spectrum from linear to highly non-linear scales, with particular emphasis on its numerical convergence and sensitivity to the thermal and reionization history of the IGM. We compute $P_{\rm 3D,\alpha}$ over a broad range of scales and model it using the phenomenological formalism of \cite{Arinyo-i-Prats_2015} over $k\leq10\,h\,\mathrm{cMpc}^{-1}$. This allows us to quantify how the measured power spectrum and the corresponding best-fitting parameters depend on simulation volume, mass resolution, extracted-grid resolution, and the assumed thermal and reionization history.

This paper is structured as follows. In Sec.~\ref{sec:sims} we present the models of the IGM based on cosmological simulations. We then describe the computation of the $P_{\mathrm{3D},\alpha}$, mitigation of cosmic variance effects, and the analytical fitting procedure in Sec.~\ref{sec:P3D_sims}, \ref{sec:ZCV} and \ref{sec:P3D_analytic}, respectively. In Sec.~\ref{sec:convergence_tests} we show the convergence tests with respect to simulation mass resolution, volume, and extracted-grid resolution. In Sec.~\ref{sec:reion_models} we investigate the impact of reionization and thermal history on $P_{\mathrm{3D},\alpha}$. We conclude our study in Sec.~\ref{sec:conclusions}.

\section{Modelling the post-reionization intergalactic medium}\label{sec:sims}

 In this work we use cosmological hydrodynamical simulations from the Sherwood\footnote{\url{https://www.nottingham.ac.uk/astronomy/sherwood/}} and Sherwood--Relics\footnote{\url{https://www.nottingham.ac.uk/astronomy/sherwood-relics/}} projects to model the IGM and its imprint on the Ly$\alpha$ forest. These simulations have been designed specifically to study the physical properties of the low-density IGM and to generate synthetic Ly$\alpha$ forest spectra that can be directly compared to observations \citep{Bolton_2017,Puchwein_2023}. Their combination of large simulation volumes and high mass resolution makes them well suited for predicting Ly$\alpha$ forest statistics across a broad range of spatial scales, including the three-dimensional clustering of transmitted flux that underlies measurements of the Ly$\alpha$ forest power spectrum.

The original Sherwood simulation suite consists of a large set of cosmological hydrodynamical simulations performed with a modified version of the Tree-PM smoothed particle hydrodynamics code \textsc{P-GADGET-3} \citep{Springel_2005}, an updated version of \textsc{GADGET-2}. These simulations follow the gravitational evolution of dark matter and baryons together with hydrodynamical processes in the gas, allowing the formation of the filamentary cosmic web that gives rise to the Ly$\alpha$ forest. The simulations adopt a $\Lambda$CDM cosmology consistent with \textit{Planck} constraints, with parameters $\Omega_{\rm m}=0.308$, $\Omega_\Lambda=0.692$, $\Omega_{\rm b}=0.0482$, $h=0.678$, $\sigma_8=0.829$, and $n_{\rm s}=0.961$ \citep{planck2014}, and primordial helium fraction by mass $Y=0.24$ \citep[][]{Hsyu_2020}. The simulations used in this study span box sizes from $L_{\rm box}=40$ to $160\,h^{-1}\,\mathrm{cMpc}$ with particle numbers from $N_{\rm part}=2\times512^3$ to $2\times2048^3$, comprising equal numbers of gas and dark matter particles. These are listed in Table~\ref{tab:simulations} as well as the corresponding dark matter and gas particle masses which reach a resolution of $M_{\rm DM}=5.37\times10^5\,h^{-1}\,\rm M_{\odot}$ and $M_{\rm gas}=9.97\times10^4\,h^{-1}\,\rm M_{\odot}$, respectively. This large dynamic range allows the simulations to resolve the small-scale density fluctuations responsible for Ly$\alpha$ absorption while simultaneously capturing the large-scale structure relevant for clustering statistics.

\begin{table*}
\centering
\caption{\label{tab:simulations}Summary of simulations used in this study including (from left to right) their name, box size, number of tracked particles, mass of dark matter and gas particles, thermal/reionization history model, and redshifts at which they were analysed. We define
$z_{\rm all}=\{2.4,\,2.8,\,3.2,\,3.6,\,4.2,\,4.8\}$.}
\begin{tabular}{ccccccc}
\hline
Name & $L_{\rm box}$ & $N_{\rm part}$ & $M_{\rm DM}$ & $M_{\rm gas}$ & Thermal/ & $z$ \\
 & [$h^{-1}\,\rm cMpc$] & & [$h^{-1}\,\rm M_{\odot}$] & [$h^{-1}\,\rm M_{\odot}$] & reionization & \\
 & & & & & model & \\
\hline
40\_512 & 40 & $2\times512^3$ & $3.44\times10^7$ & $6.38\times10^6$ & Sherwood & $z_{\rm all}$ \\
40\_1024 & 40 & $2\times1024^3$ & $4.30\times10^6$ & $7.97\times10^5$ & Sherwood, relics, & $z_{\rm all}$ \\
 & & & & & cold, hot, & 2.4, 3.2 \\
 & & & & & zr525, zr675, zr750 & 2.4, 3.2 \\
40\_2048 & 40 & $2\times2048^3$ & $5.37\times10^5$ & $9.97\times10^4$ & Sherwood & 2.4\\
80\_1024 & 80 & $2\times1024^3$ & $3.44\times10^7$ & $6.38\times10^6$ & Sherwood & $z_{\rm all}$ \\
80\_2048 & 80 & $2\times2048^3$ & $4.30\times10^6$ & $7.97\times10^5$ & Sherwood & 2.4\\
160\_2048 & 160 & $2\times2048^3$ & $3.44\times10^7$ & $6.38\times10^6$ & Sherwood & 2.4\\
 & & & & & Homogeneous & 4.2 \\
 & & & & & Patchy & 4.2 \\
\hline
\end{tabular}
\end{table*}

The ionization and thermal state of the gas in the Sherwood simulations is determined by photoionization and photoheating driven by a spatially homogeneous ultraviolet background (UVB), typically based on the model of \cite{Haardt_2012}. The gas is assumed to be optically thin and in ionization equilibrium, which is an excellent approximation for the low-density IGM after hydrogen reionization. In order to improve agreement with observational measurements of the IGM temperature evolution, small modifications to the He\,{\sc ii} photoheating rate are applied. Star formation is not followed explicitly; instead, dense (overdensity $\Delta>10^3$) and cold (gas kinetic temperature $T<10^5\rm\,K$) gas particles are converted into collisionless particles \citep{Viel_2004}. This approach significantly reduces computational cost while leaving the properties of the low-density IGM that produces the Ly$\alpha$ forest largely unaffected \citep{Viel_2013_galaxyform}. We label the simulations from this suite that are used in this manuscript as Sherwood.

The Sherwood--Relics simulations extend the original Sherwood project by improving the treatment of the thermal and ionization evolution of the IGM and by exploring a significantly larger parameter space including different reionization and heating histories \citep{Puchwein_2023}. In particular, these simulations incorporate a non-equilibrium thermo-chemistry solver that follows the time-dependent ionization state and temperature of the gas. This allows a more accurate modelling of the heating associated with cosmic reionization and avoids the artificial delays between photoionization and photoheating that can arise in equilibrium ionization schemes. The Sherwood--Relics suite includes simulations that vary key astrophysical and cosmological parameters relevant for Ly$\alpha$ forest studies, including the thermal history of the IGM, and the redshift of hydrogen reionization. Here we use models zr525, relics (fiducial), zr675 and zr750 in which reionization is completed by redshift $z_{\rm r}=5.25$, 6.00, 6.75 and 7.50, respectively. We also include cold (hot) models in which the \HI photoheating is decreased (increased) by a factor of two relative to the fiducial relics model. These simulations utilize spatially uniform UVB following \cite{Puchwein_2019}.

A further important extension of the Sherwood--Relics project is the inclusion of simulations that model the spatial fluctuations in the ionizing radiation field associated with patchy cosmic reionization. Instead of assuming a spatially uniform UVB, these simulations combine hydrodynamical calculations with radiative transfer (RT), namely ATON \citep{Aubert_2008,Aubert_2010}, to generate spatially varying ionizing radiation fields. This approach captures large-scale fluctuations in the IGM temperature and pressure smoothing scale that are imprinted by the inhomogeneous reionization process and can persist well into the post-reionization epoch \citep{Puchwein_2023}. Such effects can potentially influence post-reionization Ly$\alpha$ forest statistics. The patchy reionization simulations have been used to study the Ly$\alpha$ forest 1D power spectrum \citep[e.g.][]{Molaro_2022}, the correlation of the Ly$\alpha$ forest with galaxies \cite{Conaboy_2025}, Ly$\alpha$ forest damping wings \citep{Keating_2024_DWshape,Keating_2024_DWJWST,Sawyer_2025}, 21-cm forest \citep{Soltinsky_2021}, quasar near-zones \citep{Satyavolu_2023,Soltinsky_2023}, thermal state of the IGM \citep{Gaikwad_2020}, timing of reionization \citep{Bosman2022,Zhu_2024}, mean free path of Lyman-limit photons \citep{Feron_2024}, as well as to provide forecasts for the ELT-ANDES \citep{DOdorico_2024}\footnote{Note that \citep{Bosman2022} and \citep{Satyavolu_2023} used precursor simulations of Sherwood--Relics described in \citep{Kulkarni_2019,Keating_2020}.}. In this work we use a model (Patchy) calibrated to the Ly$\alpha$ forest effective optical depth, $\tau_{\rm eff}$, measurements in which the inhomogeneous reionization is finished at $z_{\rm r}=5.7$. We also include the corresponding Homogeneous model, which has an identical numerical setup but assumes spatially homogeneous reionization.

From all the simulations we draw $N_{\rm los}$ uniformly separated skewers of various fields with the number of pixels $N_{\rm bins}=N_{\rm part}^{1/3}$, resulting in pixel size of $\delta R=L_{\rm box}/N_{\rm bins}=L_{\rm box}/N_{\rm part}^{1/3}$. These include $\Delta$, $T$, neutral hydrogen fraction $x_{\rm HI}$ and peculiar velocity $v_{\rm pec}$. While other Ly$\alpha$ forest 3D power spectrum studies focused on generating a uniform grid where $N_{\rm grid}=N_{\rm los}^2N_{\rm bins}$ \citep[e.g.][]{Chabanier_2024}, we use this approach as our fiducial configuration but also vary $N_{\rm grid}$ to test its effect on the signal.

\subsection{Comparison to other simulations used for Ly$\alpha$ forest studies}

Compared to other simulation suites commonly used for Ly$\alpha$ forest studies, such as the Nyx simulations \citep{Lukic_2015}, Illustris \citep{Vogelsberger_2014_Illustris}, IllustrisTNG \citep{Nelson_2019_IllustrisTNG}, and EAGLE \citep{Schaye_2015_Eagle}, or the radiation-hydrodynamical simulations including THESAN \cite{Kannan_2022_Thesan} and CROC \citep{Gnedin_2014_CROC}, the Sherwood project prioritizes resolving the diffuse IGM with very high mass resolution while maintaining sufficiently large cosmological volumes. This resolution–volume combination has been shown to be sufficient for converged Ly$\alpha$ forest statistics \citep{Bolton_2009_convergence,Lukic_2015}.

While large galaxy-formation simulations such as Illustris, IllustrisTNG, and EAGLE include detailed models of star formation, feedback, and galaxy evolution, they typically operate at lower resolution in the low-density IGM and were not optimized for precision modelling of Ly$\alpha$ forest observables. In contrast, Sherwood simulations intentionally simplify galaxy and star formation physics and instead focus computational resources on accurately resolving the density, temperature, and velocity fields of the diffuse IGM that produce Ly$\alpha$ absorption. 

Radiation-hydrodynamical simulations such as THESAN and CROC model the reionization process self-consistently by coupling galaxy formation and radiative transfer within large cosmological volumes. For example, the THESAN simulations evolve a $(95.5\,{\rm cMpc})^3$ volume with baryonic mass resolution of $\sim5.8\times10^5\,M_\odot$ while simultaneously tracking the radiation field produced by galaxies during reionization. However, Sherwood and Sherwood--Relics simulations achieve high mass resolution in the diffuse IGM while simultaneously exploring a wide range of thermal and reionization histories.

Compared to grid-based hydrodynamical simulations such as Nyx, which use an Eulerian approach, Sherwood employs a smoothed particle hydrodynamics (SPH) approach with the \textsc{P-GADGET-3} code, leading to broadly consistent predictions for Ly$\alpha$ forest statistics while allowing exploration of a wide parameter space of thermal and reionization histories. 
While there is an overlap of the simulation volumes at $L_{\rm box}=160\,h^{-1}\,\rm cMpc$ with the Ly$\alpha$ forest 3D power spectrum study by \cite{Chabanier_2024}, who use the Nyx simulation suite, this work focuses primarily on very large scales. In contrast, the range of box sizes used here allows us to study the convergence of the signal and focus on smaller-scale modes. In addition, the Sherwood--Relics extension further distinguishes itself by including a large grid of simulations that explore variations in IGM temperature evolution and reionization history as well as inhomogeneous reionization.

Overall, the Sherwood and Sherwood--Relics simulations provide a flexible and physically motivated framework for modelling the Ly$\alpha$ forest. Their combination of large volumes, high resolution, and detailed treatments of the IGM thermal history makes them particularly suitable for predicting its statistical observables such as the 3D power spectrum and for studying the impact of astrophysical processes on its measurements.

\section{Ly$\alpha$ forest 3D power spectrum}\label{sec:Lya_P3D}

In this section we describe the numerical computation of the statistical observable of interest, the 3D power spectrum of the Ly$\alpha$ forest flux, $P_{\rm 3D,\alpha}$, based on our numerical simulations (Sec.~\ref{sec:P3D_sims}), the mitigation of cosmic variance on the signal in Sec.~\ref{sec:ZCV}, and the analytical fitting procedure (Sec.~\ref{sec:P3D_analytic}).

\subsection{Numerical simulation-based computation}\label{sec:P3D_sims}

To produce the synthetic Ly$\alpha$ forest flux spectra we use the LOS of fields extracted from the simulations as described in Sec.~\ref{sec:sims} and compute the optical depth to Ly$\alpha$ photons
\begin{equation}\label{eq:tau_Lya_integral}
    \tau_{\alpha}=\frac{\pi e^2}{m_{e}c}f_{12}\int n_{\mathrm{HI}}\phi_{\nu}dR.
\end{equation}
Here $e$ and $m_e$ are the electron charge and mass, respectively, $c$ is the speed of light, and $f_{12}$ is the Ly$\alpha$ resonance transition oscillator strength. $\phi_{\nu}$ is the line profile which we assume to be a Voigt line profile. This is computed via the Hjerting function \citep{Hjerting_1938}, $H$, approximated following \citep{TepperGarcia_2006}.

In the discrete form at pixel $i$ this is calculated as \citep[e.g.][]{Bolton_Haehnelt_2007}
\begin{equation}\label{eq:tau_Lya}
    \tau_{\alpha,i}=\frac{\sigma_{\alpha}c\delta R}{\pi^{1/2}}\sum^{N_{\rm bins}}_{j=1}\frac{n_{\mathrm{HI},j}}{b_j}H\left(\frac{\Lambda_{\alpha}\lambda_{\alpha}}{4\pi b_j},\frac{v_{\mathrm{H},i}-\left(v_{\mathrm{H},j}+v_{\mathrm{pec},j}\right)}{b_j}\right),
\end{equation}
where $\sigma_{\alpha}=4.48\times10^{-18}\,\rm cm^2$ is the cross-section of the Ly$\alpha$ transition, $n_{\rm HI}$ is the number density of neutral hydrogen, $b=\left(2k_{\rm B}T/m_{\rm H}\right)^{1/2}$ is the Doppler parameter, $k_{\rm B}$ is the Boltzmann constant, $m_{\rm H}$ is the hydrogen atom mass, $\Lambda_{\alpha}=6.265\times10^{8}\,\rm s^{-1}$ is the damping wing constant and $v_{\rm H}$ is the Hubble velocity. 

For each simulation model and redshift separately, we rescale the computed optical depths by a spatially uniform multiplicative factor, $A_{\rm eff}$, such that the resulting mean transmitted flux matches the observed redshift evolution from \citep{Viel_2013_WDM}. Specifically, we adopt the effective optical depth
\begin{equation}\label{eq:tau_viel}
    \tau_{\rm eff}=\begin{cases}
        0.751\left(\frac{1+z}{4.5}\right)^{2.90}-0.132, & z\leq4.5\\
        2.260\left(\frac{1+z}{6.2}\right)^{4.91}, & z>4.5.
    \end{cases}
\end{equation}
The normalized flux spectrum is then $F_{\alpha}=e^{-A_{\rm eff}\tau_{\alpha}}$. The rescaling factor is determined iteratively for each model such that the mean transmitted flux satisfies $\langle F_{\alpha}\rangle=\exp(-\tau_{\rm eff})$.

We compute the three-dimensional (3D) power spectrum of the estimator defined as the normalized flux spectrum fluctuations around its mean (averaged over all LOS within the simulation), i.e.
\begin{equation}\label{eq:P3D_estimator}
\delta_{\mathrm{F}}=\frac{F_{\alpha}}{\left\langle F_{\alpha}\right\rangle}-1.
\end{equation}
We construct the three-dimensional flux fluctuation field by placing individual skewers on a regular grid in the transverse plane, with the line-of-sight direction corresponding to the radial (velocity) coordinate. The resulting field $\delta_{\mathrm F}(\mathbf{r})$ is defined in terms of the comoving position $\mathbf{r} = (r_\perp, r_\parallel)$. We define the Fourier-space coordinates as $\mathbf{k} = (k_\perp, k_\parallel)$, where $k_\parallel$ and $k_\perp$ denote the components parallel and perpendicular to the line of sight, respectively. It is convenient to introduce $\mu \equiv |k_\parallel| / k$, where $k = |\mathbf{k}|$, such that the power spectrum can be expressed as $P_{\mathrm{3D}, \alpha}(k,\mu)$, where $\mu$ encodes the anisotropy with respect to the LOS. 

\begin{figure}[tbp]
\centering 
\includegraphics[width=1\textwidth]{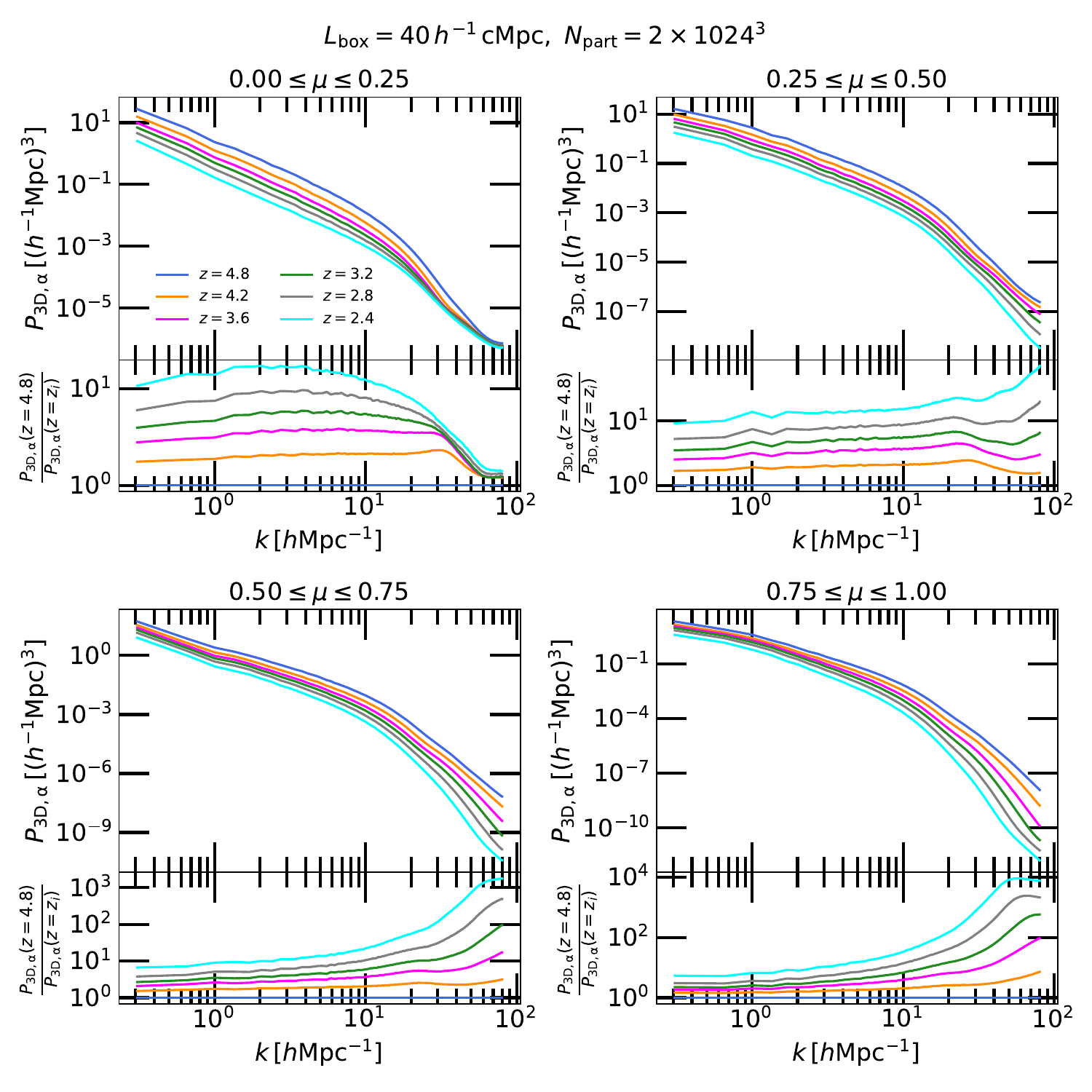}
\hfill
\caption{\label{fig:P3D_vsz}Ly$\alpha$ forest flux 3D power spectrum based on our fiducial 40\_1024 simulation and its redshift evolution (indicated by different colours) from $z=4.8$ to $2.4$. A ratio of $P_{\rm 3D,\alpha}$ at different $z$ relative to $z=4.8$ is shown too. 
}
\end{figure}

We compute its Fourier transform as
\begin{equation}\label{eq:fourier_transform}
\tilde{\delta}_{\mathrm F}(\mathbf{k}) = \int d^3 r\, \delta_{\mathrm F}(\mathbf{r}) e^{-i \mathbf{k} \cdot \mathbf{r}},
\end{equation}
where $\mathbf{k} = (k_\perp, k_\parallel)$ denotes the wavevector perpendicular and parallel to the line of sight. The three-dimensional flux power spectrum is defined through
\begin{equation}\label{eq:PS_definition}
\left\langle \tilde{\delta}_{\mathrm F}(\mathbf{k}) \tilde{\delta}_{\mathrm F}^*(\mathbf{k}') \right\rangle
= (2\pi)^3 \delta_D(\mathbf{k} - \mathbf{k}') P_{\mathrm{3D}, \alpha}(k,\mu),
\end{equation}
where $\delta_D$ is the Dirac delta function and $P_{\mathrm{3D}, \alpha}(k,\mu) \equiv P_{\mathrm{3D}, \alpha}(k_\parallel, k_\perp)$ reflects the anisotropy induced by redshift-space distortions (RSD). We bin the measurements into logarithmic $k$ bins of width $0.25\,\mathrm{dex}$ and four $\mu$ bins of width $0.25$, and adopt the same binning throughout this work.

Figure~\ref{fig:P3D_vsz} presents the evolution of $P_{\mathrm{3D},\alpha}$ for our fiducial simulation ($L_{\rm box}=40\,h^{-1}\,\mathrm{cMpc}$, $N_{\rm part}=2\times1024^3$) over the redshift range $2.4\leq z\leq4.8$. The upper panel of each subplot shows the measured power spectrum, while the lower panel presents the ratio relative to the highest-redshift output at $z=4.8$. The amplitude of $P_{\mathrm{3D},\alpha}$ increases towards lower redshift over most of the $k$-range, reflecting the continued growth of matter density fluctuations in the post-reionization IGM. The evolution is strongest for modes predominantly aligned with the LOS ($\mu\rightarrow1$), particularly on small scales, where the power increases by several orders of magnitude between $z=4.8$ and $z=2.4$. 


The angular dependence of the evolution reflects the different physical mechanisms that suppress small-scale power. For predominantly transverse modes ($\mu\lesssim0.25$), the characteristic cut-off occurs at nearly the same scale at all redshifts, indicating that it is governed primarily by the isotropic pressure-smoothing scale, which evolves only weakly over the redshift range considered. In contrast, for LOS-dominated modes thermal broadening and peculiar-velocity gradients increasingly suppress the small-scale power, leading to a much stronger redshift evolution. On large scales, redshift-space distortions (Kaiser effect) enhance the power along the LOS. This overall behaviour is consistent with previous numerical studies of the Ly$\alpha$ forest 3D power spectrum \citep[e.g.][]{Arinyo-i-Prats_2015,Chabanier_2024}.

Note that in practical observations, finite spectral resolution and pixelization introduce a window function that suppresses power on small scales, while the discrete sampling of the field by a finite number of skewers leads to an additional contribution analogous to shot noise that depends on the transverse skewer density. For completeness, and to facilitate comparison with previous Ly$\alpha$ forest studies based on one-dimensional statistics, the 1D flux power spectrum is related to the 3D power spectrum via
\begin{equation}\label{eq:P1D-P3D}
P_{1\mathrm D, \alpha}(k_\parallel) = \int \frac{d^2 k_\perp}{(2\pi)^2} P_{\mathrm{3D}, \alpha}(k_\parallel, k_\perp).
\end{equation}

The statistical uncertainties on the Ly$\alpha$ forest 3D power spectrum were estimated assuming a Gaussian covariance matrix. In this approximation, the covariance matrix is assumed to be diagonal, with the variance of each $(k,\mu)$ bin given by
\begin{equation}
    C_{ii}=2N_i^{-1}P_i^2,
\end{equation}
where $P_i$ is the measured Ly$\alpha$ forest power spectrum and $N_i$ is the number of independent Fourier modes contributing to the $i$-th bin. The $1\sigma$ uncertainties shown in Fig.~\ref{fig:P3D_AiPfit} (orange) correspond to the square root of the diagonal elements of this covariance matrix.\footnote{We also estimated the covariance using a jackknife resampling procedure and found no significant difference in the inferred best-fitting parameters or their uncertainties. We therefore adopt the Gaussian covariance throughout this work.}


\subsection{Reducing the effect of cosmic variance}\label{sec:ZCV}

The convergence tests presented in this work are based on individual simulations of finite volume and are therefore affected by sample (cosmic) variance. This is particularly important on the largest scales, where the number of available Fourier modes is small and the resulting fluctuations can obscure genuine differences between simulation configurations. To reduce this source of uncertainty we employ the Zel'dovich control variate (ZCV) technique \citep{Kokron_2022,DeRose_2023}, recently applied to the Ly$\alpha$ forest by \citep{Hadzhiyska_2025}. Throughout this work, the ZCV correction is applied to all comparisons between different simulation configurations, unless stated otherwise.

The ZCV method reduces sample variance by exploiting a computationally inexpensive field that is highly correlated with the observable of interest and whose ensemble mean is known analytically. As the control field we use the Zel'dovich approximation (ZA) matter field, generated from the same initial conditions as each hydrodynamical simulation using the \textsc{N-GenIC} code\footnote{\url{https://www.h-its.org/2014/11/05/ngenic-code/}} \citep{Springel_2005,Angulo_2012}. The corrected Ly$\alpha$ forest 3D power spectrum is then given by
\begin{equation}\label{eq:ZCV}
P_{\mathrm{3D},\alpha}^{\mathrm{ZCV}} = P_{\mathrm{3D},\alpha}^{\mathrm{sim}} - \beta_{\rm ZCV}\left(P_{\mathrm{ZA}}^{\mathrm{sim}} - P_{\mathrm{ZA}}\right),
\end{equation}
where $P_{\mathrm{ZA}}^{\mathrm{sim}}$ is the power spectrum measured from the ZA realization and $P_{\mathrm{ZA}}$ is its sample-variance-free ensemble mean, computed using the \texttt{ZeNBu} code\footnote{\url{https://github.com/sfschen/ZeNBu}}.

The coefficient $\beta_{\rm ZCV}$ is, in principle, arbitrary. We adopt the optimal value that minimises the variance of the corrected estimator,
\begin{equation}
\beta_{\rm ZCV} \equiv \beta_{\rm ZCV}^\star =
\frac{\mathrm{Cov}\left[P_{\mathrm{3D},\alpha}^{\mathrm{sim}}, P_{\mathrm{ZA}}^{\mathrm{sim}}\right]} {\mathrm{Var}\left[P_{\mathrm{ZA}}^{\mathrm{sim}}\right]},
\label{eq:beta_equation}
\end{equation}
which follows directly from the standard control-variate formalism. We estimate the covariance and variance entering Eq.~\ref{eq:beta_equation} using the Gaussian covariance approximation. Following \cite{DeRose_2023}, the resulting $\beta_{\rm ZCV}(k,\mu)$ is smoothed using a third-order Savitzky--Golay filter with a window length of 21 bins before applying the correction, and is subsequently kept fixed. This smoothing suppresses the small bias that can arise when $\beta_{\rm ZCV}$ is estimated from the same realization to which the control-variate correction is applied. With this choice, the strong correlation between the non-linear Ly$\alpha$ forest field and the ZA control field is used to subtract the correlated component of the sample variance, substantially reducing the uncertainties on large scales while leaving the estimator unbiased.

In App.~\ref{app:ZCV_Lbox} we demonstrate that the ZCV correction significantly improves the convergence of $P_{\mathrm{3D},\alpha}$ with simulation volume, allowing the underlying numerical trends to be distinguished from realization-to-realization fluctuations. The same implementation is adopted in our companion EFT analysis \cite{Autieri_2026}, while further applications of the control-variate technique to Ly$\alpha$ forest statistics and covariance estimation are presented by \citep{Hadzhiyska_2025,Hadzhiyska_2026}.

\subsection{Analytical form fitting}\label{sec:P3D_analytic}

\begin{figure}[tbp]
\centering 
\includegraphics[width=1\textwidth]{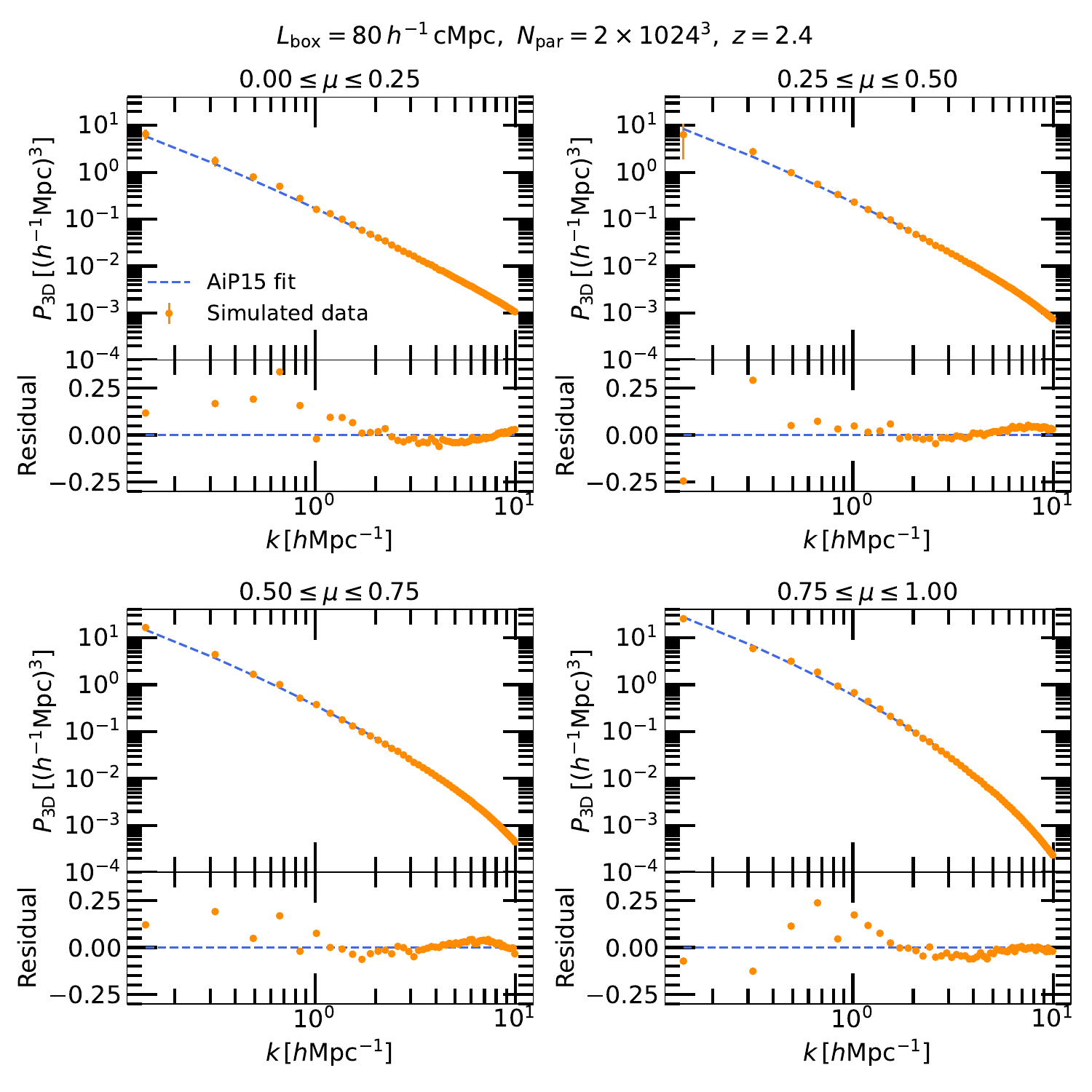}
\hfill
\caption{\label{fig:P3D_AiPfit}Analytic function fitting procedure. Simulated Ly$\alpha$ forest flux 3D power spectrum from the $80\_1024$ simulation at $z=2.4$ after the ZCV correction with uncertainties derived from the Gaussian covariance matrix (orange dots). The dashed blue curves are the best-fit functions based on AiP15 model. Different $\mu$ bins are presented in individual panels with the bottom parts presenting fractional residuals.}
\end{figure}

To interpret the measured three-dimensional flux power spectrum, we adopt a parametric model that separates linear-theory contributions from non-linear corrections, following the formalism introduced by \citep{Arinyo-i-Prats_2015} and subsequently adopted in a range of simulation-based and observational analyses of the Ly$\alpha$ forest power spectrum (e.g. \citep{duMasdesBourboux_2020, deSainteAgathe_2019, Chabanier_2024, Karim_2024}). In this approach, the flux power spectrum is written as
\begin{equation}\label{eq:P3D_AiP}
    P_{\rm{AiP15}}(k,\mu) = (b_{\rm{F}} + b_{\eta} f\mu^2)^2 P_{\rm{lin}}(k)\, D_{\mathrm{NL}}(k,\mu).
\end{equation}
where $P_{\rm{lin}}(k)$ is the linear matter power spectrum computed using \textsc{CAMB}\footnote{\url{https://camb.readthedocs.io/en/latest/}} software \citep{Lewis_2011_CAMB,Lewis_2013_CAMB}. An example of simulated $P_{\mathrm{3D},\alpha}$ and fitted $P_{\mathrm{AiP15}}$ from the $L_{\rm box}=80\,h^{-1}\,\rm cMpc$ and $N_{\rm part}=2\times1024^3$ simulation at $z=2.4$ is shown in Fig.~\ref{fig:P3D_AiPfit}. Furthermore, $f$ is the linear growth rate, and $b_{\rm{F}}$ and $b_{\eta}$ are bias parameters describing the response of the flux contrast to density and velocity-gradient fluctuations, respectively. These are defined as
\begin{equation}\label{eq:bias_factors}
    b_{\rm{F}} = \frac{\partial \delta_{\rm{F}}}{\partial \delta},\quad
    b_{\eta} = \frac{\partial \delta_{\rm{F}}}{\partial \eta},\quad
    \eta = -\frac{1}{aH}\frac{\partial v_{\rm pec}}{\partial r_\parallel},
\end{equation}
where $\delta_{\rm{F}}$ is the transmitted flux fluctuation, $\delta$ is the matter overdensity, and $\eta$ is the dimensionless velocity gradient along the LOS. This form captures the large-scale linear behaviour, including redshift-space distortions analogous to the Kaiser effect.

On smaller scales, non-linear structure formation, thermal broadening, and peculiar velocities introduce significant deviations from linear theory. These effects are incorporated through a multiplicative correction factor. Following \citep{Arinyo-i-Prats_2015}, the non-linear term is parameterized as
\begin{equation}\label{eq:P3D_nlfactor}
    D_{\mathrm{NL}}(k,\mu)=\exp\left\{\left[q_1\Delta^2(k)+q_2\Delta^4(k)\right]\left[1-\left(\frac{k}{k_{\mathrm{v}}}\right)^{a_{\mathrm{v}}}\mu^{b_{\mathrm{v}}}\right]-\left(\frac{k}{k_{\mathrm{p}}}\right)^2\right\},
\end{equation}
where
\begin{equation}
    \Delta^2(k) = \frac{k^3}{2\pi ^2}P_{\mathrm{lin}}(k).
\end{equation}
This parametrization captures three key physical effects. First, the terms proportional to $\Delta^2(k)$ and $\Delta^4(k)$ describe the enhancement of power due to non-linear growth near the transition from linear to non-linear scales. Second, the anisotropic suppression term depending on $\mu$ accounts for LOS velocity effects and thermal broadening, which dampen fluctuations more strongly along the LOS. Finally, the exponential cutoff controlled by $k_{\mathrm{p}}$ represents the isotropic suppression of power below the effective Jeans scale due to gas pressure smoothing.

Note that the first term in Eq.~\ref{eq:P3D_AiP} is equivalent to the standard form $b_{\rm F}^2(1+\beta_{\rm Kaiser} \mu^2)^2$, with $\beta_{\rm Kaiser} = b_\eta f / b_{\rm F}$. This flexible analytical form has been shown to provide an accurate description of the three-dimensional Ly$\alpha$ forest power spectrum measured in hydrodynamical simulations \citep{Arinyo-i-Prats_2015}, and has been widely adopted in both simulation-based studies and observational analyses. In particular, it provides a convenient separation between the large-scale cosmological contribution encoded in $P_{\rm{lin}}(k)$ and small-scale astrophysical effects captured by the non-linear parameters.

To constrain the parameters of the analytical model described above, we perform a likelihood-based fit to the simulated $P_{\rm 3D,\alpha}$. The fitting is carried out simultaneously across all $(k,\mu)$ bins, allowing us to capture the full anisotropic information encoded in the data.


The parameters of the analytical model (Eq.~\ref{eq:P3D_AiP}) are constrained using Hamiltonian Monte Carlo (HMC). We sample the posterior distribution with the No-U-Turn Sampler (NUTS) algorithm \citep{Hoffman_2011}, implemented in the \texttt{NumPyro} library \citep{Bingham_2018, Phan_2019}. Assuming Gaussian-distributed uncertainties, the likelihood is
\begin{equation}
    -2\ln\mathcal{L} = \Delta P_{\mathrm{3D},\alpha}^{\mathrm T}\cdot C^{-1}\cdot\Delta P_{\mathrm{3D},\alpha},
    \label{eq:likelihood}
\end{equation}
where $\Delta P_{\mathrm{3D},\alpha}\equiv P_{\rm AiP15}-P_{\mathrm{3D},\alpha}^{\mathrm{sim}}$ is the difference between the model prediction and the measured Ly$\alpha$ forest 3D power spectrum. The covariance matrix $C$ is computed assuming Gaussian statistics, as described in Section~\ref{sec:P3D_sims}.

We adopt wide uniform (top-hat) priors on all free model parameters. The prior ranges are adjusted between fits when necessary to improve sampling efficiency, while remaining sufficiently broad that the posterior distributions are not artificially truncated by the prior boundaries. We fix $q_2=0$ and adopt $k_{\max}=10\,h\,\mathrm{cMpc}^{-1}$ as our fiducial fitting choices. The sampling configuration is adjusted between individual fits as needed to achieve efficient convergence of the posterior distributions. The parameter space is first explored using a gradient-based minimization of $\chi^2$ to obtain a maximum-likelihood estimate, which is then used to initialize the NUTS sampler. The final parameter constraints are obtained from the marginalized posterior distributions, quoting the median and the 16th and 84th percentiles as credible intervals. In the case of our $80\_1024$ model at $z=2.4$ this results in $b_{\rm F}=-0.1016^{+0.0010}_{-0.0009}$, $b_{\eta}=-0.1707^{+0.0030}_{-0.0031}$, $q_1=1.034\pm0.018$, $q_2=0$, $k_{\mathrm{v}}=1.635^{+0.062}_{-0.065}\,h\,\rm cMpc^{-1}$, $a_{\mathrm{v}}=0.487\pm0.012$, $b_{\mathrm{v}}=1.540\pm0.006$ and $k_{\rm p}=15.84^{+0.21}_{-0.23}\,h\,\rm cMpc^{-1}$, with the fitted analytic function indicated by the dashed blue curves in Fig.~\ref{fig:P3D_AiPfit}.


\begin{figure}[tbp]
\centering 
\begin{minipage}{\textwidth}
\centering 
\includegraphics[width=.49\textwidth]{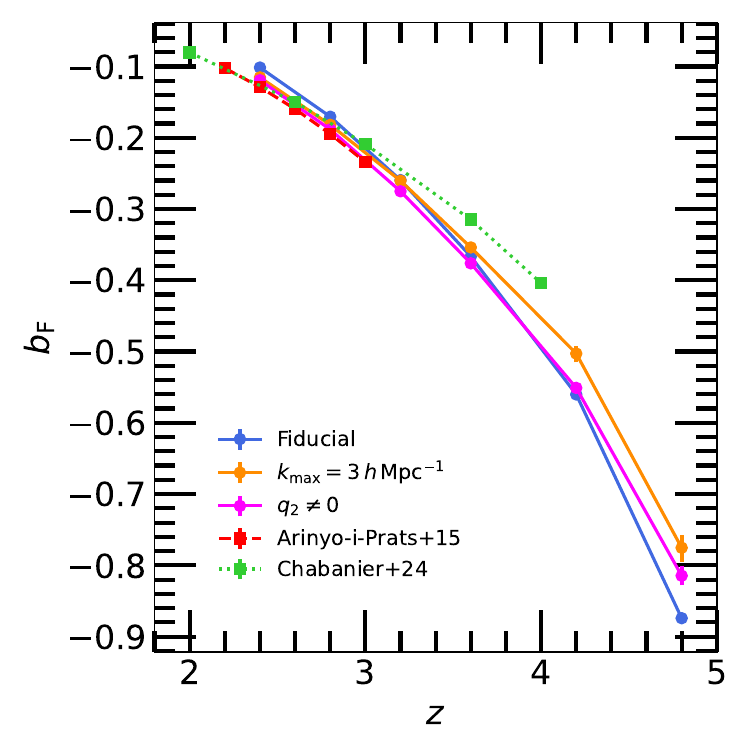}
\includegraphics[width=.49\textwidth]{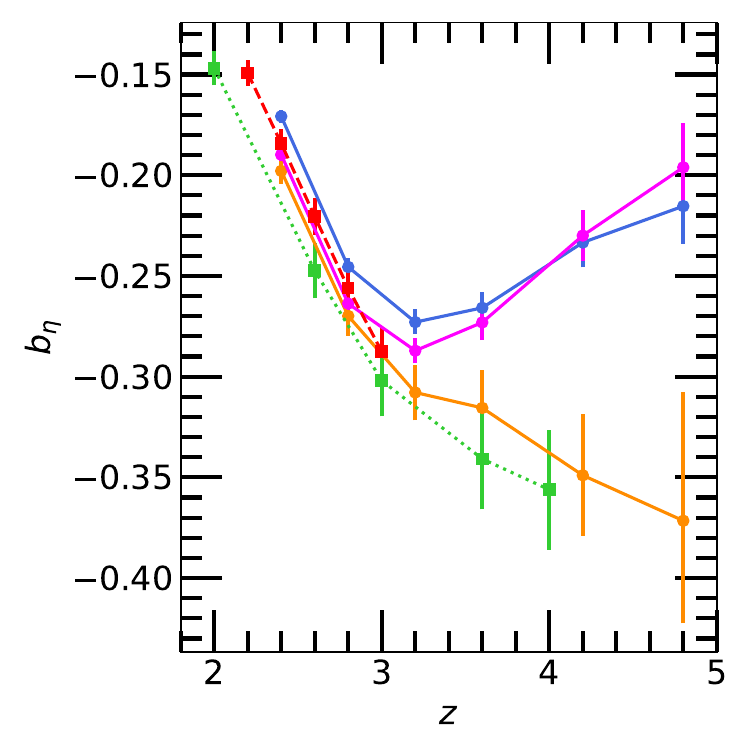}
\end{minipage}
\hfill
\caption{\label{fig:bias_zevo_test_kmax_q2}The redshift evolution of the best-fit Ly$\alpha$ forest flux bias, $b_{\rm F}$ (left), and velocity-gradient bias, $b_{\eta}$ (right), for the $L_{\rm box}=80\,h^{-1}\,\mathrm{cMpc}$ and $N_{\rm part}=2\times1024^3$ simulation. Our fiducial fits, with $k_{\max}=10\,h\,\mathrm{cMpc}^{-1}$ and $q_2=0$, are shown by the solid blue curves. We compare these with fits restricted to $k_{\max}=3\,h\,\mathrm{cMpc}^{-1}$ (solid orange curves) and fits in which $q_2$ is allowed to vary freely (solid pink curves). For comparison, the results from the Fiducial simulation of \cite{Arinyo-i-Prats_2015} and the 160R25 simulation of \cite{Chabanier_2024} are shown by the dashed red and dotted green curves, respectively.}
\end{figure}

We test the robustness of the inferred bias parameters to our choices of $q_2$ and fitting range in Fig.~\ref{fig:bias_zevo_test_kmax_q2}. Allowing $q_2$ to vary does not qualitatively alter the redshift evolution of either $b_{\rm F}$ or $b_{\eta}$, although the preferred values of several AiP15 parameters shift due to parameter degeneracies. In contrast, the inferred $b_{\eta}$ is substantially more sensitive to the range of scales included in the fit. Restricting the analysis to $k_{\max}=3\,h\,\mathrm{cMpc}^{-1}$ removes the high-redshift upturn found in our fiducial $k_{\max}=10\,h\,\mathrm{cMpc}^{-1}$ fits and brings the inferred evolution into better agreement with \cite{Arinyo-i-Prats_2015} and \cite{Chabanier_2024}. The restricted fits also agree well with the accompanying minimal effective-field-theory (mEFT) analysis of the same simulations \cite{Autieri_2026}. Inspired by EFT approaches to Ly$\alpha$ forest clustering \citep{Ivanov_2024,Chadaykin_2025,deBelsunce_2025}, the mEFT model describes the Ly$\alpha$ flux auto-power spectrum and its cross-power spectrum with the dark-matter density field using a tree-level bias model supplemented by the leading counterterms and stochastic contributions. The accompanying work explores the fidelity of this approach, including the usable $k$-range and number of fitting parameters. The change in $b_{\eta}$ with the AiP15 fitting range, together with the systematic shifts in the non-linear parameters, highlights degeneracies between the large-scale bias and non-linear model parameters when modes deep in the non-linear regime are included. Since the AiP15 model is a phenomenological description calibrated over a broad range of scales, the bias parameters inferred from such full-shape fits should likewise be interpreted as effective model parameters rather than strictly large-scale bias measurements. These degeneracies may also be relevant for cosmological full-shape analyses based on the AiP15 parametrization, since shifts in the fitted bias and non-linear parameters could propagate into inferred cosmological parameters. We nevertheless retain $k_{\max}=10\,h\,\mathrm{cMpc}^{-1}$ as our fiducial choice because our primary aim is to characterize $P_{\rm 3D,\alpha}$ over the full range of scales considered here. We investigate the dependence of the AiP15 results on $q_2$ and $k_{\max}$ in more detail in Appendix~\ref{app:kmaxq2_test}.

\section{Simulation convergence tests}\label{sec:convergence_tests}

In this section we investigate the numerical convergence of the Ly$\alpha$ forest 3D power spectrum. We first study the dependence on simulation box size (Sec.~\ref{sec:conv_box_size}), followed by mass resolution (Sec.~\ref{sec:conv_res}). We then derive a correction for finite mass resolution (Sec.~\ref{sec:res_cor}) and finally investigate the resolution of the extracted grid used to construct the flux field (Sec.~\ref{sec:conv_grid}).

\begin{figure}[tbp]
\centering 
\includegraphics[width=.9\textwidth]{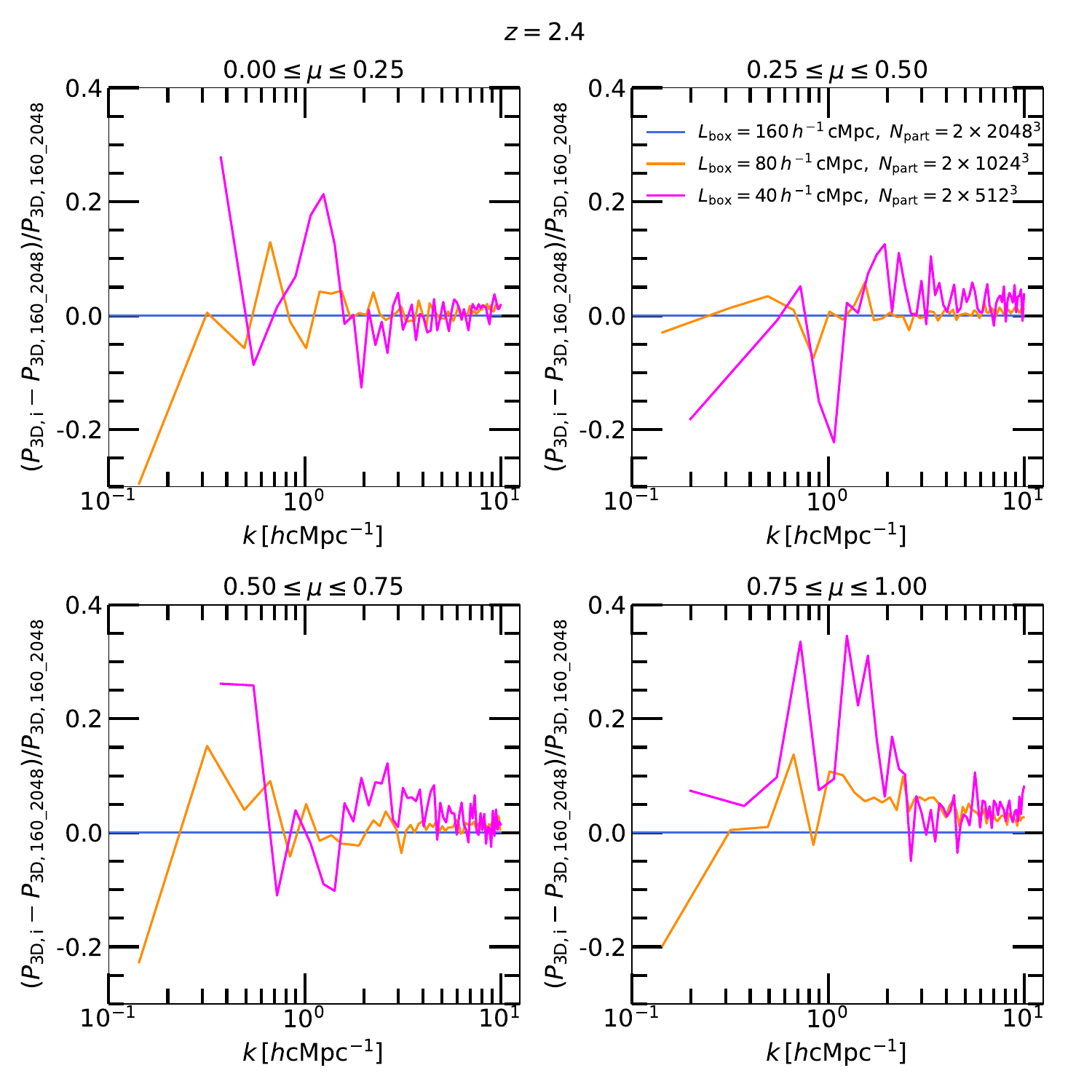}
\hfill
\caption{\label{fig:P3D_vsLbox} Simulation volume convergence test of the Ly$\alpha$ forest flux 3D power spectrum at $z=2.4$ in terms of fractional residuals compared to the $L_{\mathrm{box}}=160\,h^{-1}\,\mathrm{cMpc}$, $N_{\rm part}=2\times2048^3$ simulation. Different colours represent various $L_{\rm box}$ with adjusted $N_{\rm part}$ such that the mass resolution is fixed. In the high $\mu$-bin the smallest box has up to $\sim35\%$ higher $P_{\rm 3D,\alpha}$ at the $0.7\,h\,\mathrm{cMpc}^{-1}\lesssim k \lesssim 2\,h\,\mathrm{cMpc}^{-1}$ than the largest simulated box.}
\end{figure}

\begin{figure}[tbp]
\centering 
\includegraphics[width=.49\textwidth]{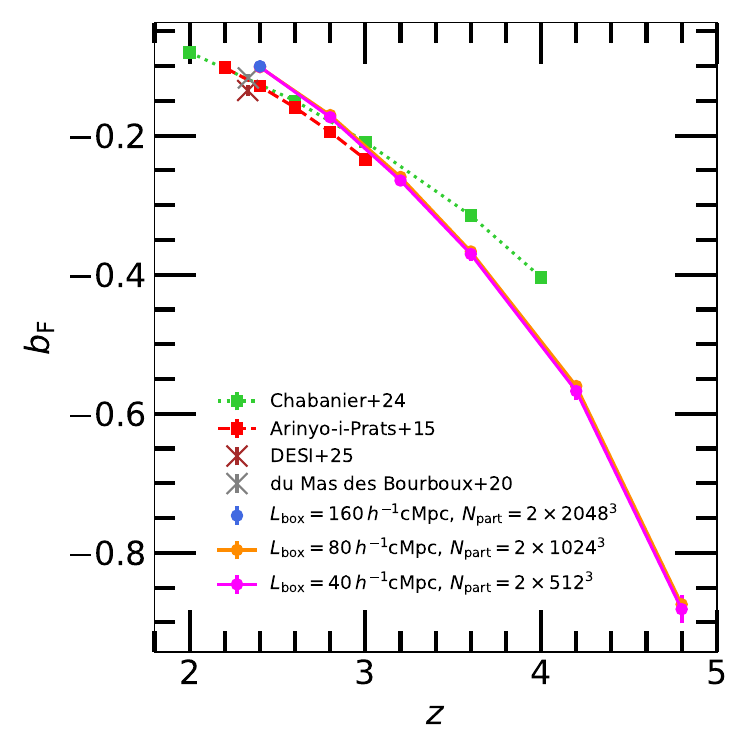}
\hfill
\includegraphics[width=.49\textwidth]{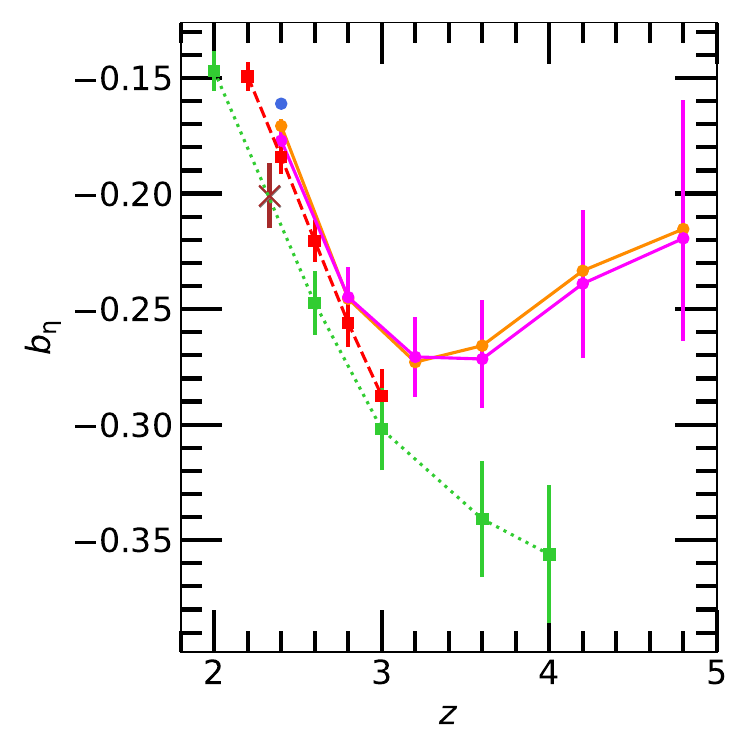}
\caption{\label{fig:bias_zevo_vsLbox} The redshift evolution of the Ly$\alpha$ forest flux (left panel) and velocity (right panel) bias. Our simulations are shown as solid curves with colours corresponding to the models presented in Fig.~\ref{fig:P3D_vsLbox}, i.e. different box sizes and numbers of particles such that the mass resolution is kept constant. The crosses are measured values by \cite{duMasdesBourboux_2020} (grey) and \cite{DESI_2025} (brown). The dotted curves are taken from simulation-based works of \cite[Fiducial,][]{Arinyo-i-Prats_2015} (red) and \cite[160R25,][]{Chabanier_2024} (green).}
\end{figure}

\subsection{The effect of simulation box size on $P_{\rm 3D,\alpha}$}\label{sec:conv_box_size}

Firstly, we test the effect of the simulation box size, $L_{\rm box}$. To isolate this effect as much as possible, we keep the mass resolution constant by decreasing the number of particles in the simulation when decreasing $L_{\rm box}$ such that $M_{\rm DM}=3.44\times10^7\,h^{-1}\,\rm M_{\odot}$ and $M_{\rm gas}=6.38\times10^6\,h^{-1}\,\rm M_{\odot}$ are fixed. In Fig.~\ref{fig:P3D_vsLbox} we compare $L_{\rm box}=40\,h^{-1}\rm cMpc$ (pink) and $80\,h^{-1}\rm cMpc$ (orange) to the largest volume considered, specifically $160\,h^{-1}\rm cMpc$ (blue). After applying the Zel'dovich control variate correction, the fractional residuals show broad consistency between the different volumes over the full range of scales for $\mu\leq0.75$. The main exception is the most line-of-sight-aligned bin, $0.75\leq\mu\leq1.00$, where the $40\,h^{-1}\,\rm cMpc$ simulation differs from the $160\,h^{-1}\,\rm cMpc$ simulation by up to $\sim35\%$ at $k\lesssim3\,h\,\rm cMpc^{-1}$.

Since each box size is represented by a single realization, the residuals in Fig.~\ref{fig:P3D_vsLbox} should be interpreted as a combination of finite-volume effects and realization variance, rather than as a pure box-size convergence test. Smaller boxes miss long-wavelength density and velocity modes, contain fewer independent Fourier modes in each low-$k$ bin and sample a smaller range of cosmic environments. These effects can lead to coherent offsets in the measured $P_{\rm 3D,\alpha}$, especially for modes close to the line of sight, where peculiar velocities and redshift-space distortions play a larger role. The remaining $\sim35\%$ discrepancy in the highest-$\mu$ bin is therefore consistent with residual sensitivity to limited volume and sample variance. At $k\gtrsim3\,h\,\mathrm{cMpc}^{-1}$ in this bin, and across all scales for lower-$\mu$ bins, the agreement improves substantially, indicating that the ZCV-corrected $P_{\rm 3D,\alpha}$ is relatively insensitive to the simulation volume for these modes.

The impact of the simulation volume on the fitted bias parameters is shown in Fig.~\ref{fig:bias_zevo_vsLbox}. For comparison, we also show the Fiducial simulation of \cite{Arinyo-i-Prats_2015} (dashed red curves), with $L_{\rm box}=60\,h^{-1}\,\mathrm{cMpc}$ and $512^3$ particles, and the $160\mathrm{R}25$ simulation of \cite{Chabanier_2024} (dotted green curves), with $L_{\rm box}=160\,h^{-1}\,\mathrm{cMpc}$ and a spatial resolution of $25\,h^{-1}\,\mathrm{ckpc}$, together with observational measurements from \cite{duMasdesBourboux_2020} and \cite{DESI_2025}. The three Sherwood simulations produce nearly identical values of the flux bias, $b_{\rm F}$, over the redshift range where they overlap, with differences typically below a few per cent. Our measurements are also broadly consistent with previous simulation-based results, particularly at $z\lesssim3.2$, although some differences emerge towards higher redshift. Differences between the simulation suites may additionally arise from their different thermal and reionization histories and hydrodynamical implementations.

The velocity-gradient bias, $b_{\eta}$, shows a stronger sensitivity to simulation volume than $b_{\rm F}$, with differences reaching $\sim14\%$ at the highest redshifts where multiple volumes are available. Nevertheless, the overall redshift evolution is broadly consistent between the different volumes. Our fiducial fits exhibit a minimum around $z\sim3.4$ followed by an upturn towards less negative $b_{\eta}$ at higher redshift, whereas \cite{Chabanier_2024} find a continued evolution towards more negative values. As discussed in Section~\ref{sec:P3D_analytic} and investigated further in Appendix~\ref{app:kmaxq2_test}, this difference is primarily associated with the range of scales included in the AiP15 fit rather than with simulation volume. The best-fit values of both bias parameters across the simulations and redshifts considered are listed in Appendix~\ref{app:parameter_tables}, particularly in Table~\ref{tab:parameters_bias}.

The fitted non-linear parameters show a moderate dependence on simulation volume, with the magnitude varying between parameters. Comparing the $80\,h^{-1}\,\mathrm{cMpc}$ and $40\,h^{-1}\,\mathrm{cMpc}$ simulations, the differences are at most $\sim3.4\%$ for $q_1$, $\sim13\%$ for $a_{\rm v}$, $\sim19\%$ for $k_{\rm v}^{a_{\rm v}}$, $\sim3.3\%$ for $b_{\rm v}$, and $\sim2.3\%$ for $k_{\rm p}$ over the redshift range considered. At $z=2.4$, where the $160\,h^{-1}\,\mathrm{cMpc}$ simulation is also available, its parameters are similarly consistent with those of the $80\,h^{-1}\,\mathrm{cMpc}$ simulation, with differences of $\lesssim8\%$. The redshift evolution of these parameters is shown in Fig.~\ref{fig:DNL_Lbox}, while their best-fitting values are listed in Table~\ref{tab:parameters_DNL}, both in Appendix~\ref{app:parameter_tables}.

\begin{figure}[tbp]
\centering 
\includegraphics[width=.9\textwidth]{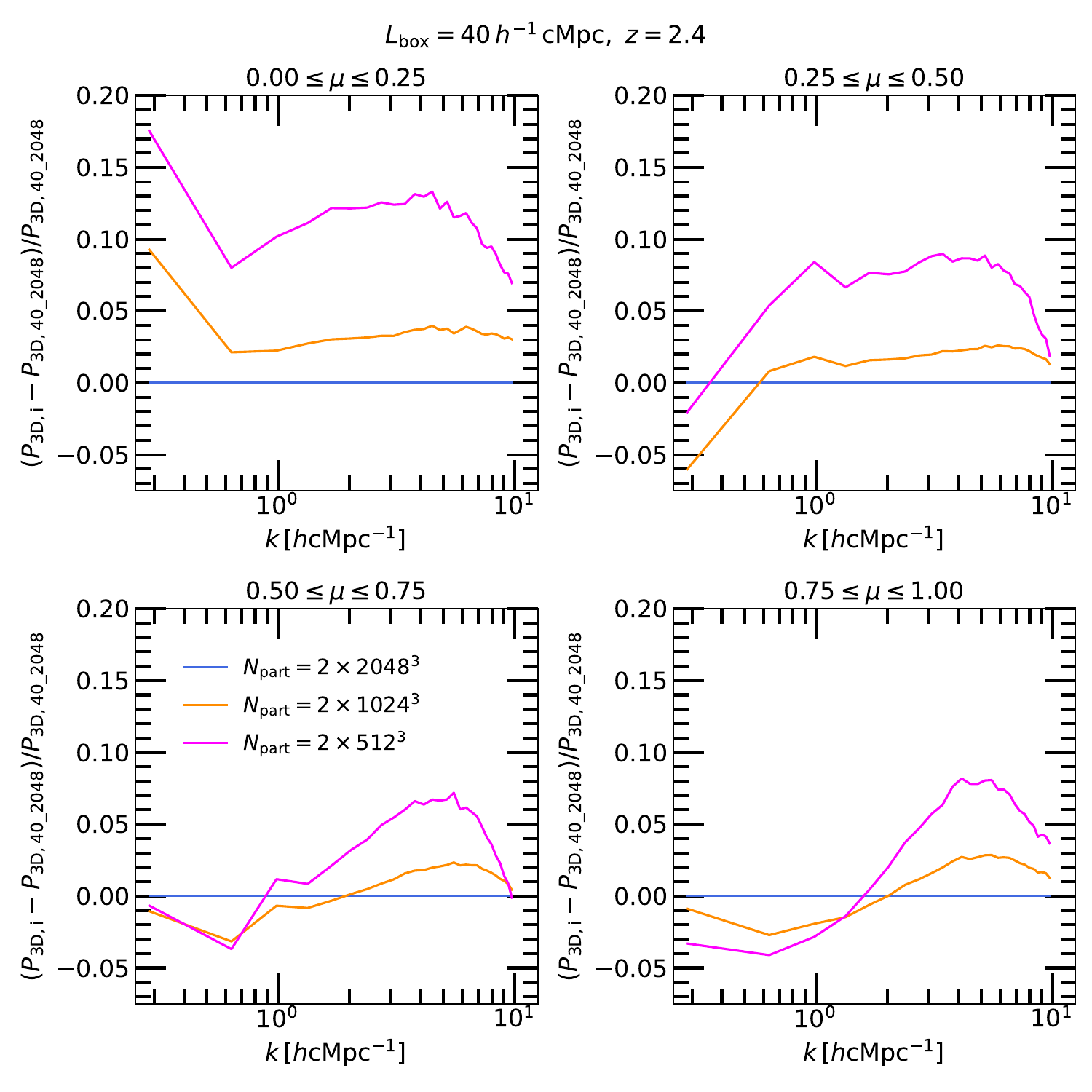}
\hfill
\caption{\label{fig:P3D_vsNpar} Simulation mass-resolution convergence test. Same as Fig.~\ref{fig:P3D_vsLbox}, but fixing the simulation volume to $L_{\mathrm{box}}=40\,h^{-1}\,\mathrm{cMpc}$ and varying the mass resolution. At $z=2.4$, the lowest-resolution simulation ($N_{\rm part}=2\times512^3$) differs from the highest-resolution model by up to $\sim13\%$, with the largest differences occurring in the lowest-$\mu$ bin. The intermediate-resolution simulation ($N_{\rm part}=2\times1024^3$) shows substantially smaller differences of up to $\sim4\%$ in this bin, while the scale dependence becomes more complex towards larger $\mu$.}
\end{figure}

\begin{figure}[tbp]
\centering 
\includegraphics[width=.49\textwidth]{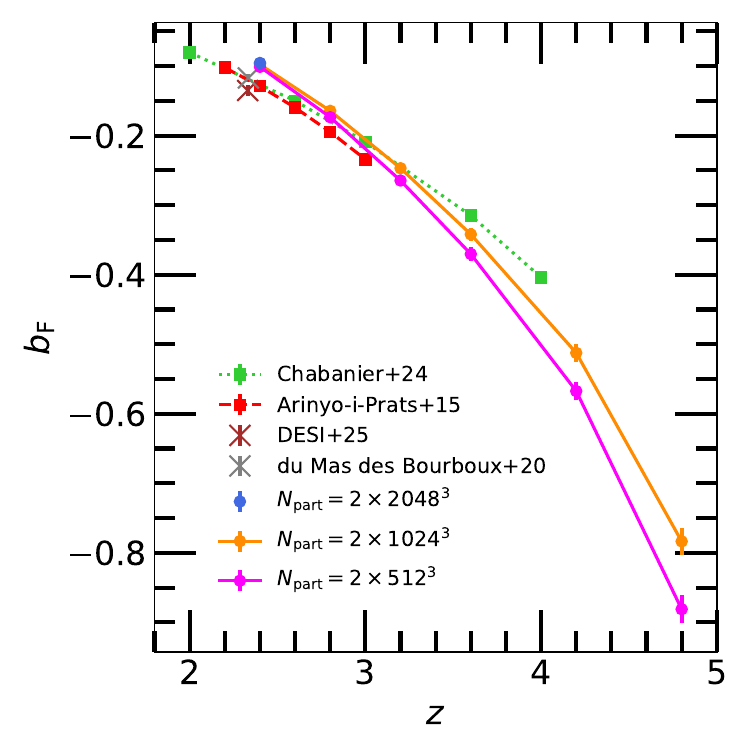}
\hfill
\includegraphics[width=.49\textwidth]{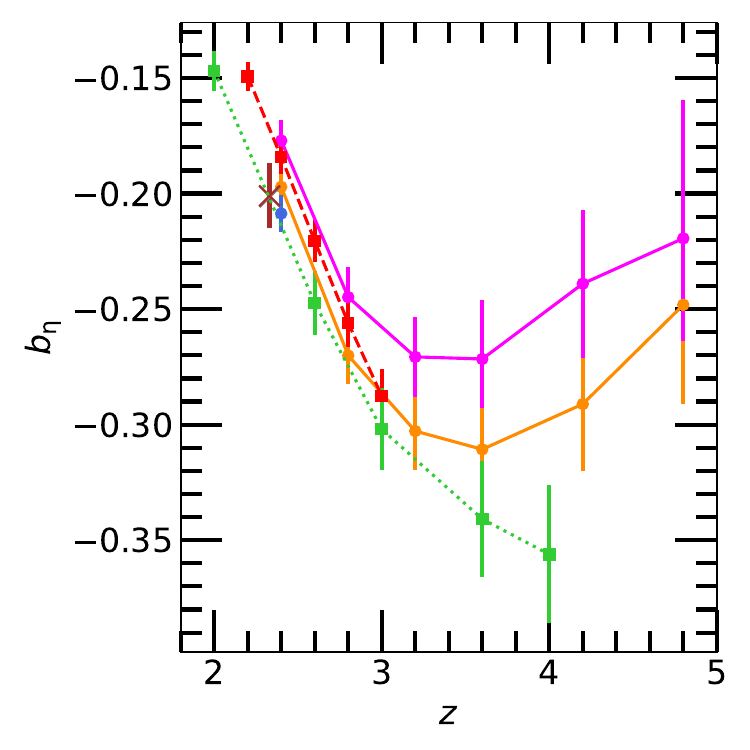}
\caption{\label{fig:bias_zevo_vsNpar} The redshift evolution of the Ly$\alpha$ forest flux (left panel) and velocity (right panel) bias in our $L_{\rm box}=40\,h^{-1}\,\rm cMpc$ simulations (solid curves) with varying number of particles (i.e. mass resolution), specifically $N_{\rm part}=2\times2048^3$ (blue), $2\times1024^3$ (orange) and $2\times512^3$ (pink). Similarly to Fig.\ref{fig:bias_zevo_vsLbox}, the dotted curves represent results from simulation-based works of \cite[Fiducial,][]{Arinyo-i-Prats_2015} (red) and \cite[160R25,][]{Chabanier_2024} (green) while the crosses indicate measured values by \cite{duMasdesBourboux_2020} (grey) and \cite{DESI_2025} (brown).}
\end{figure}

\subsection{The effect of simulation mass resolution on $P_{\rm 3D,\alpha}$}\label{sec:conv_res}

We next test convergence with respect to mass resolution by fixing the simulation volume to $L_{\rm box}=40\,h^{-1}\,\mathrm{cMpc}$ and varying the number of particles. The resulting fractional residuals are shown in Fig.~\ref{fig:P3D_vsNpar}. In contrast to the box-size test, the effect of mass resolution is visible over a wider range of scales and $\mu$. The lowest-resolution simulation ($N_{\rm part}=2\times512^3$) differs from the highest-resolution model by up to $\sim13\%$ in the lowest $\mu$-bin at $k\gtrsim0.6\,h\,\mathrm{cMpc}^{-1}$, with the power enhanced over essentially the full range of scales considered. The intermediate-resolution simulation ($N_{\rm part}=2\times1024^3$) shows the same general behaviour, but with substantially smaller differences of up to $\sim4\%$. The scale and angular dependence become more complex towards larger $\mu$: the residuals are mildly suppressed at low $k$ and enhanced at intermediate scales, with differences reaching $\sim8\%$ for the lowest-resolution simulation. Overall, the magnitude of the resolution dependence is largest for the most transverse modes.

The fact that the resolution dependence extends well beyond the smallest resolved scales indicates that insufficient mass resolution does not simply modify the power close to the nominal resolution scale. Instead, under-resolving the gas density, temperature, and velocity fields can modify the mapping between the underlying matter distribution and the transmitted Ly$\alpha$ flux over a broad range of scales. The dependence on $\mu$ further indicates that these numerical effects are anisotropic, with the strongest differences generally occurring for predominantly transverse modes.

Overall, the mass-resolution dependence is more pervasive across scales and orientations than the box-size dependence, although the latter reaches larger differences in the most LOS-dominated modes. This is qualitatively consistent with the findings of \citep{Chabanier_2024}, who also found a significant sensitivity of Ly$\alpha$ forest 3D power spectrum modelling to physical resolution. Our Sherwood simulations likewise show that mass resolution affects a broad range of scales and orientations. This dependence is also reflected in the fitted bias parameters shown in Fig.~\ref{fig:bias_zevo_vsNpar}: while $b_{\rm F}$ remains relatively stable at $z\lesssim3$, both $b_{\rm F}$ at high redshift and especially $b_{\eta}$ show noticeable shifts as the mass resolution is degraded.

The non-linear parameters exhibit a somewhat stronger dependence on mass resolution than on simulation volume, consistent with the resolution dependence of $P_{\rm 3D,\alpha}$ and the fitted bias parameters. Comparing the $2\times1024^3$ and $2\times512^3$ simulations, the differences reach up to $\sim13\%$ for $q_1$ and $a_{\rm v}$, $\sim15\%$ for $k_{\rm v}^{a_{\rm v}}$, and $\sim5\%$ for $b_{\rm v}$. The strongest resolution dependence is found for $k_{\rm p}$, for which the difference increases systematically with redshift from $\sim12\%$ at $z=2.4$ to $\sim29\%$ at $z=4.8$. At $z=2.4$, the higher-resolution $2\times2048^3$ simulation shows considerably better agreement with the $2\times1024^3$ model, with differences below $\sim9\%$ for all parameters. The specific values of the non-linear parameters are listed in Appendix~\ref{app:parameter_tables}, particularly in Table~\ref{tab:parameters_DNL}, and visualized in Fig.~\ref{fig:DNL_Npar}.

\subsection{Resolution correction}\label{sec:res_cor}

\begin{figure}[tbp]
\centering 
\includegraphics[width=.9\textwidth]{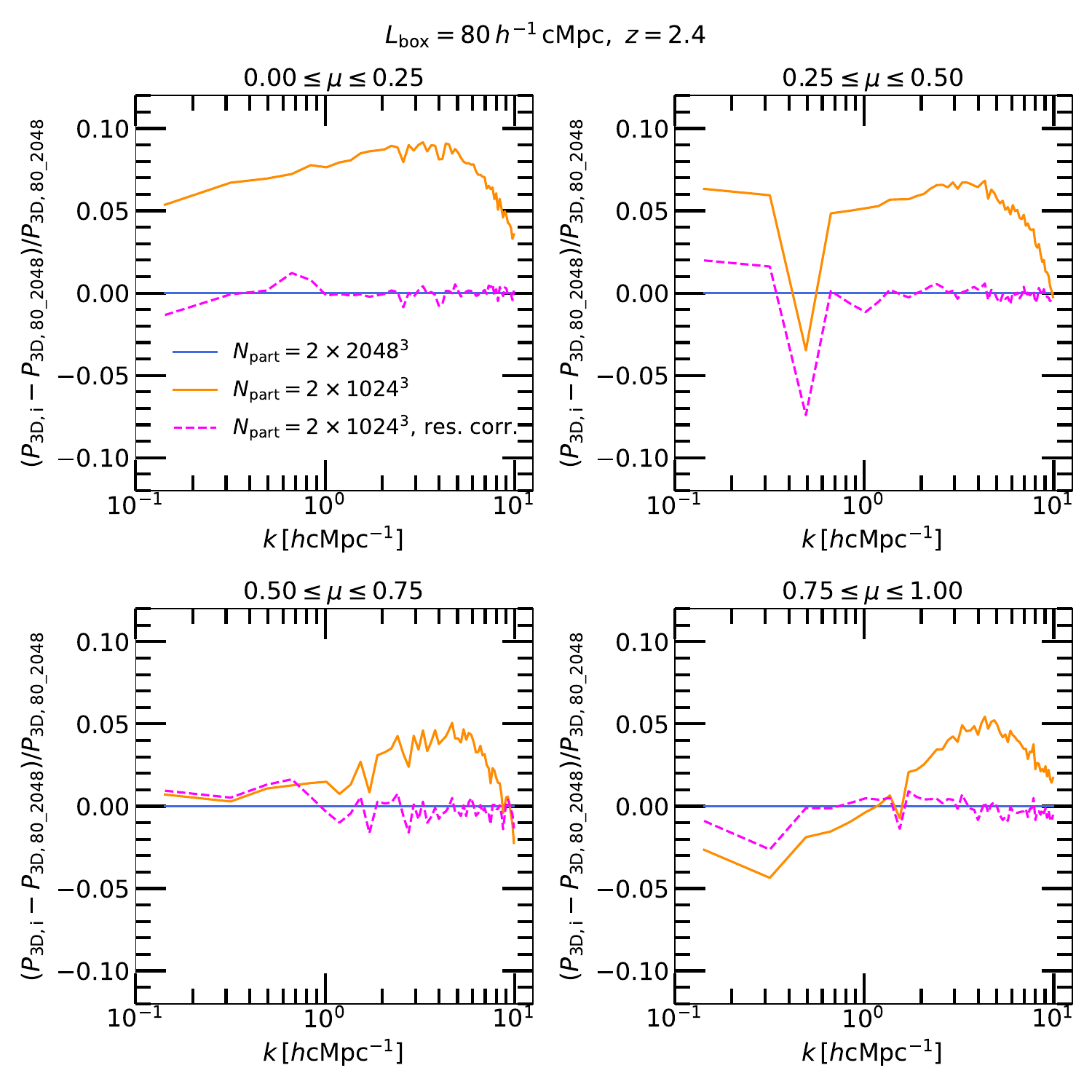}
\hfill
\caption{\label{fig:P3D_rescorr} Resolution corrected Ly$\alpha$ forest flux 3D power spectrum of the $L_{\mathrm{box}}=80\,h^{-1}\,\mathrm{cMpc}$, $N_{\rm part}=2\times1024^3$ simulation (dashed pink curves) compared to the original 3D power spectrum (solid orange curves) and a higher resolution simulation ($N_{\rm part}=2\times2048^3$, solid blue curves). The simulations are taken at $z=2.4$. The fractional residuals are relative to the higher resolution simulation.}
\end{figure}

\begin{figure}[tbp]
\centering 
\begin{minipage}{\textwidth}
\centering 
\includegraphics[width=.32\textwidth]{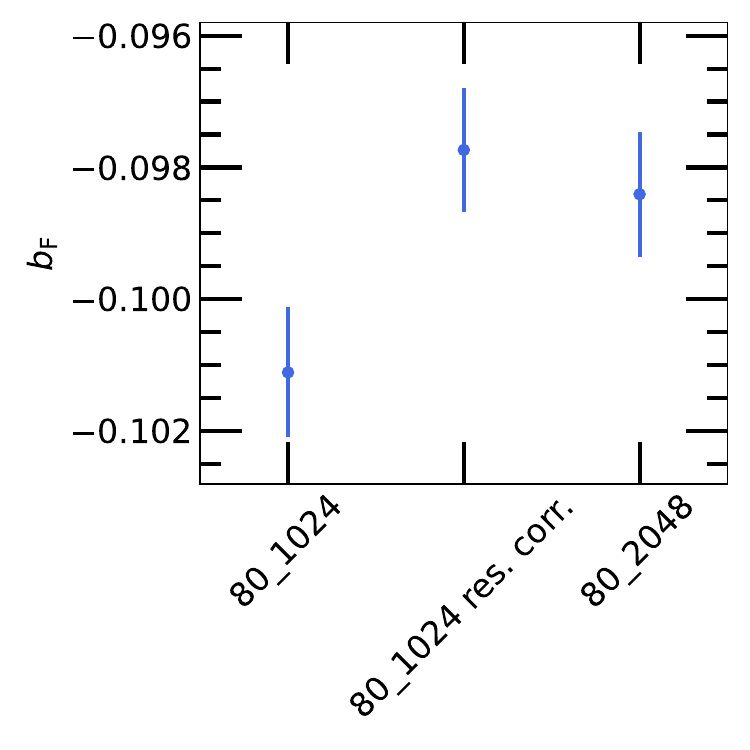}
\includegraphics[width=.32\textwidth]{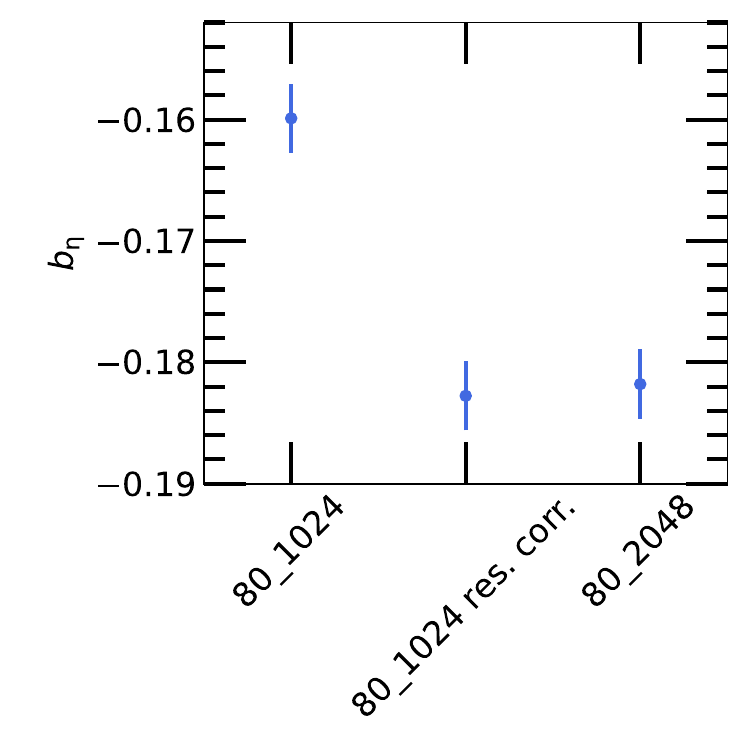}
\end{minipage}
\begin{minipage}{\textwidth}
\centering 
\includegraphics[width=.32\textwidth]{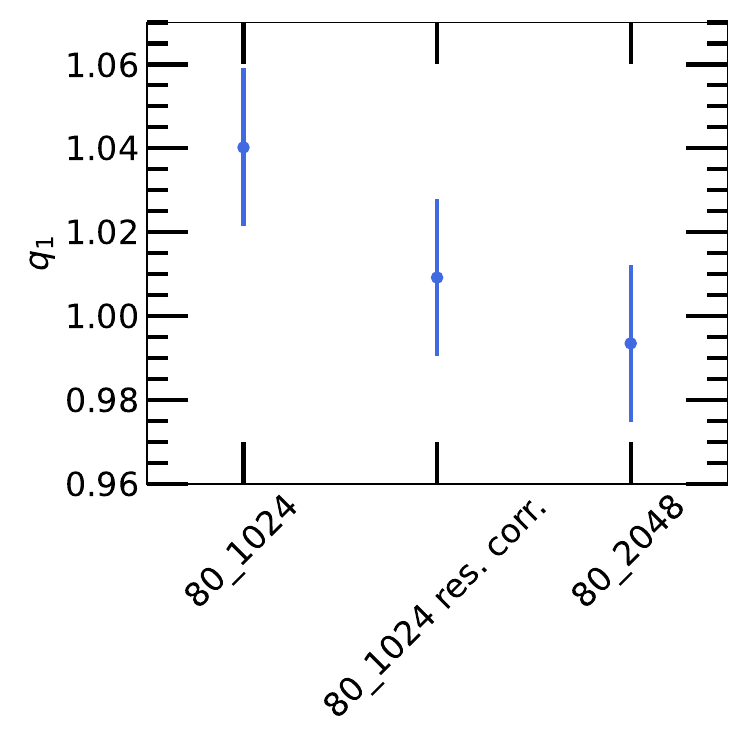}
\includegraphics[width=.32\textwidth]{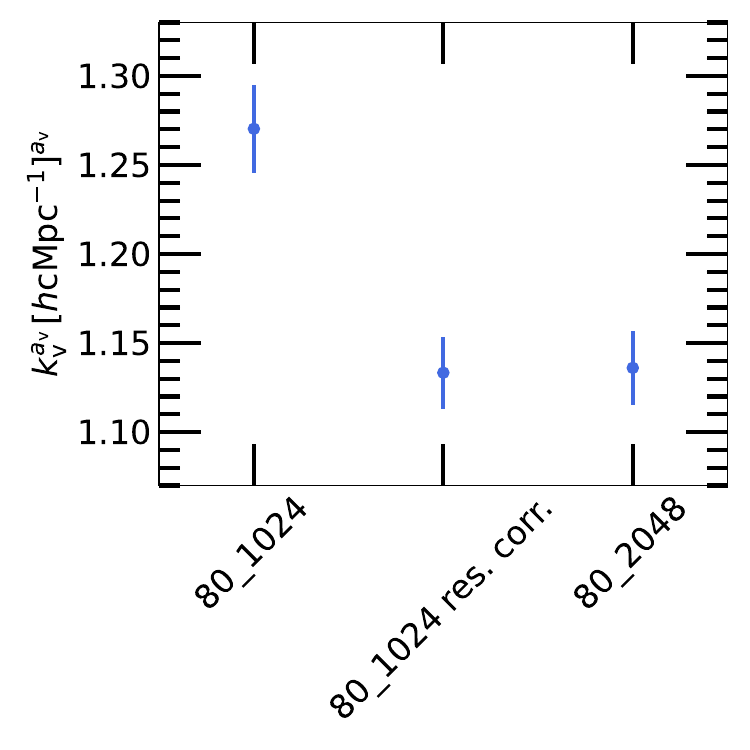}
\includegraphics[width=.32\textwidth]{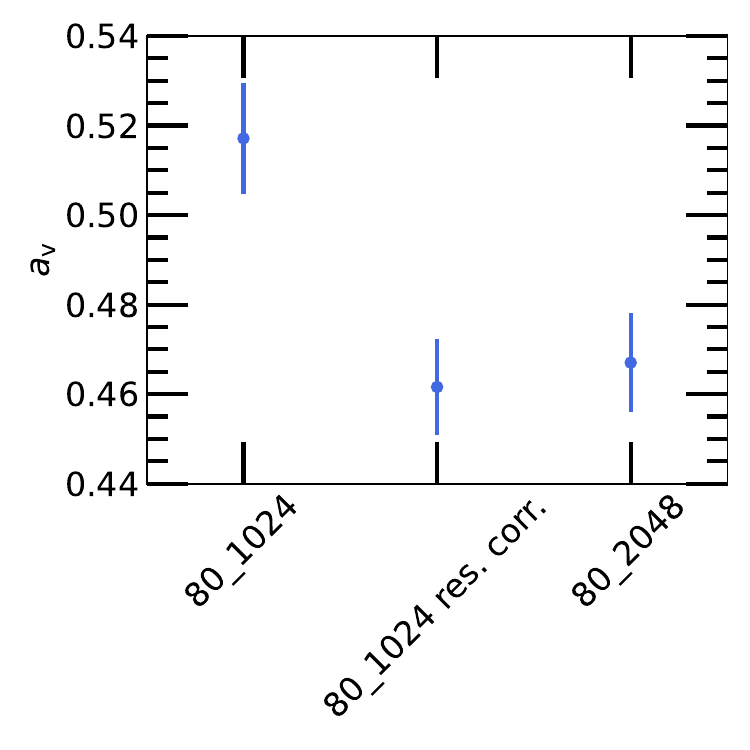}
\end{minipage}
\begin{minipage}{\textwidth}
\centering 
\includegraphics[width=.32\textwidth]{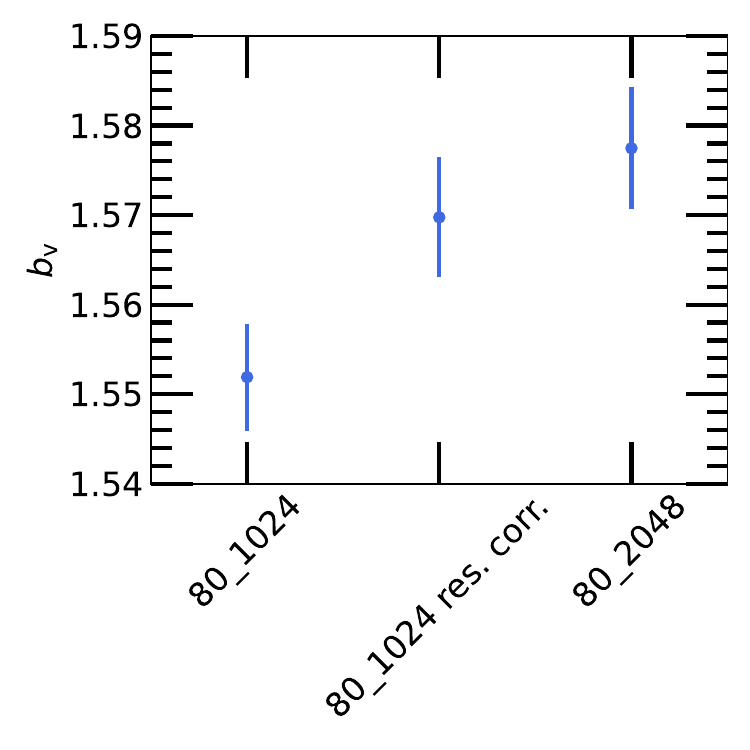}
\includegraphics[width=.32\textwidth]{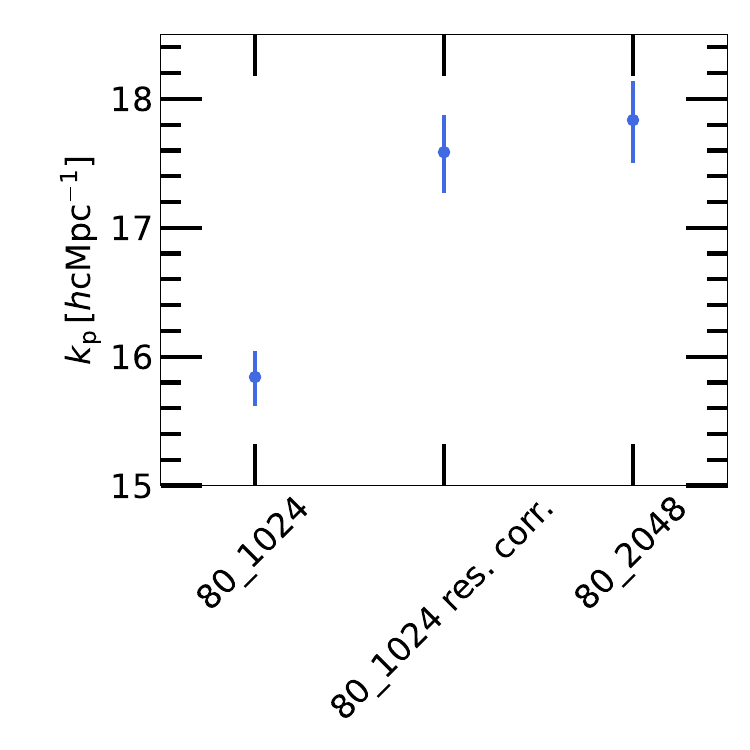}
\end{minipage}
\hfill
\caption{\label{fig:allparams_rescorr} The effect of the resolution correction on the best-fit bias (top row) and non-linear (i.e. $D_{\rm NL}$ from Eq.~\ref{eq:P3D_nlfactor}, middle and bottom rows) parameters in $L_{\rm box}=80\,h^{-1}\,\rm cMpc$ and $N_{\rm part}=2\times1024^3$ simulation. For comparison we show the values from the higher resolution simulation with $2\times2048^3$ particles.
}
\end{figure}


In the previous sections we have shown that mass resolution produces a systematic dependence of $P_{\rm 3D,\alpha}$ over a broad range of scales and orientations. In general, correcting for resolution is easier than correcting for box size, since the simulation box size mostly affects large-scale modes that are most affected by the simulation's sample variance. One can correct for resolution by combining a large-box, low-resolution simulation with a pair of small-box simulations with low and high resolution. To test the effectiveness of this approach, we apply the splicing technique of \cite{McDonald_2003}, which was also employed by \cite{Chabanier_2024}, using the $L_{\rm box}=40\,h^{-1}\,\mathrm{cMpc}$ simulations with $N_{\rm part}=2\times512^3$ and $N_{\rm part}=2\times1024^3$ to estimate the resolution correction needed for the $L_{\rm box}=80\,h^{-1}\,\mathrm{cMpc}$, $N_{\rm part}=2\times1024^3$ simulation to match the resolution of the $L_{\rm box}=80\,h^{-1}\,\mathrm{cMpc}$, $N_{\rm part}=2\times2048^3$ simulation. We do not apply the $k>k_{\rm max}$ branch of the original prescription, since $k_{\rm max}=k_{\rm Nyq}/4\simeq10\,h\,\mathrm{cMpc}^{-1}$ for the $80\_1024$ simulation, corresponding approximately to the maximum wavenumber considered in our analysis. We adopt $k_{\rm min}=0.5\,h\,\mathrm{cMpc}^{-1}$ and find that varying this choice has a negligible impact on the resulting correction.


The splicing technique reduces the mass-resolution dependence of $P_{\mathrm{3D},\alpha}$. Figure~\ref{fig:P3D_rescorr} compares the original (solid orange curves) and resolution-corrected (dashed pink curves) $P_{\rm 3D,\alpha}$ of the $80\_1024$ simulation with the directly simulated higher-resolution $80\_2048$ result (solid blue curves) at $z=2.4$. While the uncorrected spectrum differs from the higher-resolution result by up to $\sim10\%$ (except on the largest scales in the $\mu\geq0.75$ bin), the resolution correction reduces the residuals to $\lesssim2\%$ over essentially the entire range of scales and orientations considered. The main exceptions, both occurring on large scales, are the bins at $k\simeq0.5\,h\,\mathrm{cMpc}^{-1}$ for $0.25\leq\mu\leq0.50$, where the residual reaches $\approx7.4\%$, and at $k\simeq0.3\,h\,\mathrm{cMpc}^{-1}$ for $0.75\leq\mu\leq1.00$, where it reaches $\approx2.6\%$. The effectiveness of the correction is also reflected in the parameters of the AiP15 model. As shown in Fig.~\ref{fig:allparams_rescorr}, all fitted bias and non-linear parameters obtained from the resolution-corrected spectrum are consistent within $1\sigma$ with those measured directly from the $80\_2048$ simulation. This demonstrates that, at $z=2.4$, the resolution dependence identified in Sec.~\ref{sec:convergence_tests} can be substantially reduced using the lower-resolution simulations, both at the level of the 3D flux power spectrum itself and of its fitted model parameters.

Previous convergence studies of Ly$\alpha$ forest statistics have shown that numerical resolution requirements become increasingly stringent towards higher redshift \citep{Bolton_2009_convergence,Lukic_2015,Doughty_2023}. This is primarily because Ly$\alpha$ transmission at high redshift increasingly arises from underdense regions of the IGM, which are more challenging to resolve, while the shorter time available for pressure smoothing following reionization can further preserve small-scale gas structure \citep{Doughty_2023}. Consistent with this expectation, Fig.~\ref{fig:bias_zevo_vsNpar} shows that the sensitivity of the fitted bias parameters to mass resolution increases towards higher redshift, and we therefore expect the required resolution correction to become larger at higher $z$. Nevertheless, the results at $z=2.4$ demonstrate that the established resolution-correction procedure can substantially reduce the numerical resolution dependence and recover both $P_{\rm 3D,\alpha}$ and its fitted parameters to good accuracy at this redshift. Extending this validation explicitly to higher redshift will be important for future high-precision $P_{\rm 3D,\alpha}$ modelling.

\subsection{The effect of extracted grid coarseness on $P_{\rm 3D,\alpha}$}\label{sec:conv_grid}

\begin{figure}[tbp]
\centering 
\includegraphics[width=.9\textwidth]{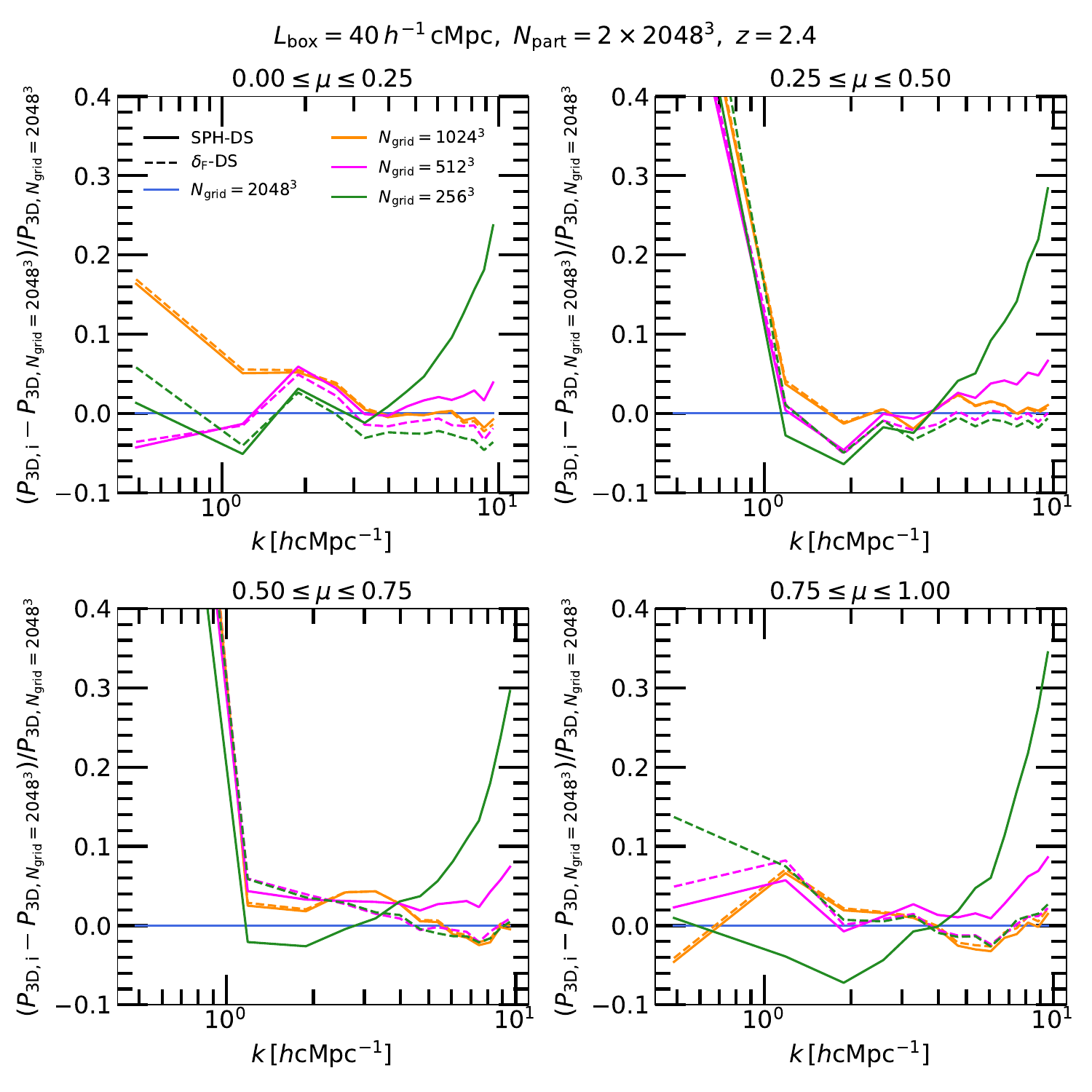}
\hfill
\caption{\label{fig:P3D_vsNgrid}
Extracted-grid convergence test of the Ly$\alpha$ forest flux 3D power spectrum at $z=2.4$. Fractional residuals are shown relative to the finest extracted grid ($N_{\rm grid}=2048^3$) for the $L_{\rm box}=40\,h^{-1}\,\mathrm{cMpc}$, $N_{\rm part}=2\times2048^3$ simulation. Solid curves correspond to the SPH-DS method and dashed curves show the $\delta_{\rm F}$-DS method in which the optical depth and Ly$\alpha$ forest fluctuation field are first computed on the finest grid before degrading the resulting $\delta_{\rm F}$ field. The SPH-DS method introduces an artificial enhancement of small-scale power that becomes increasingly significant for $N_{\rm grid}\leq512^3$, whereas the $\delta_{\rm F}$-DS approach substantially reduces the resolution dependence and avoids this systematic small-scale enhancement.
}
\end{figure}

While for all simulations we extract a uniformly spaced grid with the number of pixels $N_{\rm grid}=N_{\rm los}^2N_{\rm bins}=N_{\rm part}/2$, here we explore the effect of downsampling the grid, particularly on the finest mass resolution simulation we consider in this study ($L_{\rm box}=40\,h^{-1}\,\rm cMpc$, $N_{\rm part}=2\times2048^3$). The downsampling is done in two ways for $N_{\rm grid}=[2048^3,1024^3,512^3,256^3]$. Firstly, we extract gas-field grids of each size directly from the simulated particle dataset through SPH interpolation. We then compute the optical depth based on these grids. We label this method as SPH downsampling (SPH-DS). In the second method we start with the finest grid of $N_{\rm grid}=2048^3$ extracted using the SPH interpolation routine and compute the optical depth and corresponding $F_{\rm Ly\alpha}$. Then we construct coarser grids by averaging the Ly$\alpha$ forest fluctuation field, $\delta_{\rm F}$, over neighbouring pixels such that we obtain the desired $N_{\rm grid}$. This method is labelled as $\delta_{\rm F}$-DS. We do not apply the ZCV correction in these tests.

Since in the SPH-DS approach the optical depth is computed directly on already downsampled gas-field grids, the line profile is not sampled finely enough when the grid pixels become too large. This produces artificial small-scale wiggles in the Ly$\alpha$ flux spectra, similar to the discretization artefacts that can arise when the LOS is not sampled sufficiently finely in optical-depth calculations. These numerical wiggles add spurious small-scale structure to the flux field and therefore boost $P_{\rm 3D,\alpha}$ close to the grid scale, as can be seen in Fig.~\ref{fig:P3D_vsNgrid}. This figure shows the residual 3D power spectra for various $N_{\rm grid}$ relative to the finest grid. Specifically, at $k\gtrsim4\,h\,\rm cMpc^{-1}$, $P_{\rm 3D,\alpha}$ calculated from the SPH-DS grids (solid curves) is enhanced by up to $\sim24\%$ ($\sim4\%$) in the lowest $\mu$-bin when downsampling to $N_{\rm grid}=256^3$ ($512^3$), increasing to $\sim35\%$ ($\sim9\%$) in the highest $\mu$-bin.

In contrast, the $\delta_{\rm F}$-DS approach first computes the optical depth and the corresponding Ly$\alpha$ forest fluctuation field, $\delta_{\rm F}$ (Eq.~\ref{eq:P3D_estimator}), on the highest-resolution grid. The coarser grids are then obtained by averaging neighbouring $\delta_{\rm F}$ cells. Since the optical-depth calculation is performed before the degradation, the resulting power spectra show substantially better convergence, particularly at small scales, and do not exhibit the systematic enhancement of small-scale power present in the SPH-DS approach.

Comparing the resulting $P_{\rm 3D,\alpha}$ from these two methods, one can see the enhanced power at small scales in the SPH-DS method more clearly. Conversely, the power is suppressed in this method relative to the $\delta_{\rm F}$-DS at $k\lesssim4\,h\,\rm cMpc^{-1}$. This effect is non-negligible for $N_{\rm grid}\leq512^3$. Based on these results, we recommend computing the optical depth on the finest available grid before subsequently degrading the Ly$\alpha$ forest fluctuation field, rather than downsampling the gas fields prior to the optical-depth calculation.

Overall, the effects discussed in this section are numerical rather than astrophysical in origin. Nevertheless, they provide practical guidelines for generating high-fidelity Ly$\alpha$ forest models from hydrodynamical simulations of the post-reionization IGM. In particular, our tests indicate that the mass-resolution dependence is more pervasive across scales and orientations than the simulation-volume dependence, while the optical depth should be computed on the highest available grid resolution before any subsequent degradation of the Ly$\alpha$ forest fluctuation field. We note, however, that our highest-resolution mass- and extracted-grid convergence tests, reaching $N_{\rm part}=2\times2048^3$ and $N_{\rm grid}=2048^3$, respectively, are performed only at $z=2.4$ and therefore do not by themselves establish the same level of convergence at higher redshifts.

\section{The effect of reionization on post-reionization $P_{\rm 3D,\alpha}$}\label{sec:reion_models}

Having quantified the numerical sensitivity of $P_{\rm 3D,\alpha}$, we now investigate its response to different reionization and thermal histories of the IGM. The models considered in this section share the same numerical setup, with $L_{\rm box}=40\,h^{-1}\,\mathrm{cMpc}$ and $N_{\rm part}=2\times1024^3$. As demonstrated in Sec.~\ref{sec:convergence_tests}, the simulation volume has a relatively limited impact on the ZCV-corrected $P_{\rm 3D,\alpha}$ over most of the scales and orientations considered here. Mass resolution produces a more pervasive dependence across scales and orientations, although the $N_{\rm part}=2\times1024^3$ simulation shows substantially better convergence than the $N_{\rm part}=2\times512^3$ model considered in our tests. Since our aim here is to characterize the relative imprint of different IGM histories rather than to make precision predictions of the absolute power spectrum, we compare models with identical numerical resolution and volume. The following analysis explores how variations in the timing and morphology of reionization and the thermal evolution of the IGM affect the scale and angular dependence of $P_{\rm 3D,\alpha}$ after reionization has completed. These IGM models were produced within the Sherwood--Relics simulation suite \cite{Puchwein_2023}. The fiducial model selected here follows a scenario in which reionization is completed by $z_{\rm r}=6$.

\begin{figure}[tbp]
\centering 
\begin{minipage}{\textwidth}
\centering 
\includegraphics[width=.32\textwidth]{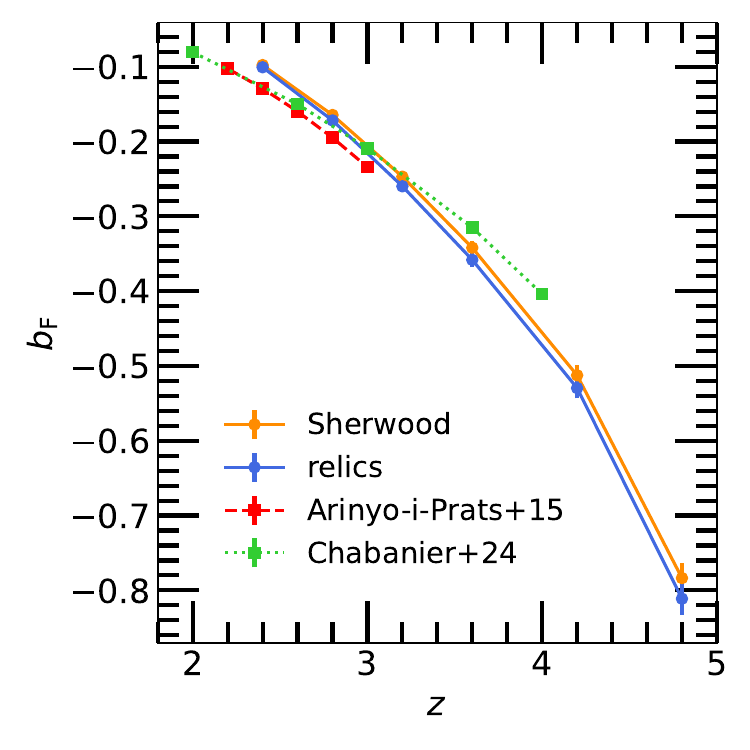}
\includegraphics[width=.32\textwidth]{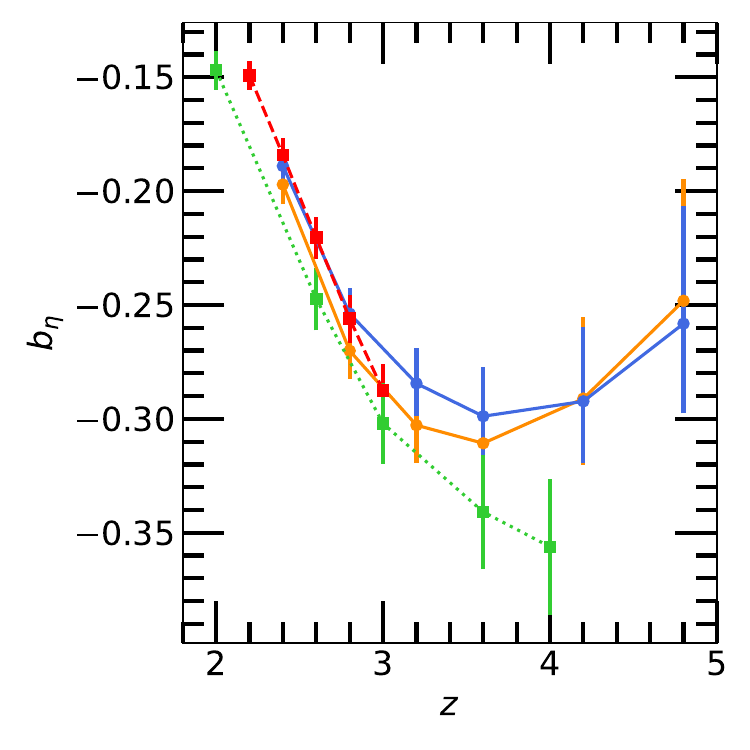}
\end{minipage}
\begin{minipage}{\textwidth}
\centering 
\includegraphics[width=.32\textwidth]{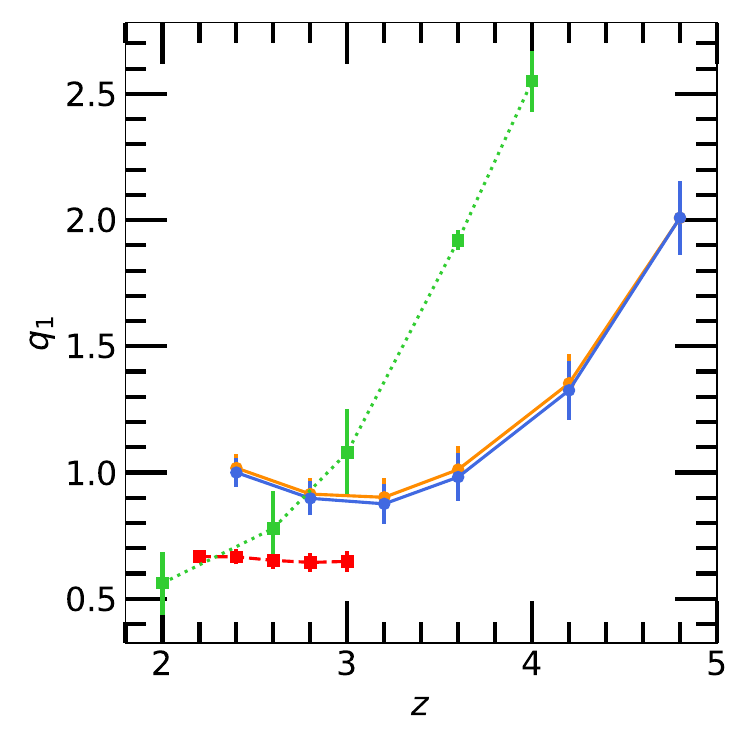}
\includegraphics[width=.32\textwidth]{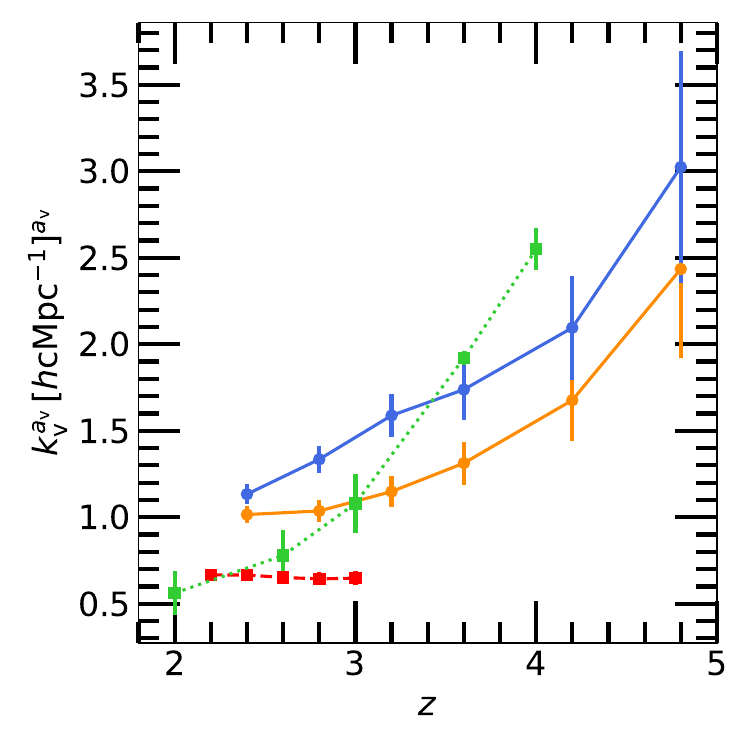}
\includegraphics[width=.32\textwidth]{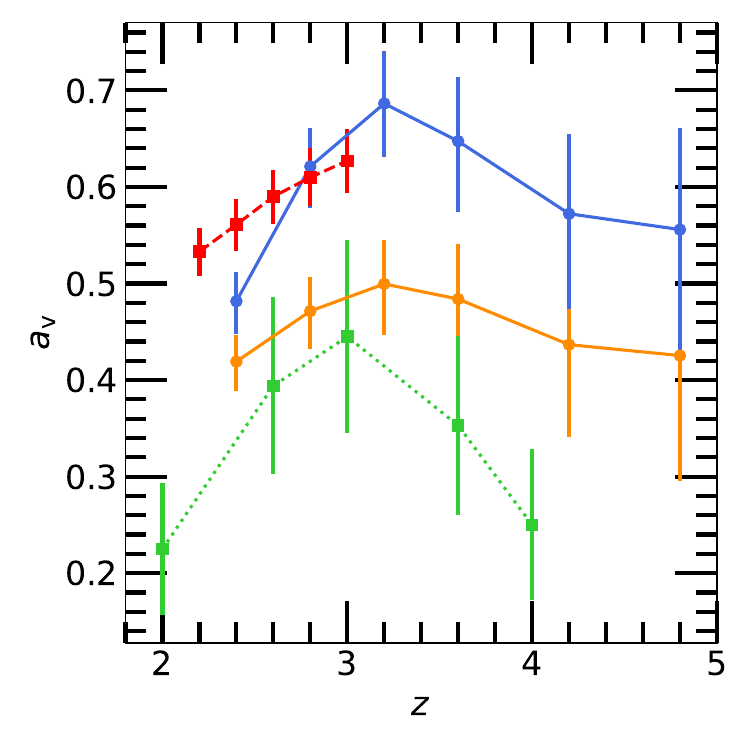}
\end{minipage}
\begin{minipage}{\textwidth}
\centering 
\includegraphics[width=.32\textwidth]{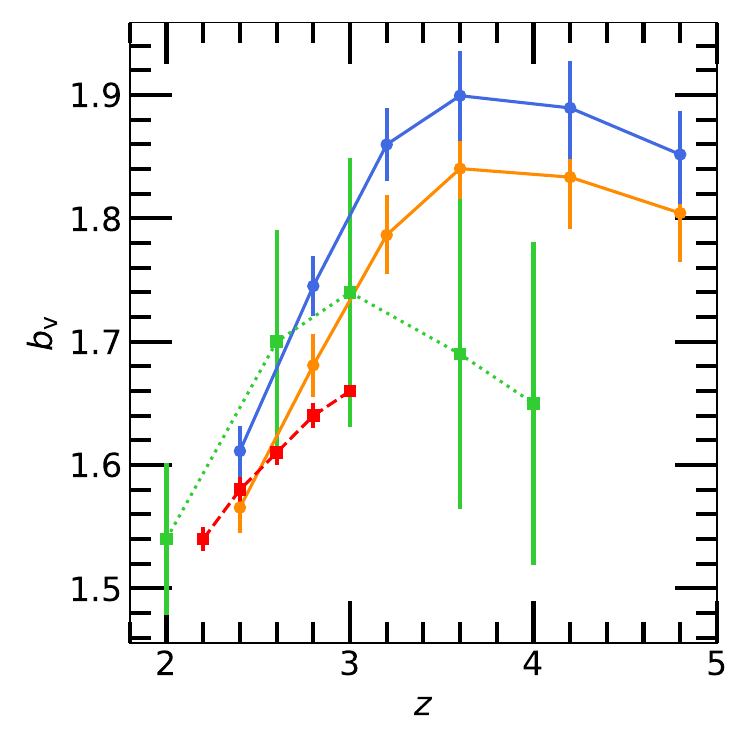}
\includegraphics[width=.32\textwidth]{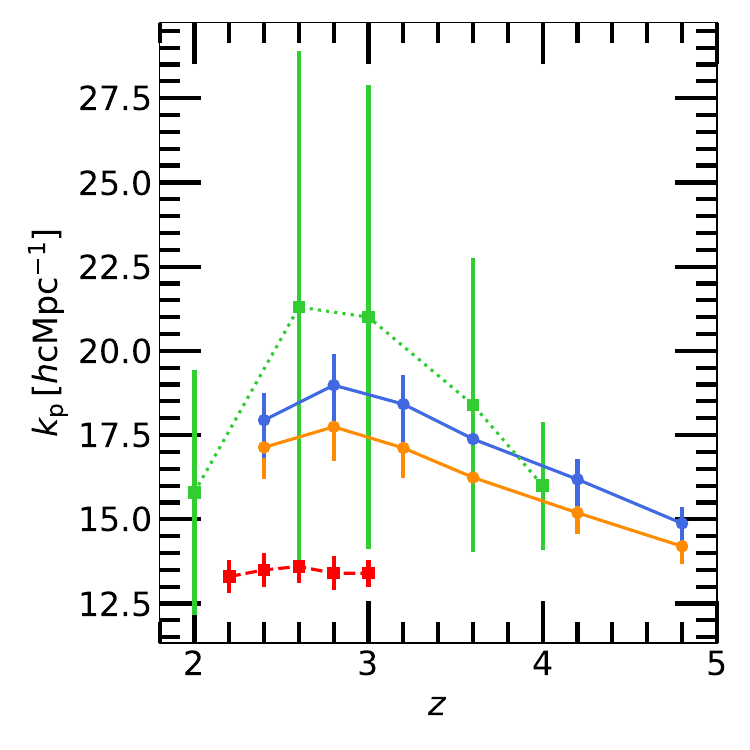}
\end{minipage}
\hfill
\caption{\label{fig:allparams_vsrelics}The redshift evolution of the best-fit bias (top row) and non-linear (i.e. $D_{\rm NL}$ from Eq.~\ref{eq:P3D_nlfactor}, middle and bottom rows) parameters from $L_{\rm box}=40\,h^{-1}\,\rm cMpc$ and $N_{\rm part}=2\times1024^3$ Sherwood (solid orange curves) and relics (solid blue curves) simulations. The results from Fiducial simulation of \cite{Arinyo-i-Prats_2015} and the 160R25 simulation from \cite{Chabanier_2024} are shown for comparison by the dotted red curves and the dashed green curves, respectively.
}
\end{figure}

\begin{figure}[tbp]
\centering 
\includegraphics[width=.9\textwidth]{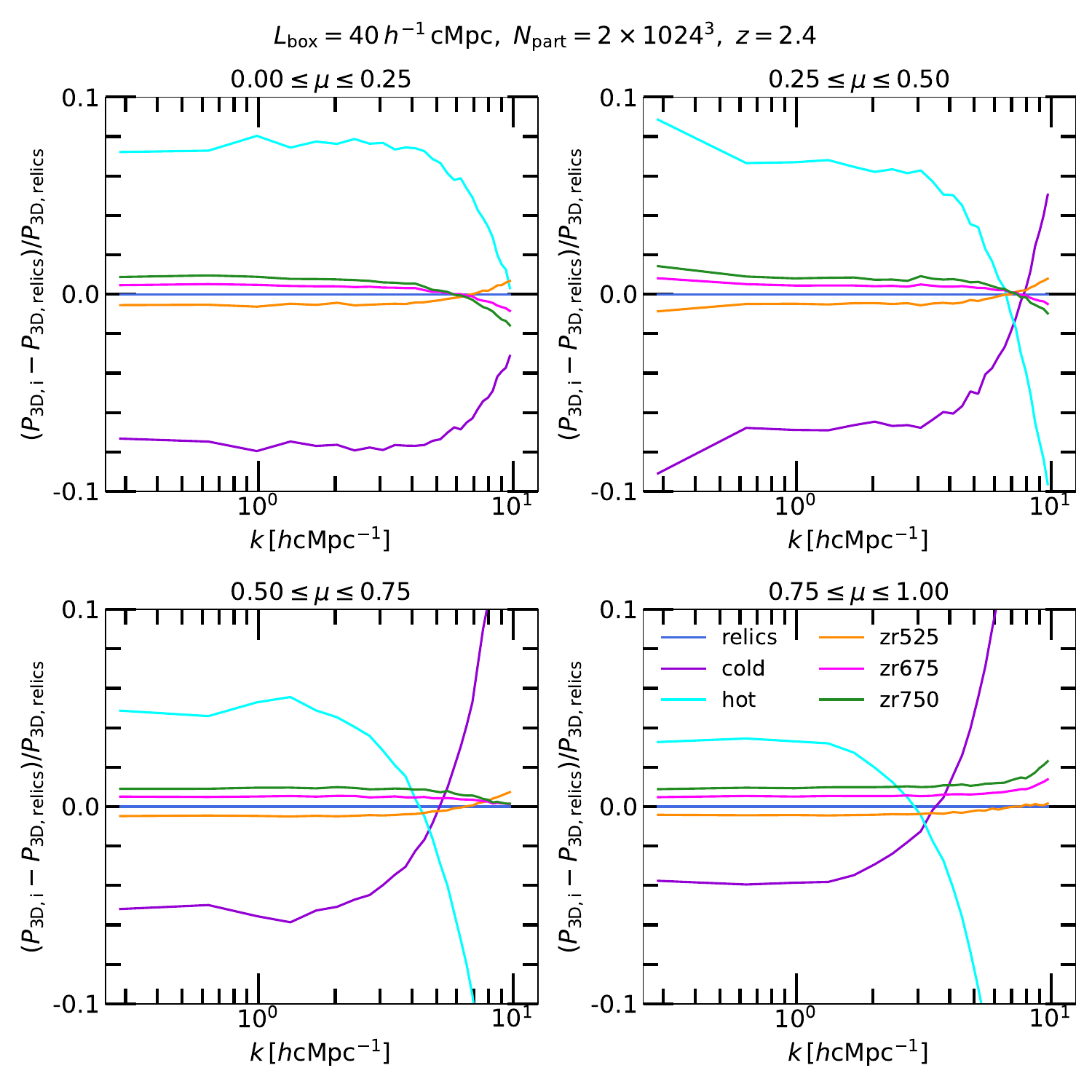}
\hfill
\caption{\label{fig:P3D_vsreionhist}Fractional residuals of $P_{\rm 3D,\alpha}$ across different thermal and reionization histories relative to the fiducial Sherwood--Relics simulation, for $L_{\rm box}=40\,h^{-1}\,\mathrm{cMpc}$, $N_{\rm part}=2\times1024^3$, and $z=2.4$. Models with different reionization redshifts differ by $\lesssim1\%$. Doubling (halving) the photoheating during reionization enhances (suppresses) the large-scale power by up to $\sim8\%$ in the lowest-$\mu$ bin, with a smaller response towards LOS-dominated modes. On the smallest scales the sign of the difference reverses.}
\end{figure}

Before examining variations within the Sherwood--Relics suite, we first compare its fiducial model with the Sherwood simulation of the same $L_{\rm box}=40\,h^{-1}\,\mathrm{cMpc}$ and $N_{\rm part}=2\times1024^3$ numerical configuration. Figure~\ref{fig:allparams_vsrelics} shows the redshift evolution of the best-fitting AiP15 parameters for the two simulations, together with the results of \cite{Arinyo-i-Prats_2015} and \cite{Chabanier_2024}. The flux and velocity-gradient biases, $b_{\rm F}$ and $b_{\eta}$, are broadly consistent between Sherwood and Sherwood--Relics, as is $q_1$. More noticeable systematic differences occur in the remaining non-linear parameters, with the relics model generally yielding larger $a_{\rm v}$, $b_{\rm v}$, $k_{\rm p}$, and $k_{\rm v}^{a_{\rm v}}$. Despite differences in their absolute values, some of the non-linear parameters exhibit qualitatively similar redshift evolution to previous studies. In particular, $a_{\rm v}$ shows a turnover at a similar redshift, $z\sim3$, in our simulations and \cite{Chabanier_2024}. Furthermore, $k_{\rm v}^{a_{\rm v}}$ increases towards higher redshift in both our simulations and \cite{Chabanier_2024}. Nevertheless, the absolute values and detailed redshift evolution of the individual $D_{\rm NL}$ parameters can differ substantially between simulation suites, indicating their sensitivity to the underlying simulation and modelling choices. We now use the fiducial Sherwood--Relics model as the reference for exploring variations in the thermal and reionization history; the corresponding fractional residuals in $P_{\mathrm{3D},\alpha}$ are presented in Fig.~\ref{fig:P3D_vsreionhist}.

\begin{figure}[tbp]
\centering 
\begin{minipage}{\textwidth}
\centering 
\includegraphics[width=.32\textwidth]{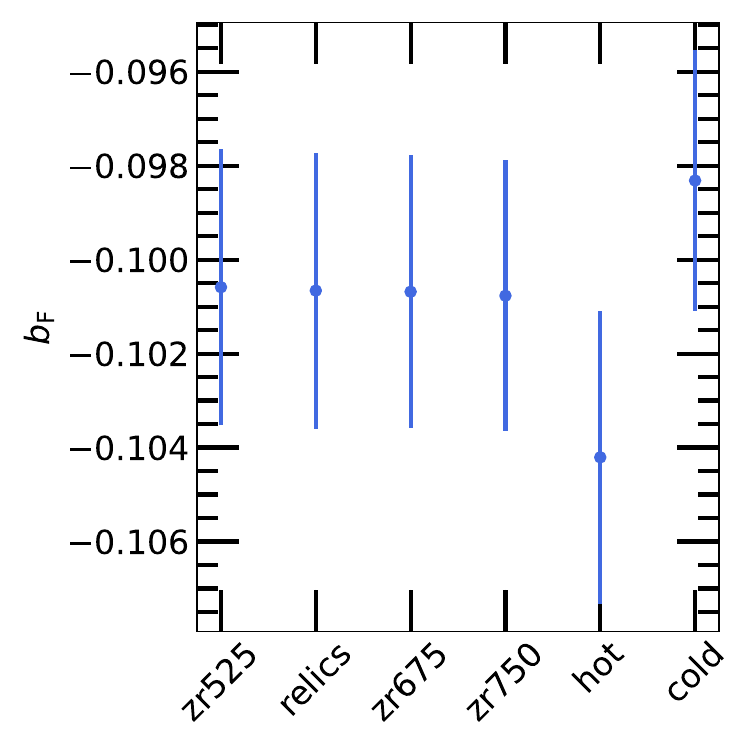}
\includegraphics[width=.32\textwidth]{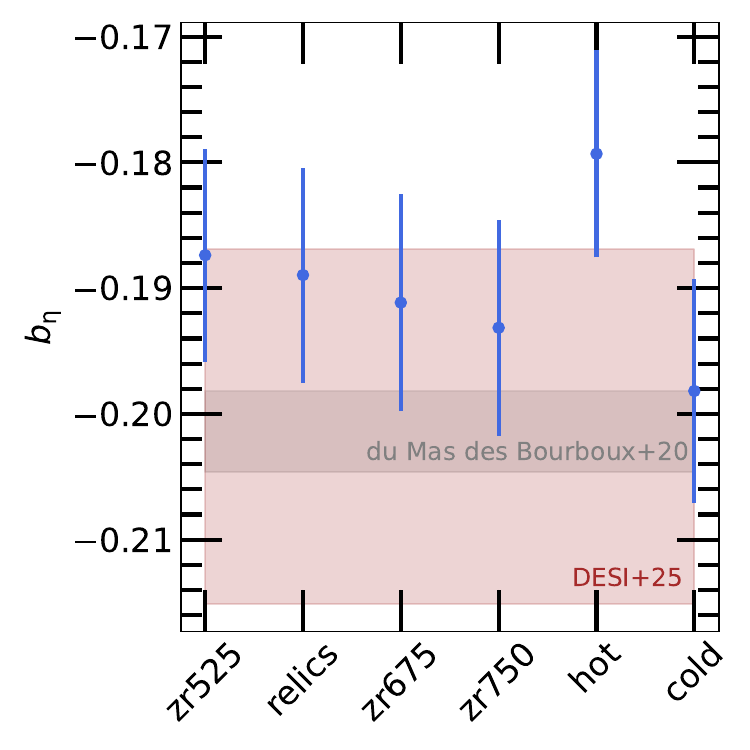}
\end{minipage}
\begin{minipage}{\textwidth}
\centering 
\includegraphics[width=.32\textwidth]{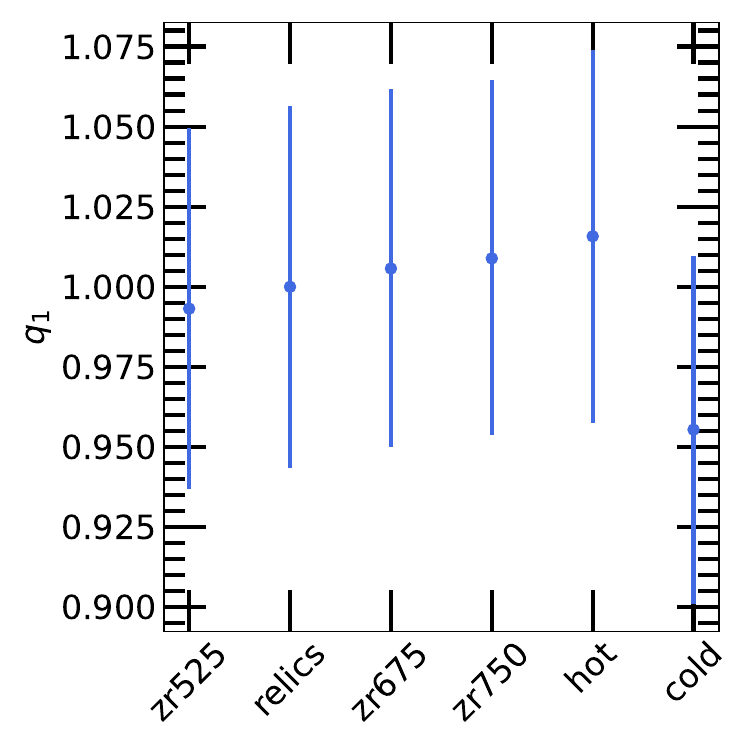}
\includegraphics[width=.32\textwidth]{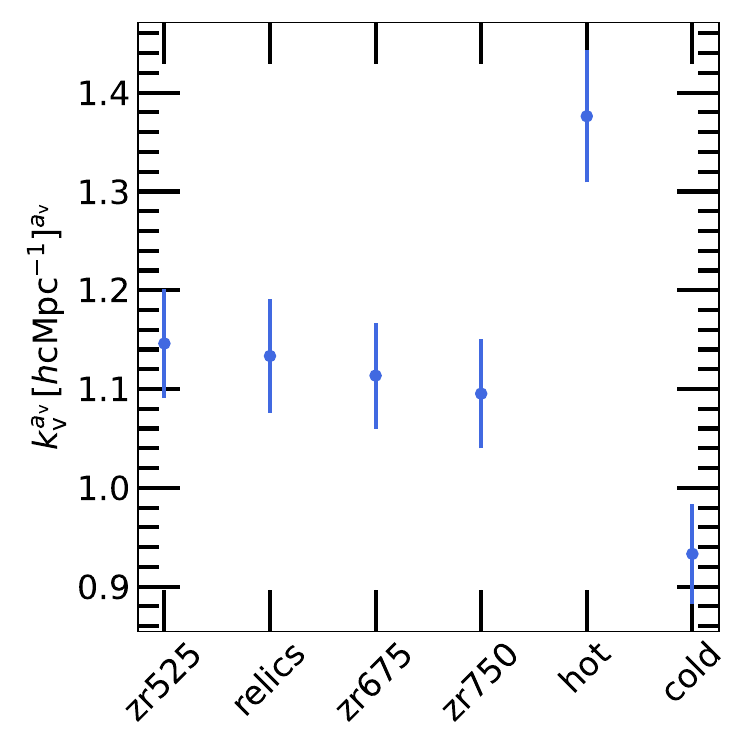}
\includegraphics[width=.32\textwidth]{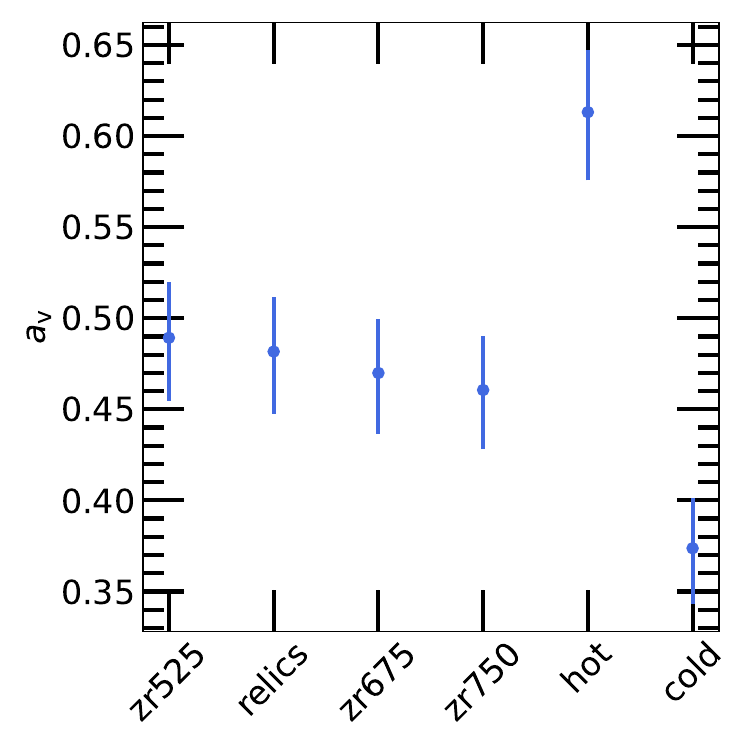}
\end{minipage}
\begin{minipage}{\textwidth}
\centering 
\includegraphics[width=.32\textwidth]{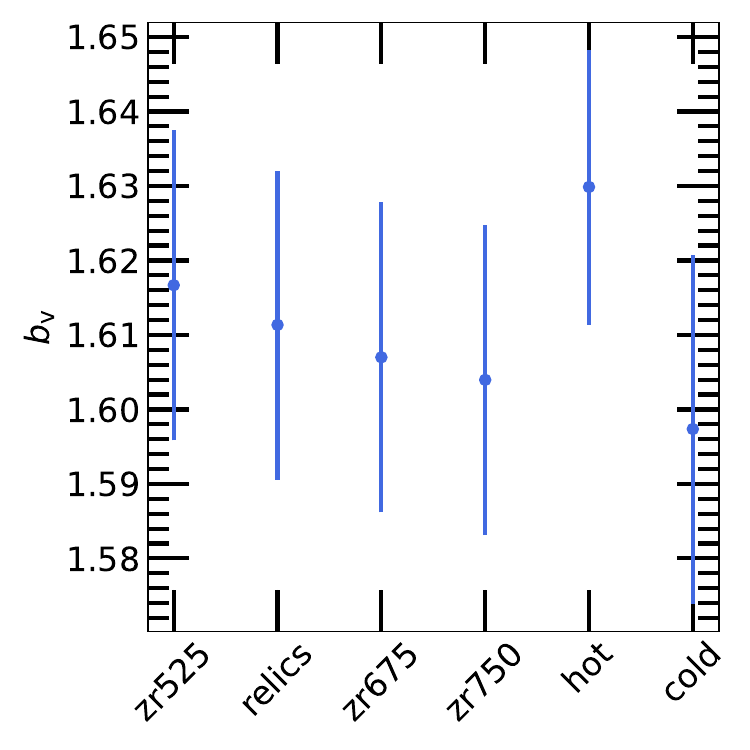}
\includegraphics[width=.32\textwidth]{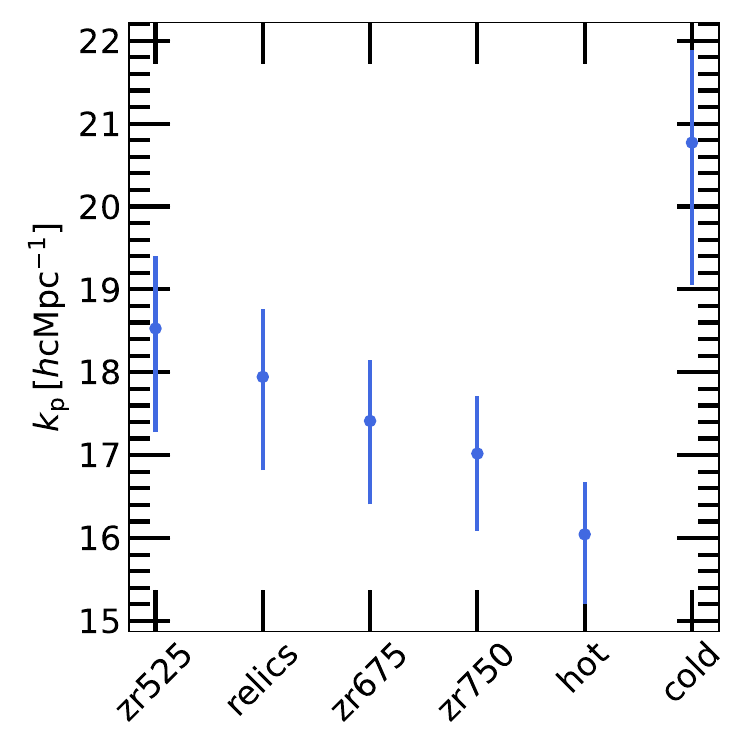}
\end{minipage}
\hfill
\caption{\label{fig:parameters_vsreionhist} Best-fitting AiP15 bias (top row) and non-linear (middle and bottom rows) parameters at $z=2.4$ for different thermal and reionization histories. The horizontal shaded regions in the top-right panel indicate observational measurements of the velocity-gradient bias, $b_{\eta}$, at $z\simeq2.33$ from \cite{duMasdesBourboux_2020} (grey) and \cite{DESI_2025} (red).}
\end{figure}

Firstly, we explore the effect of the timing of reionization on $P_{\rm 3D,\alpha}$. To this end, we consider two early-ending and one late-ending reionization models, in which reionization is completed at $z_{\rm r}=7.50$ (green), $6.75$ (pink), and $5.25$ (orange), respectively. We find a small enhancement (suppression) of power over most of the explored $k$-range when reionization ends earlier (later). This behaviour reflects the thermal memory of reionization retained by the post-reionization IGM: gas that is reionized earlier has more time to cool and for the pressure-smoothing imprint of photoheating to evolve before the redshift of observation. Nevertheless, the resulting differences are only of order $\sim1\%$ at $z=2.4$, owing to the long time elapsed since the end of reionization, and show negligible dependence on $\mu$. The differences increase towards higher redshift and are approximately twice as large at $z=3.2$ in the lowest-$\mu$ bin.

The weak dependence on reionization timing is also reflected in the fitted bias parameters. At $z=2.4$, both $b_{\rm F}\approx-0.1006$ and $b_{\eta}\approx-0.19$ show a mild monotonic trend with $z_{\rm r}$, as shown in Fig.~\ref{fig:parameters_vsreionhist}. Earlier-ending reionization models yield slightly more negative values of the bias parameters. The fractional differences in $b_{\rm F}$ increase from only $\lesssim0.1\%$ at $z=2.4$ to $\lesssim0.8\%$ at $z=3.2$, while the differences in $b_{\eta}$ are somewhat larger, reaching up to $\sim2.1\%$ and $\sim1.5\%$ at $z=2.4$ and $3.2$, respectively (Table~\ref{tab:parameters_reion_bias}). The non-linear parameters likewise show only weak sensitivity to $z_{\rm r}$ at $z=2.4$, as shown in Fig.~\ref{fig:parameters_vsreionhist}, with a similarly weak dependence found at $z=3.2$ (Table~\ref{tab:parameters_reion_DNL}). This suggests that the percent-level imprint of reionization timing visible in $P_{\rm 3D,\alpha}$ translates only into weak shifts of the fitted AiP15 parameters at $z\leq3.2$.

The weak sensitivity to $z_{\rm r}$ found here is particularly relevant given the growing observational evidence for reionization ending at $z_{\rm r}<5.6$, including observed large scatter in Ly$\alpha$ forest transmission between sightlines \citep{Becker_2015,Kulkarni_2019,Bosman_2022}, Ly$\alpha$ forest transmission spikes at $z>5$ \citep{Gaikwad_2020}, low abundance of Ly$\alpha$ emitting galaxies around long Ly$\alpha$ absorption troughs \citep{Kashino_2020,Keating_2020,Christenson_2021}, clustering of Ly$\alpha$ emitters \citep{Weinberger_2019}, Ly$\alpha$ and Ly$\beta$ forest spectra containing long dark gaps \citep{Zhu_2021,Zhu_2022,Maity_2026}, presence of damping-wing absorption at $z<5.6$ \citep{Becker_2024,Spina_2024,Zhu_2024,Sawyer_2025}, mean free path of ionizing photons at $z=6$ \citep{Becker_2021,Cain_2021,Zhu_2023,Gaikwad_2023}, as well as various metal absorption and emission lines \citep{Becker_2019,Sebastian_2024,Kakiichi_2025}. While the exact timing is uncertain, our results suggest that differences in the end redshift of reionization within the range explored here leave only a percent-level imprint on $P_{\rm 3D,\alpha}$ at $z\lesssim3.2$. This effect is therefore subdominant to the numerical modelling uncertainties identified in this work.

Spatially inhomogeneous thermal histories can also leave long-lived signatures in the Ly$\alpha$ forest: X-ray pre-heating prior to \HI reionization can modify the subsequent reionization relic in both the 3D and 1D flux power spectra \citep{Montero-Camacho_2024}, while patchy \HeII reionization produces additional temperature fluctuations and several-percent effects in one-dimensional Ly$\alpha$ forest statistics \citep{Upton_Sanderbeck_2020}. In this work, we find that variations in the amount of photoheating during reionization leave a much stronger and longer-lasting imprint on $P_{\rm 3D,\alpha}$ than variations in the timing of reionization. In the hot model (cyan in Fig.~\ref{fig:P3D_vsreionhist}), in which the photoheating during reionization is doubled relative to the fiducial model, enhanced thermal broadening and pressure smoothing suppress power on small scales, while the power is enhanced on larger scales. The scale at which the residual changes sign shifts towards lower $k$ with increasing $\mu$, i.e. from predominantly transverse to predominantly LOS modes. This angular dependence is consistent with the increasing importance of thermal broadening towards LOS-dominated modes, in addition to the three-dimensional smoothing of the gas distribution caused by the integrated pressure response to enhanced photoheating. The large-scale enhancement reaches $\sim8\%$ in the lowest $\mu$-bin and gradually decreases to $\sim4\%$ towards the most LOS-dominated modes. The cold model exhibits approximately the opposite behaviour, with the residuals broadly mirroring those of the hot model.

The stronger sensitivity to thermal history is also reflected in the fitted bias parameters, with differences of up to $\sim3.4\%$ and $\sim5.4\%$ in $b_{\rm F}$ and $b_{\eta}$ at $z=2.4$, respectively, relative to the fiducial model, and $\sim3.8\%$ and $\sim2.8\%$ at $z=3.2$ (Table~\ref{tab:parameters_reion_bias}). The non-linear parameters are also substantially more sensitive to the thermal history than to the timing of reionization, as illustrated at $z=2.4$ in Fig.~\ref{fig:parameters_vsreionhist}. For example, $a_{\rm v}$ changes from $0.482_{-0.034}^{+0.030}$ in the fiducial model to $0.374_{-0.031}^{+0.028}$ and $0.613_{-0.037}^{+0.034}$ in the cold and hot models, respectively (Table~\ref{tab:parameters_reion_DNL}).

The fitted bias parameters may be compared with current observational measurements. The predicted values of $b_{\eta}$ from all models are consistent within $1\sigma$ with the DESI DR1 measurement at $z=2.33$ \citep{DESI_2025}. In contrast, agreement with the eBOSS DR16 measurement \citep{duMasdesBourboux_2020} is obtained only for the models with earlier reionization (zr675 and zr750) or reduced photoheating (cold). However, none of our models reproduces the observed Ly$\alpha$ flux bias, $b_{\rm F}$, which is systematically more negative than predicted. Consequently, the agreement in $b_{\eta}$ alone should not be interpreted as evidence favouring an earlier end to reionization or a colder thermal history.

\begin{figure}[tbp]
\centering 
\includegraphics[width=.9\textwidth]{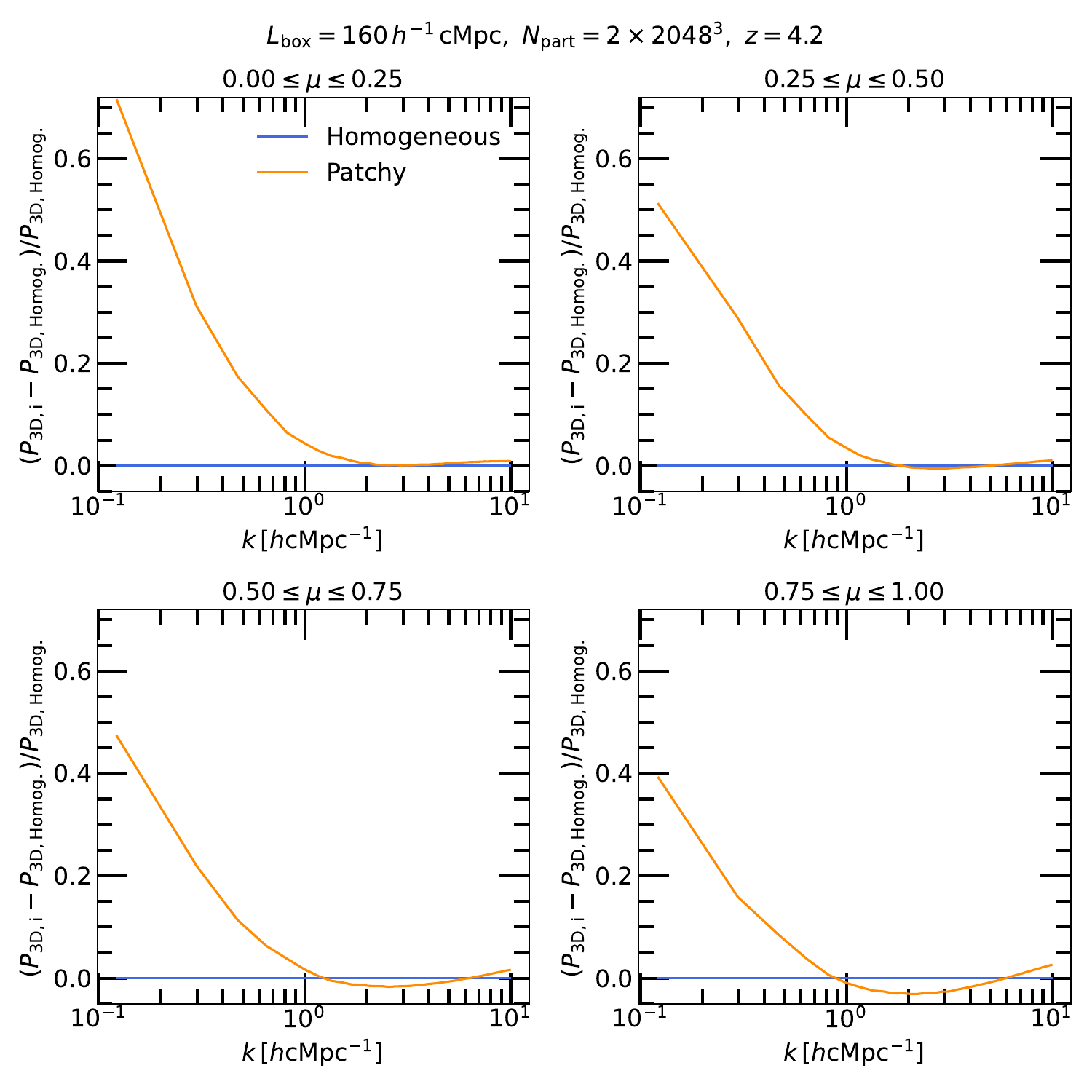}
\hfill
\caption{\label{fig:P3D_patchy}Same as Fig.~\ref{fig:P3D_vsreionhist}, but comparing homogeneous and inhomogeneous (patchy) reionization while keeping the global reionization history fixed. On the largest scales, patchy reionization boosts the power by up to $\sim70\%$ in the most transverse modes, decreasing to $\sim40\%$ towards the most LOS-dominated modes.}
\end{figure}

Finally, we explore the effect of the spatial morphology of inhomogeneous \HI reionization on the Ly$\alpha$ forest 3D power spectrum. Previous studies have shown that large-scale temperature fluctuations associated with patchy reionization can leave substantial imprints on Ly$\alpha$ forest clustering. Focusing on \HeII reionization, \citep{McQuinn_2011} found order-unity changes in the 3D Ly$\alpha$ forest power spectrum at $k\lesssim0.1\,\rm cMpc^{-1}$, while \citep{Greig_2015} found an enhancement of $20$--$30\%$ at $k\sim0.02\,\rm cMpc^{-1}$. In contrast, here we focus specifically on the imprint of patchy \HI reionization and therefore analyse the models at $z=4.2$, before the expected onset of strong \HeII reionization heating.

For \HI reionization, \citep{Montero-Camacho_2019} found deviations between inhomogeneous and homogeneous reionization models of $19$--$36\%$ at $z=4$ and $k=0.14\,\rm cMpc^{-1}$, decreasing to $2.0$--$4.1\%$ by $z=2$. Their approach combined small-box hydrodynamical simulations, used to capture the small-scale response of the Ly$\alpha$ forest, with large-volume semi-numerical reionization fields generated using \textsc{21cmFAST} \citep{Mesinger_2011_21CMFAST} to model the large-scale morphology of inhomogeneous reionization. More recently, \citep{Zheng_2026} incorporated this reionization-memory contribution into the modelling of the 3D Ly$\alpha$ forest correlation function measured by eBOSS, while \citep{Ma_2026_WEAVE-QSO} investigated the prospects for detecting the relic signature of patchy reionization in the 1D Ly$\alpha$ forest power spectrum with WEAVE-QSO.

Here, we revisit the imprint of patchy \HI reionization using the Sherwood--Relics simulations, in which the reionization morphology and its thermal imprint are followed using the radiative-transfer code \textsc{ATON} \citep{Aubert_2008} within large-volume hydrodynamical simulations. This provides a more self-consistent treatment of the coupling between the spatially inhomogeneous reionization history and the subsequent evolution of the gas responsible for the Ly$\alpha$ forest. To isolate the effect of spatially inhomogeneous reionization, we compare two large Sherwood--Relics simulations with identical numerical setups, $L_{\rm box}=160\,h^{-1}\,\mathrm{cMpc}$ and $N_{\rm part}=2\times2048^3$, and the same mean photoionization and photoheating histories. The Homogeneous model assumes spatially uniform ionization and thermal evolution, while the Patchy model follows the spatially inhomogeneous radiation field produced by the \textsc{ATON} radiative-transfer calculation \citep{Aubert_2008}. Since the mean-flux rescaling is spatially uniform and is performed separately for each model, as explained in Sec.~\ref{sec:P3D_sims}, differences in the volume-averaged transmitted flux are removed, while spatial fluctuations associated with the inhomogeneous ionization and thermal fields are preserved. Such a rescaling would need to be treated with caution during reionization, when regions of the IGM remain significantly neutral and differences in the mean transmission contain physical information about the ionization state that would be artificially removed by matching the mean flux. At $z=4.2$, however, \HI reionization is complete in both models. The comparison therefore captures the combined imprint of inhomogeneous reionization, including spatial fluctuations in the neutral fraction and ionizing radiation field, as well as the associated thermal-memory and pressure-smoothing effects, while controlling for the volume-averaged photoionization and photoheating histories.

Figure~\ref{fig:P3D_patchy} shows that patchy \HI reionization produces a strong enhancement of $P_{\rm 3D,\alpha}$ on large scales. The effect is largest at the lowest wavenumbers considered, reaching $\sim40$--$70\%$ depending on $\mu$, and decreases rapidly towards smaller scales. At $k\gtrsim1$--$2\,h\,\mathrm{cMpc}^{-1}$, the Patchy and Homogeneous models become broadly consistent. The enhancement is strongest for predominantly transverse modes and gradually decreases towards LOS-dominated modes. Our results are qualitatively consistent with previous predictions of a large-scale enhancement of the 3D Ly$\alpha$ forest power spectrum associated with inhomogeneous \HI reionization \citep{Montero-Camacho_2019}. The present comparison measures the total imprint of the spatially inhomogeneous reionization history; separating the contributions from instantaneous ionization-rate/neutral-fraction fluctuations and from the accumulated thermal and hydrodynamical response would require an additional decomposition test.

\section{Conclusions}\label{sec:conclusions}

In this study, we have investigated the sensitivity of the three-dimensional Ly$\alpha$ forest flux power spectrum of the post-reionization IGM ($2.4\leq z\leq4.8$) to the numerical configuration of cosmological simulations. For these tests, we base our calculations on the Sherwood simulation suite, considering simulated volumes of up to $160\,h^{-1}\,\rm cMpc$. We employ the Zel'dovich control variate (ZCV) correction to reduce sample variance on large scales arising from the availability of only a single realization of the IGM fields for each simulation configuration. We model the scale and angular dependence of $P_{\rm 3D,\alpha}$ using the analytical parametrization of \citep{Arinyo-i-Prats_2015}, allowing us to additionally quantify the sensitivity of the flux and velocity-gradient bias parameters, $b_{\rm F}$ and $b_{\eta}$, to the numerical configuration and IGM history. The same simulated power spectra are also being modelled within an effective field theory (EFT) framework \cite{Autieri_2026}, which will enable a direct comparison between phenomenological and perturbative descriptions of $P_{\rm 3D,\alpha}$. Our main results can be summarized as follows.

\begin{itemize}
    \item After applying the ZCV correction, $P_{\rm 3D,\alpha}$ at $z=2.4$ is broadly consistent between the different simulation volumes over the full wavenumber range considered in this work ($k\leq10\,h\,\rm cMpc^{-1}$). There is an exception for $\mu>0.75$ at $k\lesssim3\,h\,\rm cMpc^{-1}$, where the power differs by up to $\sim35\%$ ($\sim20\%$) between the $L_{\rm box}=40$ (80) and $160\,h^{-1}\,\rm cMpc$ simulations.
    \item The mass-resolution dependence of $P_{\rm 3D,\alpha}$ is more pervasive across scales and orientations than the simulation-volume dependence, although the latter reaches larger differences in the most LOS-dominated modes. This is qualitatively consistent with the findings of \cite{Chabanier_2024} based on the Eulerian \texttt{Nyx} simulation suite. At $z=2.4$, the lowest-resolution $2\times512^3$ simulation differs from the $2\times2048^3$ simulation by up to $\sim13\%$, with the largest differences occurring for predominantly transverse modes. The intermediate-resolution $2\times1024^3$ simulation shows substantially smaller differences of up to $\sim4\%$ in the lowest-$\mu$ bin.
    \item The numerical sensitivity of $P_{\rm 3D,\alpha}$ is also reflected in the fitted flux and velocity-gradient biases, $b_{\rm F}$ and $b_{\eta}$. Our measurements are broadly consistent with previous simulation-based results \citep{Arinyo-i-Prats_2015,Chabanier_2024} at $z\lesssim3$. The sensitivity of the fitted bias parameters to both simulation volume and mass resolution generally increases towards higher redshift. Nevertheless, $b_{\rm F}$ exhibits a few per cent level convergence with simulation volume across the full redshift range considered, whereas $b_{\eta}$ is more sensitive to finite-volume effects. Allowing $q_2$ to vary shifts the preferred values of several AiP15 parameters through parameter degeneracies, but does not qualitatively alter the redshift evolution of the inferred bias parameters or remove the high-redshift upturn in $b_{\eta}$. In contrast, the inferred $b_{\eta}$ is substantially more sensitive to the range of scales included in the AiP15 fit: restricting the analysis to $k_{\max}=3\,h\,\mathrm{cMpc}^{-1}$ removes the high-redshift upturn found in our fiducial $k_{\max}=10\,h\,\mathrm{cMpc}^{-1}$ fits, highlighting degeneracies between the large-scale bias and non-linear model parameters. The bias parameters obtained from our fiducial full-shape AiP15 fits should therefore be regarded as effective phenomenological parameters, particularly when modes deep in the non-linear regime are included.
    \item At $z=2.4$, we demonstrate that the mass-resolution dependence can be effectively corrected using the splicing technique of \cite{McDonald_2003}. Applying a resolution correction derived from the $40\,h^{-1}\,\mathrm{cMpc}$ simulations to the $L_{\rm box}=80\,h^{-1}\,\mathrm{cMpc}$, $N_{\rm part}=2\times1024^3$ simulation reduces the differences in $P_{\rm 3D,\alpha}$ relative to the directly simulated $N_{\rm part}=2\times2048^3$ result to $\lesssim2\%$ over almost all scales and orientations considered. The corresponding best-fitting bias and non-linear parameters are all consistent with the higher-resolution simulation within $1\sigma$.
    \item The resolution of the grid on which the optical depth is calculated represents an additional source of numerical uncertainty. If $P_{\rm 3D,\alpha}$ is required on a coarse grid, we recommend first extracting the gas fields and computing $\tau_{\alpha}$ on the finest available grid, followed by downsampling of the resulting Ly$\alpha$ forest fluctuation field through averaging of neighbouring cells. Computing $\tau_{\alpha}$ directly from already downsampled gas fields can introduce artificial small-scale structure in the flux spectra and enhance the small-scale power by up to $\sim35\%$, depending on the downsampling factor and $\mu$.
\end{itemize}

Furthermore, we explore the long-lasting imprints of different reionization and thermal histories on $P_{\rm 3D,\alpha}$ at $z=2.4$ using the Sherwood--Relics simulations. We additionally investigate the effect of the spatial morphology of \HI reionization at $z=4.2$. Our main findings are listed below.

\begin{itemize}
    \item Later-ending reionization results in a systematically lower Ly$\alpha$ forest 3D power spectrum. The effect is approximately scale-independent over most of the $k$-range considered, but the differences are only of order $\sim1\%$ at $z=2.4$, increasing towards $z=3.2$. We find $b_{\rm F}\approx-0.1006$ and $b_{\eta}\approx-0.19$ for the fiducial model. Both bias parameters become slightly more negative with increasing reionization redshift, although the differences between the models remain within the statistical uncertainties.
    \item Variations in the amount of photoheating during reionization leave a substantially stronger imprint. Increasing the photoheating by a factor of two enhances the large-scale power by $\sim4$--$8\%$, with the largest enhancement occurring for predominantly transverse modes. Reducing the photoheating by the same factor produces an approximately mirrored response. Relative to the fiducial model, at $z=2.4$ the $b_{\rm F}$ is $\sim3\%$ more (less) negative in the hot (cold) model, while the corresponding differences in $b_{\eta}$ reach $\sim5\%$.
    
    \item The spatial patchiness of \HI reionization leaves a substantial large-scale imprint on $P_{\rm 3D,\alpha}$ at $z=4.2$, enhancing the power at $k\lesssim1$--$2\,h\,\rm cMpc^{-1}$. The enhancement reaches $\sim40\%$ for predominantly LOS modes ($\mu\geq0.75$) and increases gradually towards transverse modes, reaching $\sim70\%$ in the lowest-$\mu$ bin. This is larger than the $19$--$36\%$ effect reported by \citep{Montero-Camacho_2019} at $z=4$ and $k=0.14\,\rm cMpc^{-1}$, although a direct quantitative comparison is complicated by differences in the scales probed and numerical methodology. Our comparison captures the total large-scale imprint of spatially inhomogeneous \HI reionization, including both ionization-field fluctuations and the associated thermal and hydrodynamical response.
\end{itemize}

There are several astrophysical effects not included in our modelling that may alter the predicted $P_{\rm 3D,\alpha}$. Firstly, absorption by metal lines contaminates the Ly$\alpha$ forest and can introduce additional power with a distinct scale and angular dependence \citep{Ma_2026_SIII}. Accurate modelling of these contaminants will therefore be required for direct comparison of our predictions, particularly on small scales, with observational measurements. In addition, the damping wings of high column density absorbing systems of \HI, which are challenging to model numerically due to a large dynamic range required for such studies, can also contaminate the $P_{\rm 3D,\alpha}$ \cite{Rogers_2018}.

Secondly, the Sherwood--Relics models considered here do not explore the impact of X-ray preheating prior to \HI reionization. X-rays can heat the neutral IGM and modify its small-scale structure before the passage of ionization fronts, thereby altering the subsequent thermal and pressure-smoothing history of the post-reionization IGM \citep{Montero-Camacho_2024}. The effect can persist to the redshifts probed by the Ly$\alpha$ forest and modify the relic signatures of inhomogeneous reionization. Given the sensitivity of $P_{\rm 3D,\alpha}$ to the amount of photoheating demonstrated in this work, future studies should investigate possible degeneracies between the thermal imprint of reionization and earlier X-ray preheating. In this context, $P_{\rm 3D,\alpha}$ can provide a complementary probe to the 21-cm forest, which is directly sensitive to the thermal state of the neutral IGM during reionization and to the level of X-ray preheating \citep[e.g.][]{Xu_2011,Soltinsky_2025}. Combining these observables may therefore help constrain the thermal evolution of the IGM across the reionization epoch and distinguish between heating occurring before and during reionization.

Finally, the simulations analysed in this work do not self-consistently follow spatially inhomogeneous \HeII reionization. This limitation is particularly relevant towards the lower-redshift end of our analysis, as quasar-driven \HeII reionization is expected to heat the IGM and generate large-scale temperature fluctuations over $2\lesssim z\lesssim5$ \citep{McQuinn_2011,Greig_2015}. Such fluctuations can substantially modify the large-scale Ly$\alpha$ forest 3D power spectrum and may overlap with the relic thermal signatures of \HI reionization studied here. Several approaches could be used to incorporate these effects in future work. Spatially inhomogeneous \HeII reionization can be included approximately in hydrodynamical simulations at low computational cost \citep{Upton_Sanderbeck_2020}, while the \textsc{ATON-HE} extension of the radiative-transfer framework used for Sherwood--Relics explicitly includes helium \citep{Asthana_2024}. Alternatively, QSO-driven \HeIII regions can be modelled and applied to Sherwood--Relics outputs \citep{Meiksin_2024}.

In summary, our results demonstrate that numerical effects can be comparable to, or even exceed, the imprints of the thermal and reionization history on $P_{\rm 3D,\alpha}$. While the impact of the finite simulation volume can be substantially mitigated by the ZCV correction, sufficient mass resolution remains essential for accurately interpreting the relic signatures of reionization imprinted on the post-reionization Ly$\alpha$ forest 3D power spectrum. At the same time, the distinct scale and angular dependence of the thermal-history and reionization-morphology signatures demonstrate the potential of $P_{\rm 3D,\alpha}$ as a complementary probe of the thermal and reionization history of the IGM. This will become increasingly important as the statistical precision of Ly$\alpha$ forest measurements improves with ongoing and future spectroscopic surveys such as DESI and WST. Fully exploiting these data will therefore require theoretical predictions in which numerical uncertainties are controlled to a level comparable to, or below, the astrophysical signatures of interest.

\newpage

7

\appendix
\section{Impact of the Zel'dovich control variate correction}
\label{app:ZCV_Lbox}

\begin{figure}[tbp]
\centering 
\includegraphics[width=.9\textwidth]{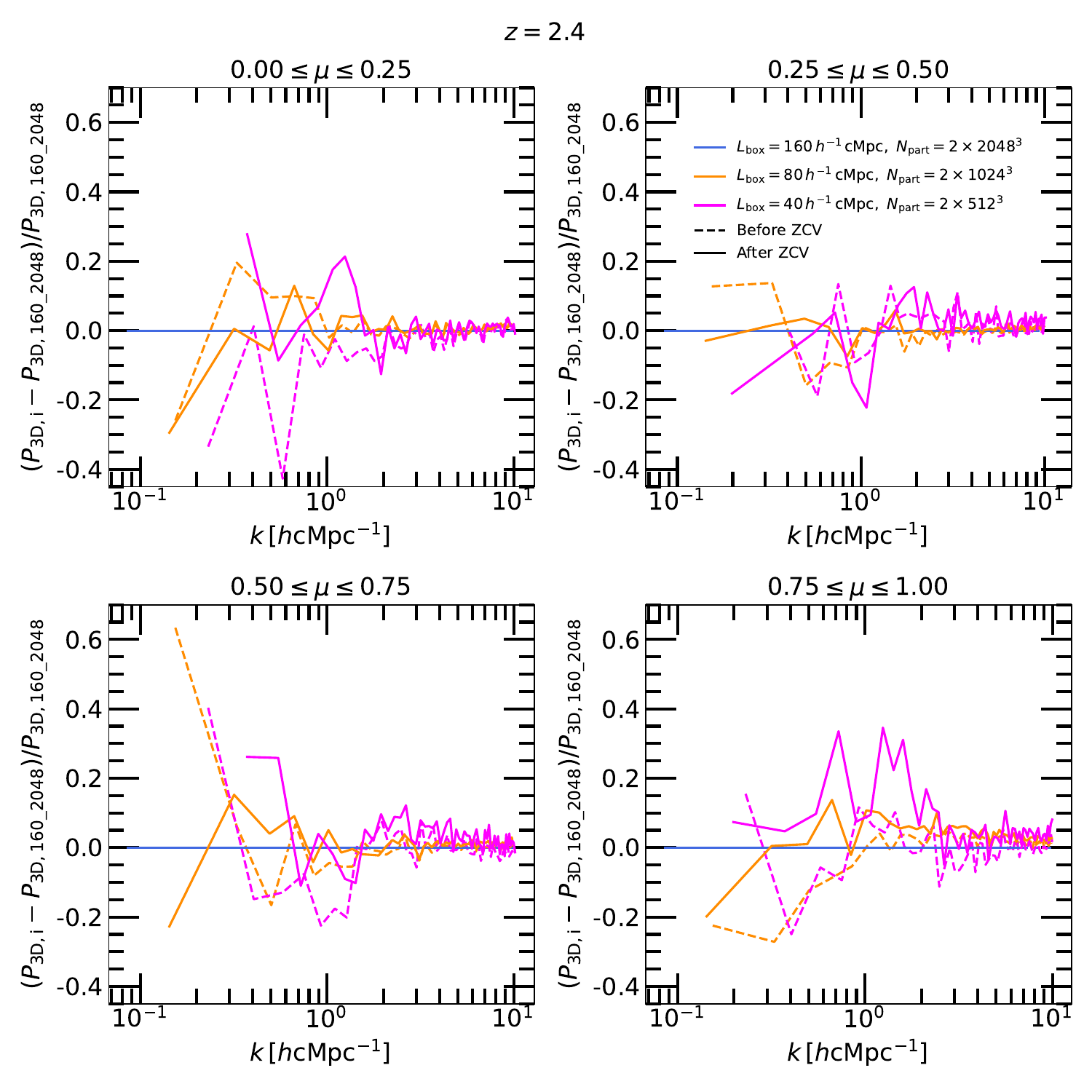}
\hfill
\caption{\label{fig:ZCV_test}Fractional difference in the Ly$\alpha$ forest 3D power spectrum relative to the reference simulation with $L_{\rm box}=160\,h^{-1}\,\mathrm{cMpc}$ and $N_{\rm part}=2\times2048^3$, shown at $z=2.4$ in four bins of $\mu$. Colours distinguish the different simulation volumes and mass resolutions. Dashed curves show the measurements before applying the Zel'dovich control variate correction, while solid curves show the corresponding ZCV-corrected measurements. The correction reduces the coherent large-scale offsets, particularly for the $80\,h^{-1}\,\mathrm{cMpc}$ simulation, although residual deviations remain for the smallest simulation volume.
}
\end{figure}

In Section~\ref{sec:ZCV}, we apply the Zel'dovich control variate (ZCV) correction to reduce the contribution of finite-volume sample variance to the measured Ly$\alpha$ forest 3D power spectrum. Here, we illustrate explicitly how this correction affects the comparison between simulations with different box sizes.

Figure~\ref{fig:ZCV_test} shows the fractional difference in $P_{\mathrm{3D},\alpha}$ relative to the largest-volume simulation with $L_{\rm box}=160\,h^{-1}\,\mathrm{cMpc}$ and $N_{\rm part}=2\times2048^3$. We compare simulations with matched mass resolution but different volumes: $L_{\rm box}=40$ and $80\,h^{-1}\,\mathrm{cMpc}$. Dashed curves show the uncorrected measurements, while solid curves show the corresponding power spectra after applying the ZCV correction.

Before applying ZCV, the smaller-volume simulations exhibit coherent deviations from the $160\_2048$ result on the largest scales. These differences are particularly apparent at $k\lesssim1\,h\,\mathrm{cMpc}^{-1}$ and vary between the different $\mu$ bins, as expected from the limited number of long-wavelength density and velocity modes contained within a finite simulation volume. After applying ZCV, the large-scale offsets are generally reduced, and the measurements cluster more closely around the largest-volume simulation result.

The improvement is most evident for the $80\,h^{-1}\,\mathrm{cMpc}$ simulation, for which the corrected power spectrum agrees closely with the $160\,h^{-1}\,\mathrm{cMpc}$ result over most of the fitted $k$-range. The $40\,h^{-1}\,\mathrm{cMpc}$ simulation retains larger fluctuations, particularly for modes with a substantial line-of-sight component. Residual deviations of up to approximately $35$ per cent remain in individual bins. These differences may arise from a combination of residual realization variance and genuine finite-volume effects that cannot be removed by the control-variate correction.

We therefore conclude that ZCV improves the stability of the box-size comparison by reducing the realization-dependent contribution from long-wavelength modes. However, it does not fully eliminate the limitations associated with the smallest simulation volume. This motivates our use of the ZCV-corrected power spectra in the convergence analysis, while retaining a conservative interpretation of the residual differences between the simulations.

\section{Best fit parameter values}
\label{app:parameter_tables}

\begin{table}
\centering
\small
\caption{Best-fitting bias parameters of the AiP15 model for the different Sherwood simulations considered in this work.}
\label{tab:parameters_bias}
\vspace{0.2cm}
\begin{tabular}{cccc}
\hline
Simulation & $z$ & $b_{\rm F}$ & $b_{\eta}$\\
\hline
$40\_512$ & 2.4 & $-0.1011_{-0.0028}^{+0.0028}$ & $-0.1771_{-0.0087}^{+0.0087}$ \\
 & 2.8 & $-0.1736_{-0.0046}^{+0.0046}$ & $-0.2448_{-0.0121}^{+0.0131}$ \\
 & 3.2 & $-0.2646_{-0.0068}^{+0.0068}$ & $-0.2707_{-0.0172}^{+0.0171}$ \\
 & 3.6 & $-0.3703_{-0.0097}^{+0.0099}$ & $-0.2716_{-0.0210}^{+0.0257}$ \\
 & 4.2 & $-0.5674_{-0.0135}^{+0.0134}$ & $-0.2390_{-0.0323}^{+0.0319}$ \\
 & 4.8 & $-0.8811_{-0.0202}^{+0.0200}$ & $-0.2194_{-0.0444}^{+0.0599}$ \\
\hline
$40\_1024$ & 2.4 & $-0.0977_{-0.0027}^{+0.0027}$ & $-0.1971_{-0.0086}^{+0.0086}$ \\
 & 2.8 & $-0.1645_{-0.0044}^{+0.0044}$ & $-0.2701_{-0.0123}^{+0.0124}$ \\
 & 3.2 & $-0.2469_{-0.0065}^{+0.0065}$ & $-0.3028_{-0.0167}^{+0.0166}$ \\
 & 3.6 & $-0.3420_{-0.0090}^{+0.0090}$ & $-0.3107_{-0.0227}^{+0.0224}$ \\
 & 4.2 & $-0.5125_{-0.0131}^{+0.0133}$ & $-0.2911_{-0.0290}^{+0.0356}$ \\
 & 4.8 & $-0.7836_{-0.0193}^{+0.0196}$ & $-0.2482_{-0.0429}^{+0.0537}$ \\
\hline
$40\_2048$ & 2.4 & $-0.0958_{-0.0025}^{+0.0025}$ & $-0.2086_{-0.0078}^{+0.0078}$ \\
\hline
$80\_1024$ & 2.4 & $-0.1016_{-0.0009}^{+0.0010}$ & $-0.1707_{-0.0031}^{+0.0030}$ \\
 & 2.8 & $-0.1705_{-0.0016}^{+0.0016}$ & $-0.2455_{-0.0043}^{+0.0043}$ \\
 & 3.2 & $-0.2595_{-0.0023}^{+0.0023}$ & $-0.2729_{-0.0058}^{+0.0064}$ \\
 & 3.6 & $-0.3666_{-0.0032}^{+0.0032}$ & $-0.2658_{-0.0080}^{+0.0080}$ \\
 & 4.2 & $-0.5602_{-0.0047}^{+0.0047}$ & $-0.2334_{-0.0124}^{+0.0124}$ \\
 & 4.8 & $-0.8741_{-0.0071}^{+0.0071}$ & $-0.2153_{-0.0187}^{+0.0205}$ \\
\hline
$80\_1024$, res. corr. & 2.4 & $-0.0977_{-0.0009}^{+0.0009}$ & $-0.1828_{-0.0028}^{+0.0028}$ \\
\hline
$80\_2048$ & 2.4 & $-0.0984_{-0.0010}^{+0.0010}$ & $-0.1818_{-0.0029}^{+0.0029}$ \\
\hline
$160\_2048$ & 2.4 & $-0.1005_{-0.0003}^{+0.0003}$ & $-0.1611_{-0.0010}^{+0.0010}$ \\
\hline
\end{tabular}
\end{table}

\begin{table*}
\centering
\small
\setlength{\tabcolsep}{4pt}
\caption{Best-fitting non-linear $D_{\rm NL}$ parameters of the AiP15 model for the different Sherwood simulations considered in this work.}
\label{tab:parameters_DNL}
\vspace{0.2cm}
\begin{tabular}{ccccccc}
\hline
Simulation & $z$ &
$q_1$ &
$k_{\mathrm v}$ &
$a_{\mathrm v}$ &
$b_{\mathrm v}$ &
$k_{\rm p}$ \\
&
&
&
[$h\,\mathrm{cMpc}^{-1}$]
&
&
&
[$h\,\mathrm{cMpc}^{-1}$] \\
\hline
$40\_512$
&2.4&$1.055_{-0.053}^{+0.053}$&$1.492_{-0.177}^{+0.162}$&$0.457_{-0.034}^{+0.030}$&$1.535_{-0.017}^{+0.017}$&$15.35_{-0.63}^{+0.50}$\\
&2.8&$0.921_{-0.063}^{+0.063}$&$1.387_{-0.187}^{+0.185}$&$0.498_{-0.042}^{+0.038}$&$1.657_{-0.023}^{+0.023}$&$15.04_{-0.57}^{+0.47}$\\
&3.2&$0.869_{-0.074}^{+0.074}$&$1.561_{-0.276}^{+0.245}$&$0.517_{-0.059}^{+0.049}$&$1.793_{-0.032}^{+0.029}$&$14.02_{-0.45}^{+0.37}$\\
&3.6&$0.952_{-0.090}^{+0.093}$&$1.971_{-0.465}^{+0.397}$&$0.484_{-0.078}^{+0.064}$&$1.882_{-0.041}^{+0.041}$&$12.91_{-0.36}^{+0.27}$\\
&4.2&$1.203_{-0.105}^{+0.104}$&$3.616_{-0.981}^{+0.843}$&$0.440_{-0.113}^{+0.070}$&$1.924_{-0.047}^{+0.041}$&$11.95_{-0.26}^{+0.23}$\\
&4.8&$1.791_{-0.127}^{+0.127}$&$7.311_{-1.311}^{+1.982}$&$0.376_{-0.125}^{+0.097}$&$1.901_{-0.045}^{+0.040}$&$11.01_{-0.21}^{+0.18}$\\
\hline
$40\_1024$
&2.4&$1.018_{-0.055}^{+0.054}$&$1.038_{-0.137}^{+0.123}$&$0.419_{-0.031}^{+0.027}$&$1.566_{-0.020}^{+0.020}$&$17.14_{-0.93}^{+0.70}$\\
&2.8&$0.915_{-0.064}^{+0.065}$&$1.079_{-0.157}^{+0.141}$&$0.471_{-0.039}^{+0.035}$&$1.681_{-0.025}^{+0.026}$&$17.75_{-1.01}^{+0.74}$\\
&3.2&$0.902_{-0.076}^{+0.076}$&$1.319_{-0.229}^{+0.202}$&$0.500_{-0.053}^{+0.045}$&$1.787_{-0.032}^{+0.032}$&$17.12_{-0.89}^{+0.67}$\\
&3.6&$1.012_{-0.092}^{+0.092}$&$1.755_{-0.383}^{+0.323}$&$0.484_{-0.072}^{+0.057}$&$1.840_{-0.040}^{+0.037}$&$16.24_{-0.77}^{+0.59}$\\
&4.2&$1.353_{-0.115}^{+0.116}$&$3.266_{-0.725}^{+0.811}$&$0.437_{-0.096}^{+0.076}$&$1.833_{-0.042}^{+0.037}$&$15.19_{-0.64}^{+0.50}$\\
&4.8&$2.009_{-0.139}^{+0.141}$&$8.101_{-1.072}^{+1.507}$&$0.425_{-0.130}^{+0.094}$&$1.804_{-0.040}^{+0.035}$&$14.20_{-0.53}^{+0.42}$\\
\hline
$40\_2048$
&2.4&$1.012_{-0.051}^{+0.050}$&$0.831_{-0.108}^{+0.096}$&$0.393_{-0.025}^{+0.025}$&$1.569_{-0.019}^{+0.019}$&$17.66_{-0.93}^{+0.74}$\\
\hline
$80\_1024$
&2.4&$1.034_{-0.018}^{+0.018}$&$1.635_{-0.065}^{+0.062}$&$0.487_{-0.012}^{+0.012}$&$1.540_{-0.006}^{+0.006}$&$15.84_{-0.23}^{+0.21}$\\
&2.8&$0.950_{-0.022}^{+0.022}$&$1.399_{-0.065}^{+0.065}$&$0.486_{-0.014}^{+0.014}$&$1.657_{-0.008}^{+0.008}$&$14.83_{-0.18}^{+0.18}$\\
&3.2&$0.909_{-0.026}^{+0.026}$&$1.570_{-0.097}^{+0.098}$&$0.496_{-0.019}^{+0.020}$&$1.799_{-0.011}^{+0.011}$&$13.74_{-0.14}^{+0.14}$\\
&3.6&$0.970_{-0.030}^{+0.030}$&$2.177_{-0.164}^{+0.165}$&$0.493_{-0.026}^{+0.026}$&$1.903_{-0.014}^{+0.014}$&$12.76_{-0.11}^{+0.11}$\\
&4.2&$1.246_{-0.038}^{+0.038}$&$4.043_{-0.357}^{+0.398}$&$0.440_{-0.040}^{+0.036}$&$1.951_{-0.018}^{+0.018}$&$11.72_{-0.09}^{+0.09}$\\
&4.8&$1.822_{-0.045}^{+0.045}$&$7.702_{-0.608}^{+0.737}$&$0.346_{-0.044}^{+0.039}$&$1.956_{-0.017}^{+0.017}$&$10.83_{-0.07}^{+0.07}$\\
\hline
$80\_1024$, res. corr.
&2.4&$1.009_{-0.019}^{+0.019}$&$1.312_{-0.050}^{+0.050}$&$0.462_{-0.011}^{+0.011}$&$1.570_{-0.007}^{+0.007}$&$17.59_{-0.32}^{+0.29}$\\
\hline
$80\_2048$
&2.4&$0.993_{-0.019}^{+0.019}$&$1.314_{-0.051}^{+0.051}$&$0.467_{-0.011}^{+0.011}$&$1.577_{-0.007}^{+0.007}$&$17.84_{-0.33}^{+0.30}$\\
\hline
$160\_2048$
&2.4&$1.060_{-0.007}^{+0.007}$&$1.724_{-0.023}^{+0.023}$&$0.479_{-0.004}^{+0.004}$&$1.565_{-0.002}^{+0.002}$&$14.94_{-0.07}^{+0.07}$\\
\hline
\end{tabular}
\end{table*}

\begin{table}
\centering
\small
\caption{Best-fitting bias parameters of the AiP15 model for the different reionization and thermal history models from the Sherwood--Relics suite. All simulations have $L_{\rm box}=40\,h^{-1}\,\mathrm{cMpc}$ and $N_{\rm part}=2\times1024^3$.}
\label{tab:parameters_reion_bias}
\vspace{0.2cm}
\begin{tabular}{cccc}
\hline
Simulation & $z$ & $b_{\rm F}$ & $b_{\eta}$ \\
\hline
relics
& 2.4
& $-0.1007_{-0.0029}^{+0.0029}$
& $-0.1890_{-0.0086}^{+0.0085}$ \\

& 2.8
& $-0.1717_{-0.0049}^{+0.0049}$
& $-0.2538_{-0.0115}^{+0.0114}$ \\

& 3.2
& $-0.2598_{-0.0072}^{+0.0071}$
& $-0.2844_{-0.0142}^{+0.0155}$ \\

& 3.6
& $-0.3581_{-0.0097}^{+0.0098}$
& $-0.2988_{-0.0188}^{+0.0218}$ \\

& 4.2
& $-0.5296_{-0.0138}^{+0.0140}$
& $-0.2923_{-0.0272}^{+0.0327}$ \\

& 4.8
& $-0.8112_{-0.0213}^{+0.0212}$
& $-0.2582_{-0.0394}^{+0.0515}$ \\
\hline

cold
& 2.4
& $-0.0983_{-0.0028}^{+0.0028}$
& $-0.1982_{-0.0089}^{+0.0089}$ \\

& 3.2
& $-0.2502_{-0.0071}^{+0.0072}$
& $-0.2927_{-0.0172}^{+0.0172}$ \\
\hline

hot
& 2.4
& $-0.1042_{-0.0031}^{+0.0031}$
& $-0.1793_{-0.0082}^{+0.0082}$ \\

& 3.2
& $-0.2685_{-0.0071}^{+0.0071}$
& $-0.2860_{-0.0134}^{+0.0135}$ \\
\hline

zr525
& 2.4
& $-0.1006_{-0.0029}^{+0.0029}$
& $-0.1874_{-0.0085}^{+0.0085}$ \\

& 3.2
& $-0.2587_{-0.0071}^{+0.0072}$
& $-0.2877_{-0.0147}^{+0.0147}$ \\
\hline

zr675
& 2.4
& $-0.1007_{-0.0029}^{+0.0029}$
& $-0.1911_{-0.0086}^{+0.0086}$ \\

& 3.2
& $-0.2610_{-0.0071}^{+0.0071}$
& $-0.2813_{-0.0143}^{+0.0157}$ \\
\hline

zr750
& 2.4
& $-0.1008_{-0.0029}^{+0.0029}$
& $-0.1931_{-0.0086}^{+0.0086}$ \\

& 3.2
& $-0.2618_{-0.0072}^{+0.0073}$
& $-0.2801_{-0.0144}^{+0.0168}$ \\
\hline
\end{tabular}
\end{table}

\begin{table*}
\centering
\small
\setlength{\tabcolsep}{4pt}
\caption{Best-fitting non-linear $D_{\rm NL}$ parameters of the AiP15 model for the different reionization and thermal history models from the Sherwood--Relics suite. All simulations have $L_{\rm box}=40\,h^{-1}\,\mathrm{cMpc}$ and $N_{\rm part}=2\times1024^3$.}
\label{tab:parameters_reion_DNL}
\vspace{0.2cm}
\begin{tabular}{ccccccc}
\hline
Simulation & $z$ &
$q_1$ &
$k_{\mathrm v}$ &
$a_{\mathrm v}$ &
$b_{\mathrm v}$ &
$k_{\rm p}$ \\
& & & [$h\,\rm{cMpc}^{-1}$] & & & [$h\,\rm{cMpc}^{-1}$] \\
\hline
relics
& 2.4
& $1.000_{-0.057}^{+0.056}$
& $1.297_{-0.134}^{+0.135}$
& $0.482_{-0.034}^{+0.030}$
& $1.611_{-0.021}^{+0.021}$
& $17.94_{-1.12}^{+0.81}$ \\

& 2.8
& $0.898_{-0.067}^{+0.067}$
& $1.592_{-0.143}^{+0.144}$
& $0.621_{-0.043}^{+0.039}$
& $1.745_{-0.024}^{+0.024}$
& $18.98_{-1.34}^{+0.92}$ \\

& 3.2
& $0.876_{-0.080}^{+0.079}$
& $1.962_{-0.196}^{+0.194}$
& $0.686_{-0.055}^{+0.055}$
& $1.860_{-0.030}^{+0.030}$
& $18.41_{-1.18}^{+0.86}$ \\

& 3.6
& $0.982_{-0.093}^{+0.094}$
& $2.349_{-0.298}^{+0.301}$
& $0.647_{-0.073}^{+0.066}$
& $1.899_{-0.036}^{+0.036}$
& $17.38_{-0.99}^{+0.71}$ \\

& 4.2
& $1.326_{-0.117}^{+0.118}$
& $3.643_{-0.543}^{+0.604}$
& $0.572_{-0.099}^{+0.083}$
& $1.890_{-0.042}^{+0.038}$
& $16.19_{-0.79}^{+0.62}$ \\

& 4.8
& $2.009_{-0.148}^{+0.147}$
& $7.321_{-0.715}^{+0.984}$
& $0.556_{-0.129}^{+0.105}$
& $1.852_{-0.040}^{+0.036}$
& $14.88_{-0.64}^{+0.50}$ \\
\hline

cold
& 2.4
& $0.955_{-0.054}^{+0.054}$
& $0.831_{-0.141}^{+0.120}$
& $0.374_{-0.031}^{+0.028}$
& $1.597_{-0.023}^{+0.023}$
& $20.77_{-1.71}^{+1.11}$ \\

& 3.2
& $0.844_{-0.082}^{+0.082}$
& $1.425_{-0.250}^{+0.220}$
& $0.518_{-0.060}^{+0.051}$
& $1.849_{-0.038}^{+0.038}$
& $21.71_{-2.04}^{+1.29}$ \\
\hline

hot
& 2.4
& $1.016_{-0.058}^{+0.058}$
& $1.683_{-0.124}^{+0.124}$
& $0.613_{-0.037}^{+0.034}$
& $1.630_{-0.018}^{+0.018}$
& $16.04_{-0.84}^{+0.63}$ \\

& 3.2
& $0.908_{-0.077}^{+0.077}$
& $2.163_{-0.153}^{+0.154}$
& $0.831_{-0.054}^{+0.049}$
& $1.840_{-0.024}^{+0.024}$
& $16.38_{-0.82}^{+0.62}$ \\
\hline

zr525
& 2.4
& $0.993_{-0.056}^{+0.057}$
& $1.321_{-0.141}^{+0.128}$
& $0.489_{-0.035}^{+0.031}$
& $1.617_{-0.021}^{+0.021}$
& $18.53_{-1.25}^{+0.87}$ \\

& 3.2
& $0.870_{-0.079}^{+0.080}$
& $1.950_{-0.191}^{+0.190}$
& $0.694_{-0.054}^{+0.054}$
& $1.869_{-0.030}^{+0.030}$
& $19.31_{-1.37}^{+0.95}$ \\
\hline

zr675
& 2.4
& $1.006_{-0.056}^{+0.056}$
& $1.257_{-0.142}^{+0.128}$
& $0.470_{-0.033}^{+0.030}$
& $1.607_{-0.021}^{+0.021}$
& $17.41_{-1.01}^{+0.74}$ \\

& 3.2
& $0.880_{-0.079}^{+0.079}$
& $1.969_{-0.201}^{+0.200}$
& $0.674_{-0.058}^{+0.052}$
& $1.853_{-0.030}^{+0.030}$
& $17.54_{-1.01}^{+0.74}$ \\
\hline

zr750
& 2.4
& $1.009_{-0.055}^{+0.056}$
& $1.219_{-0.133}^{+0.133}$
& $0.461_{-0.032}^{+0.030}$
& $1.604_{-0.021}^{+0.021}$
& $17.02_{-0.93}^{+0.69}$ \\

& 3.2
& $0.885_{-0.080}^{+0.081}$
& $1.951_{-0.213}^{+0.212}$
& $0.661_{-0.059}^{+0.055}$
& $1.847_{-0.031}^{+0.031}$
& $16.85_{-0.93}^{+0.68}$ \\
\hline
\end{tabular}
\end{table*}

\begin{figure}[tbp]
\centering 
\includegraphics[width=.32\textwidth]{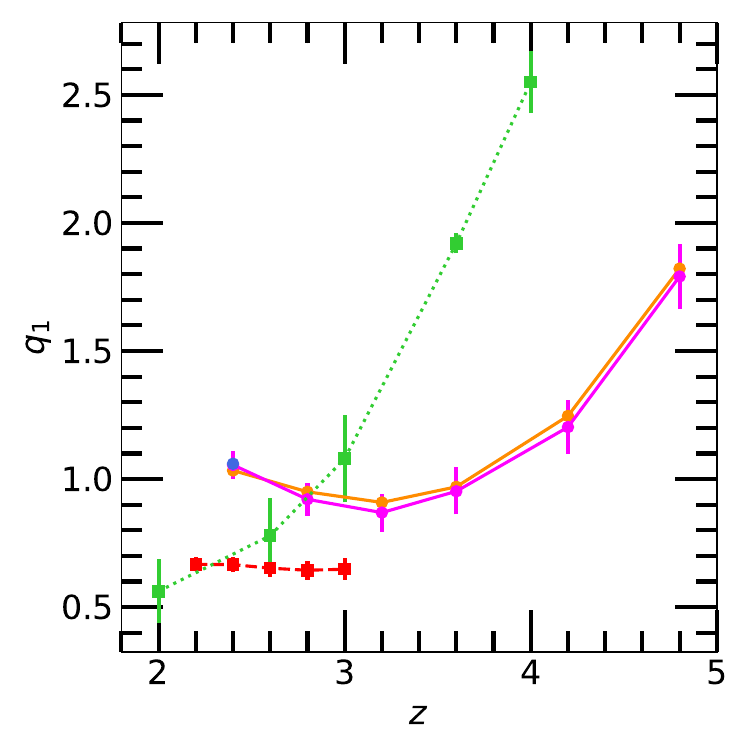}
\includegraphics[width=.32\textwidth]{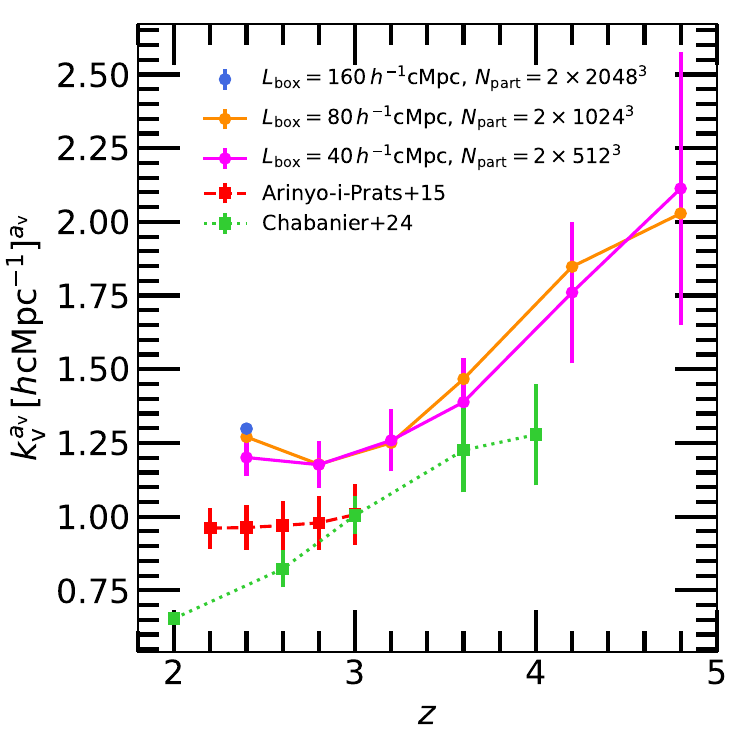}
\includegraphics[width=.32\textwidth]{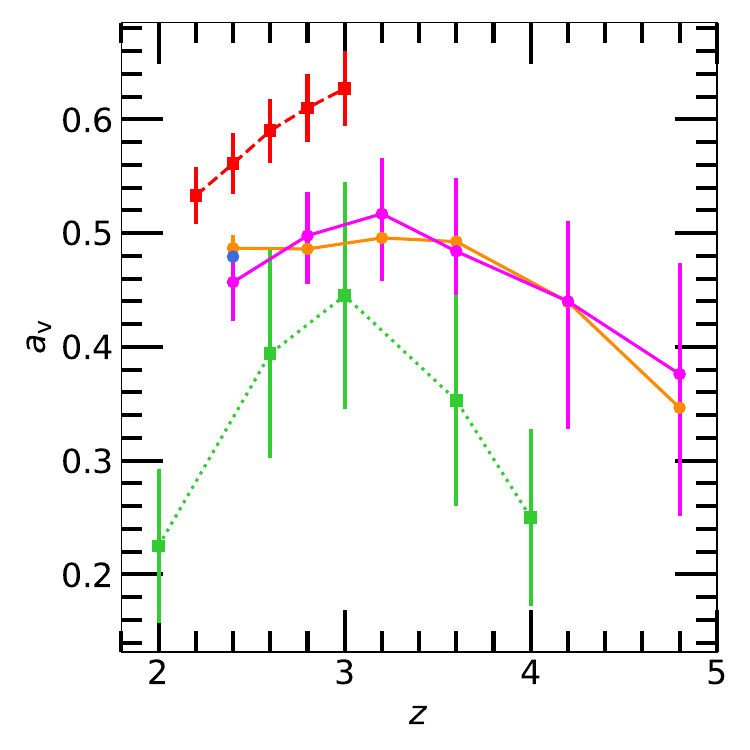}
\includegraphics[width=.32\textwidth]{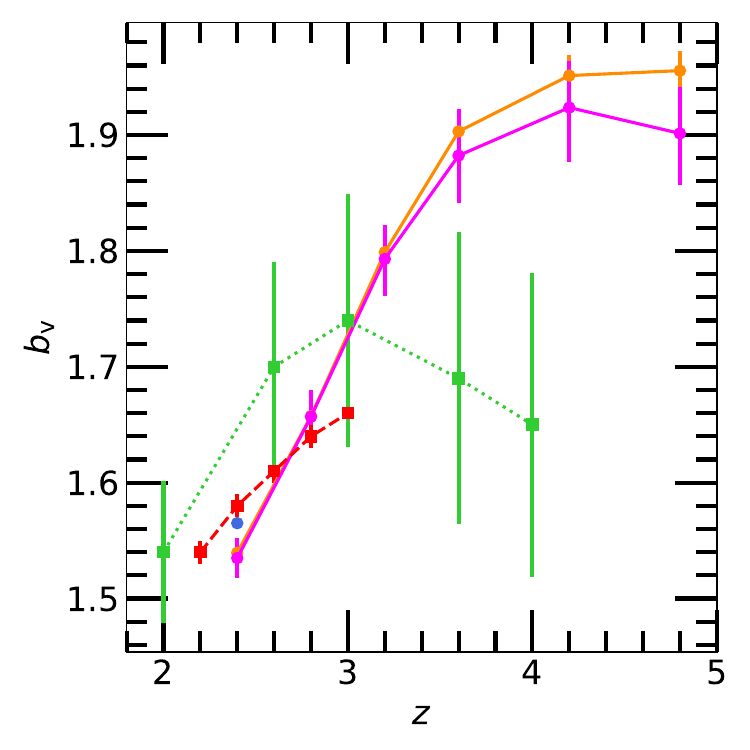}
\includegraphics[width=.32\textwidth]{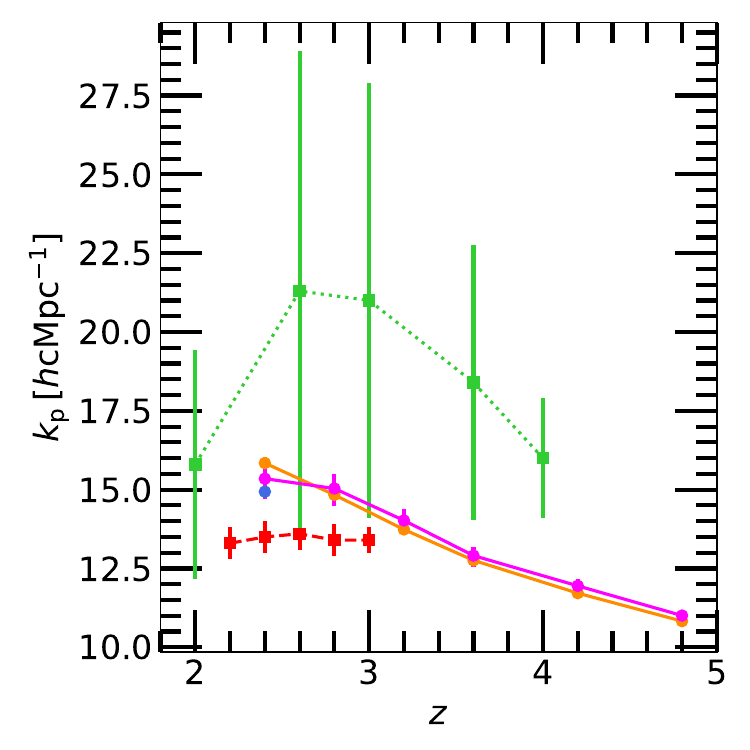}
\hfill
\caption{\label{fig:DNL_Lbox}The redshift evolution of the best-fit non-linear (i.e. $D_{\rm NL}$) parameters from Eq.~\ref{eq:P3D_nlfactor} from $L_{\rm box}=160\,h^{-1}\,\rm cMpc$ (blue), $80\,h^{-1}\,\rm cMpc$ (orange) and $40\,h^{-1}\,\rm cMpc$ (pink). In all three cases the mass resolution is fixed. The dotted red curves mark the values from the Fiducial simulation of \cite{Arinyo-i-Prats_2015} and the dashed green curves follow the 160R25 simulation from \cite{Chabanier_2024}.
}
\end{figure}

\begin{figure}[tbp]
\centering 
\includegraphics[width=.32\textwidth]{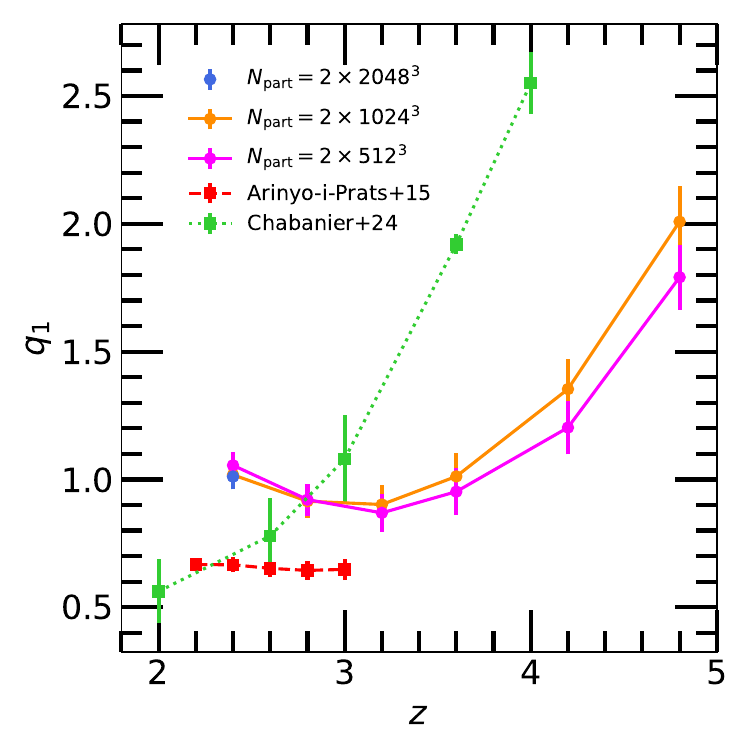}
\includegraphics[width=.32\textwidth]{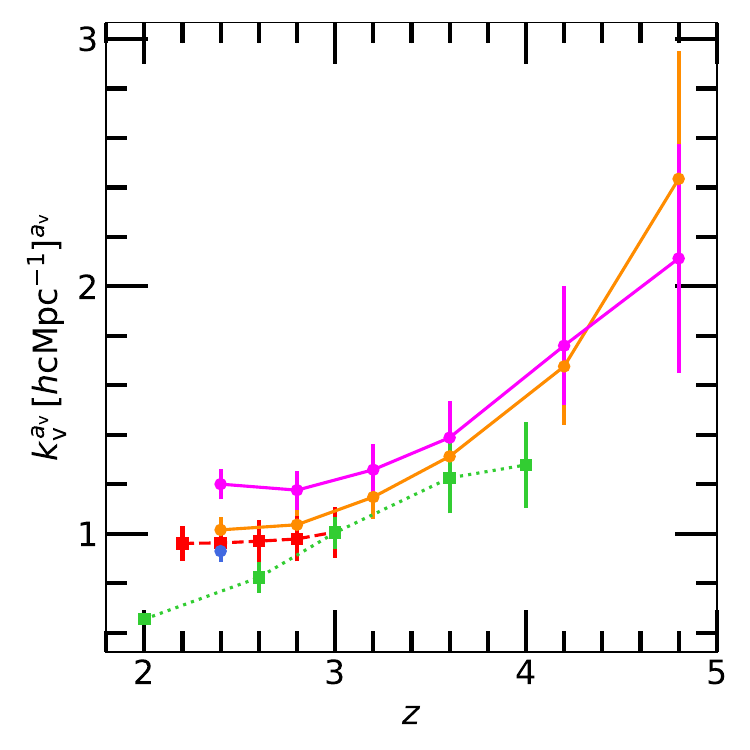}
\includegraphics[width=.32\textwidth]{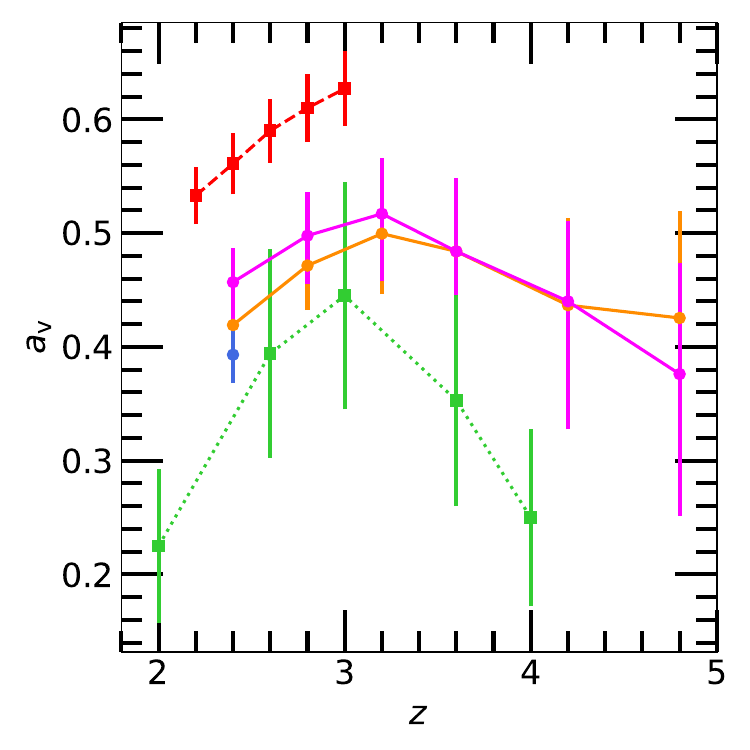}
\includegraphics[width=.32\textwidth]{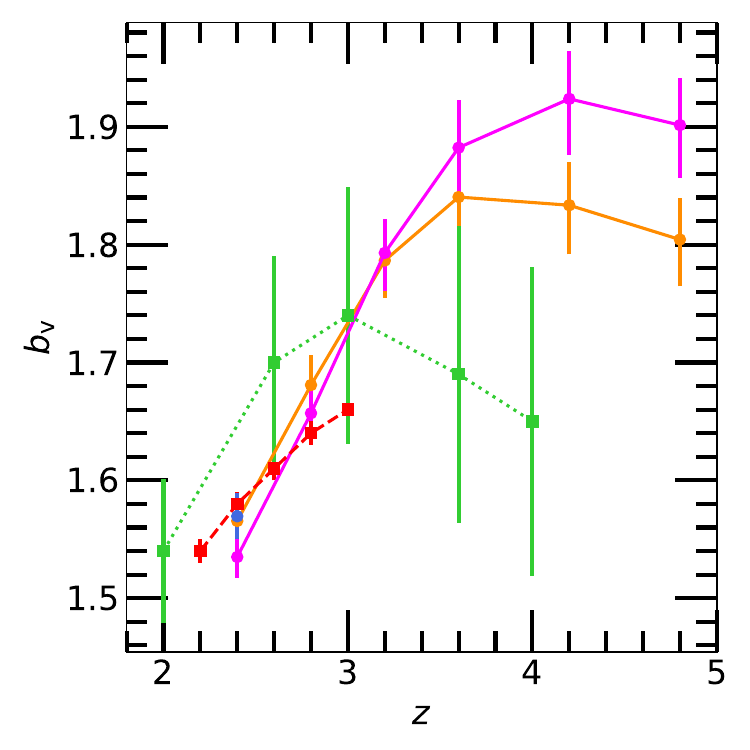}
\includegraphics[width=.32\textwidth]{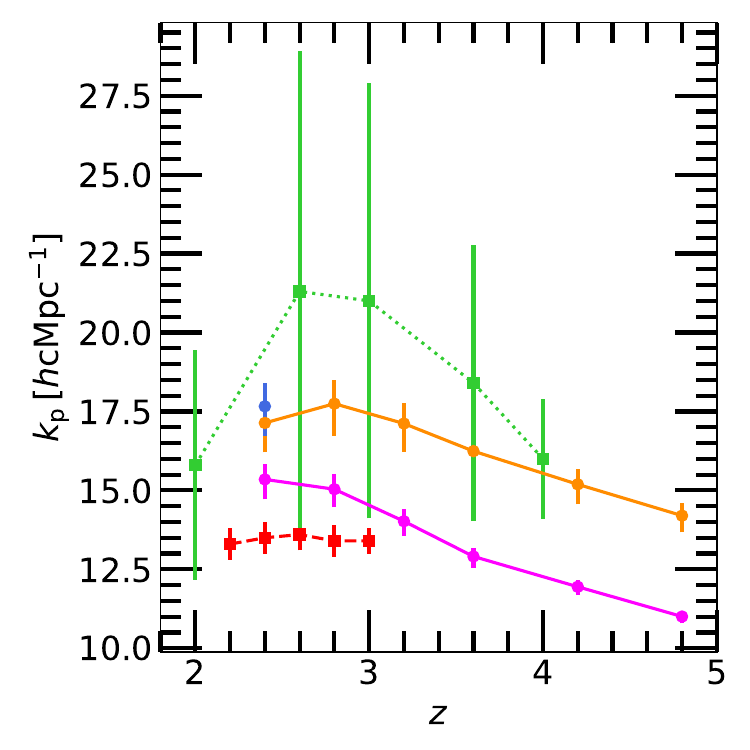}
\hfill
\caption{\label{fig:DNL_Npar}The redshift evolution of the best-fit non-linear (i.e. $D_{\rm NL}$) parameters from Eq.~\ref{eq:P3D_nlfactor} from $L_{\rm box}=40\,h^{-1}\,\rm cMpc$ with $N_{\rm part}$ varying from $2\times2048^3$ (blue) through $2\times1024^3$ (orange) to $2\times512^3$ (pink). The results from Fiducial simulation of \cite{Arinyo-i-Prats_2015} and the 160R25 simulation from \cite{Chabanier_2024} are shown for comparison by the dotted red curves and the dashed green curves, respectively.
}
\end{figure}

Here we list all of the best fit parameter values for the AiP15 function (i.e. Eq.~\ref{eq:P3D_AiP}-\ref{eq:P3D_nlfactor}) for different simulation configurations from Sherwood suite in Tables~\ref{tab:parameters_bias} and~\ref{tab:parameters_DNL} and different thermal and reionization models from Sherwood--Relics simulation suite in Tables~\ref{tab:parameters_reion_bias} and~\ref{tab:parameters_reion_DNL}. We present the level of convergence of the non-linear parameters with the simulation box size in Fig.~\ref{fig:DNL_Lbox} and with the simulation mass resolution in Fig.~\ref{fig:DNL_Npar}. Note that we do not include $q_2$ in these tables because it is fixed to zero in our fiducial fitting procedure. The impact of allowing $q_2$ to vary freely is investigated separately in Appendix~\ref{app:kmaxq2_test}.

\section{Impact of $q_2$ and the range of scales on the fitting procedure}
\label{app:kmaxq2_test}

\begin{figure}[tbp]
\centering 
\begin{minipage}{\textwidth}
\centering 
\includegraphics[width=.32\textwidth]{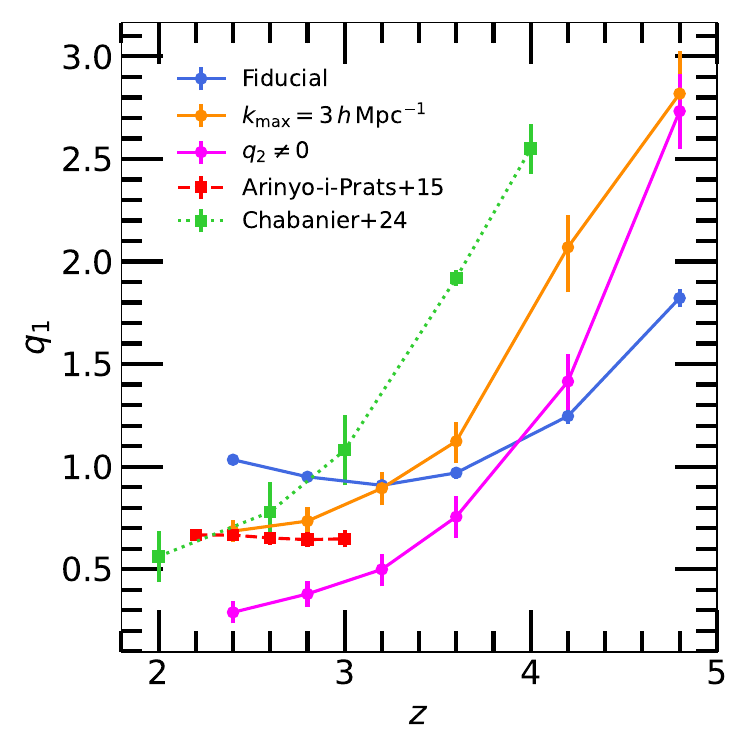}
\includegraphics[width=.32\textwidth]{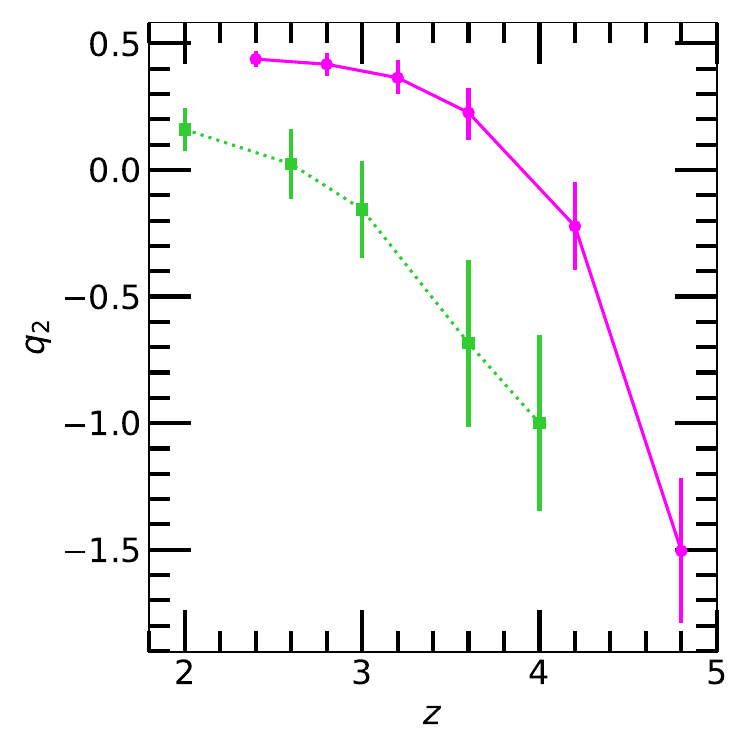}
\includegraphics[width=.32\textwidth]{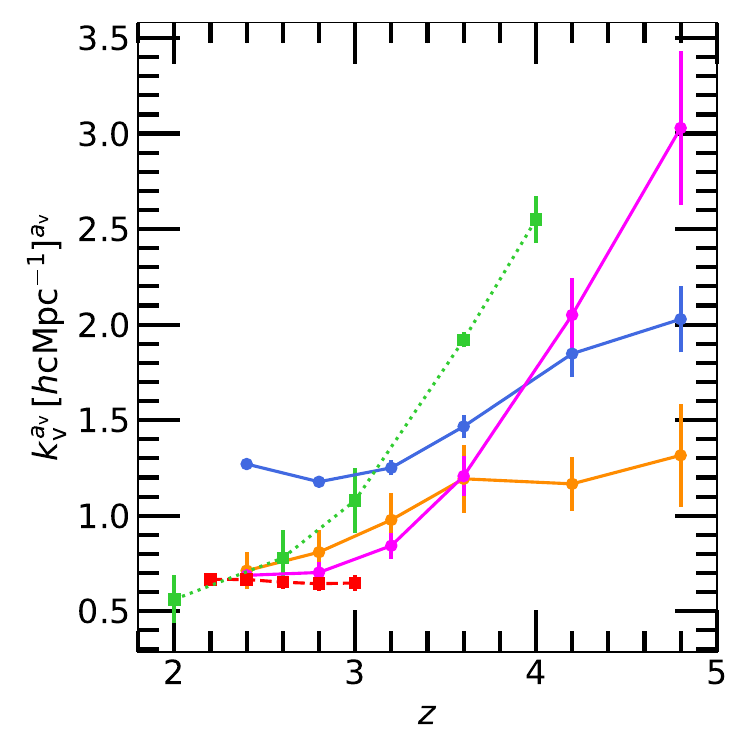}
\end{minipage}
\begin{minipage}{\textwidth}
\centering 
\includegraphics[width=.32\textwidth]{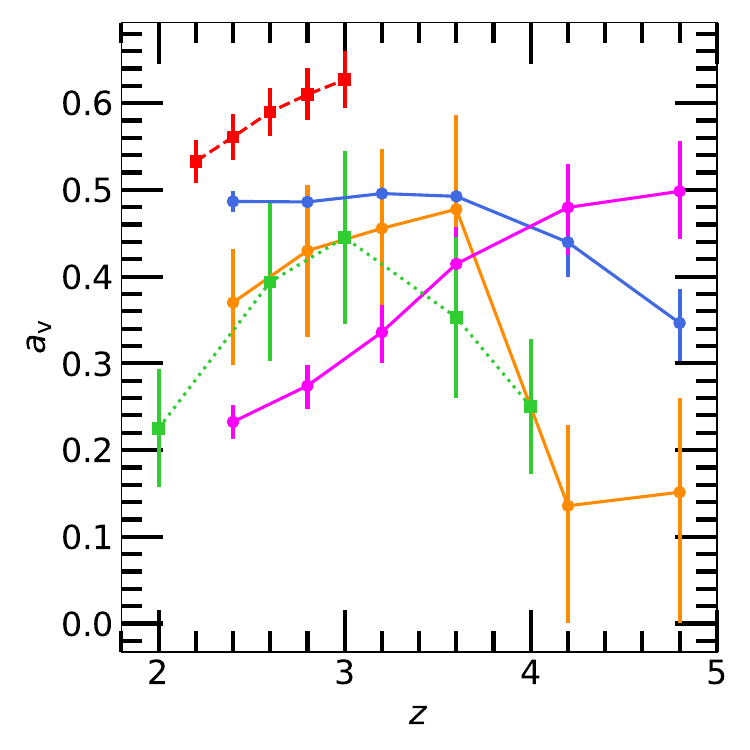}
\includegraphics[width=.32\textwidth]{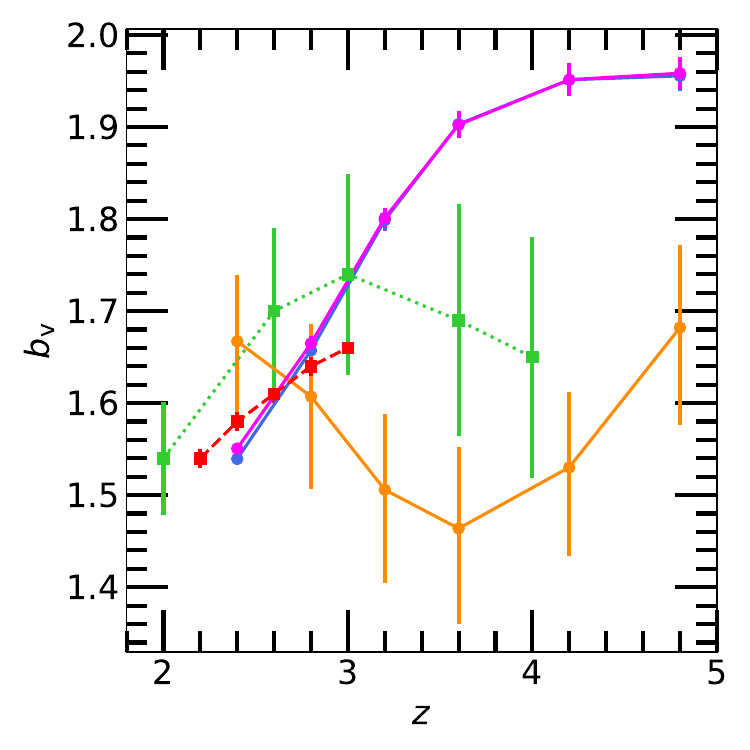}
\includegraphics[width=.32\textwidth]{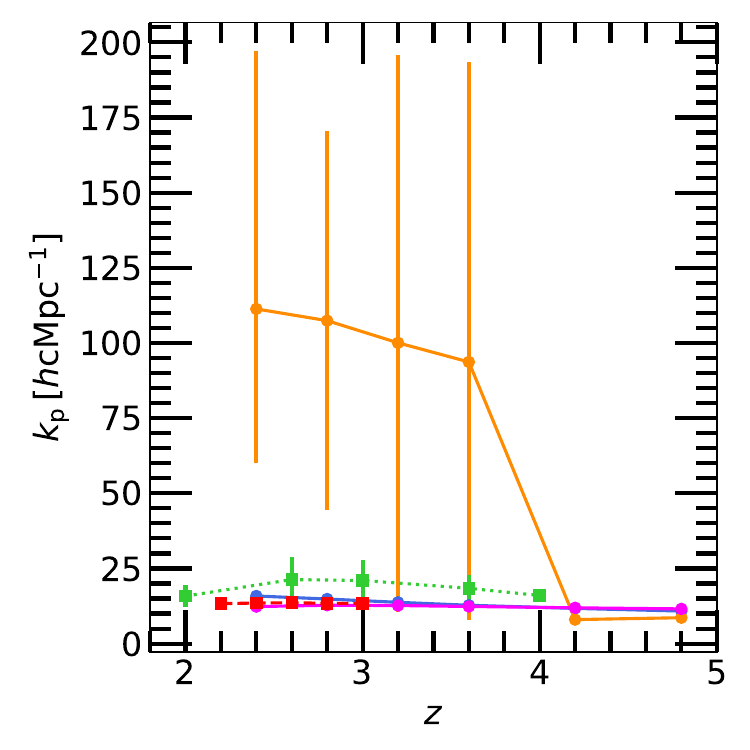}
\end{minipage}
\hfill
\caption{\label{fig:DNL_zevo_test_kmax_q2}The redshift evolution of the best-fit non-linear (including $q_2$) parameters in the $L_{\rm box}=80\,h^{-1}\,\rm cMpc$ and $N_{\rm part}=2\times1024^3$ fitting up to $k_{\rm max}=10\,h\,\rm cMpc^{-1}$ and keeping $q_2=0$ (fiducial value, solid blue curves) compared to when we fit up to $3\,h\,\rm cMpc^{-1}$ (solid orange curves) or allow $q_2$ to be a free parameter (solid pink curves). Similarly to Fig.~\ref{fig:bias_zevo_vsLbox}, the dotted curves represent the parameter evolution presented by \cite{Arinyo-i-Prats_2015} (dashed red curves) from their Fiducial simulation and \cite{Chabanier_2024} (dotted green curves) from their 160R25 simulation.
}
\end{figure}

\begin{figure}[tbp]
\centering 
\includegraphics[width=\textwidth]{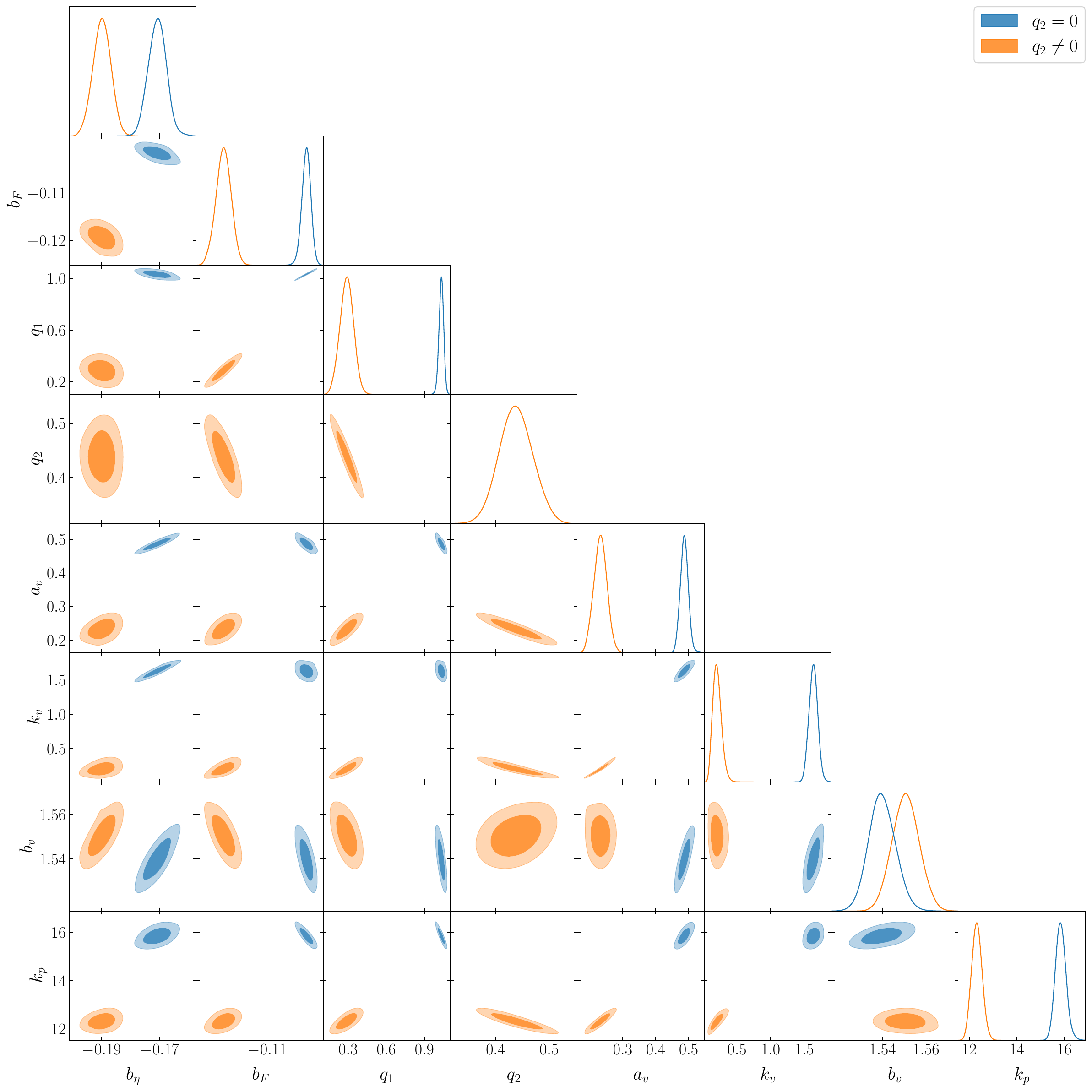}
\caption{\label{fig:triangle_q2_z24} Triangle plot for fitting $80\_1024$ Sherwood simulation at $z=2.4$. We compare posterior distributions from our fiducial AiP15 fitting procedure in which we fix $q_2=0$ (blue) with the case in which we allow $q_2$ to vary freely (orange).}
\end{figure}

\begin{figure}[tbp]
\centering 
\includegraphics[width=\textwidth]{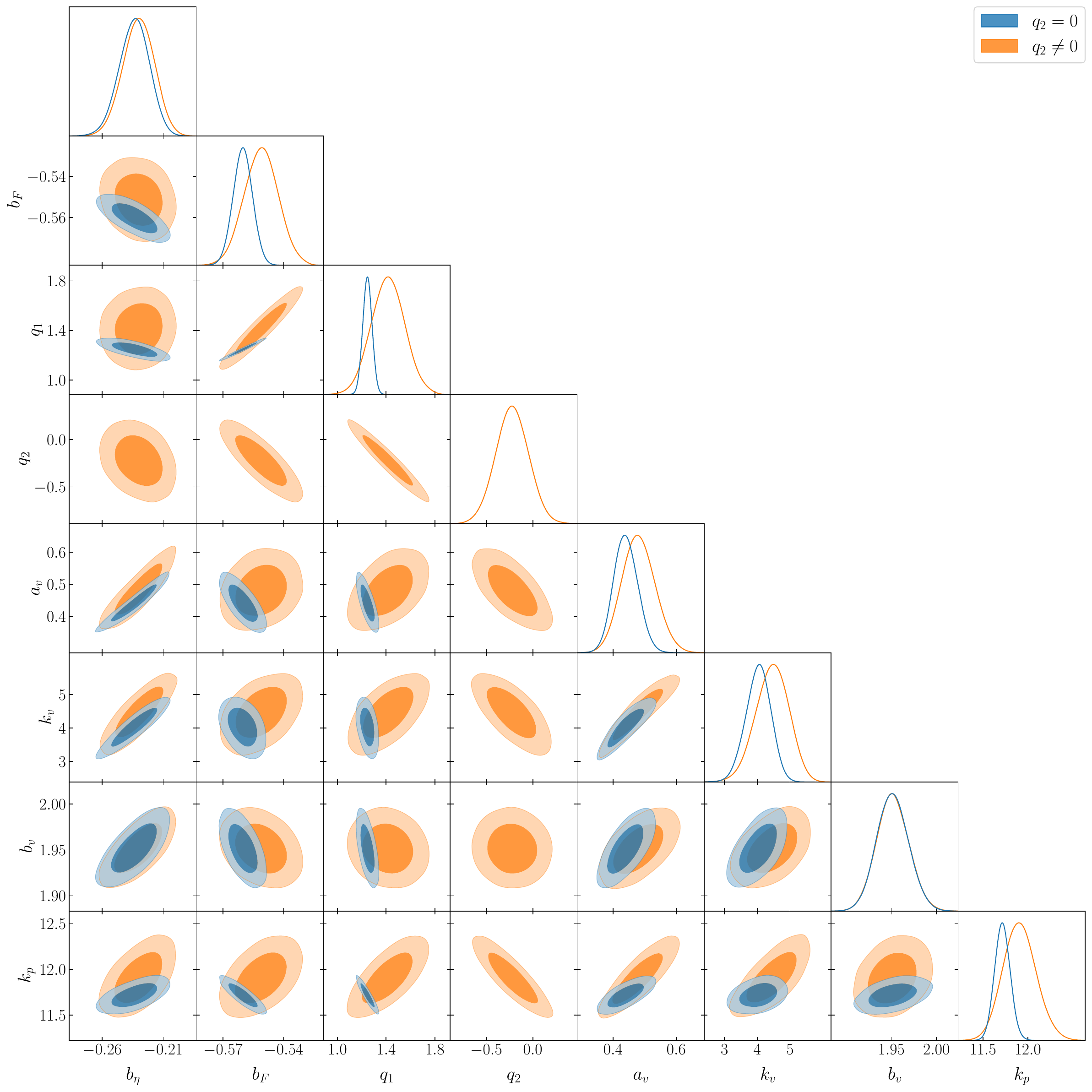}
\caption{\label{fig:triangle_q2_z42} Triangle plot for fitting the $80\_1024$ Sherwood simulation at $z=4.2$. As in Fig.~\ref{fig:triangle_q2_z24}, we compare posterior distributions obtained with $q_2=0$ (blue) with those obtained when $q_2$ is allowed to vary freely (orange).}
\end{figure}

\begin{figure}[tbp]
\centering 
\includegraphics[width=\textwidth]{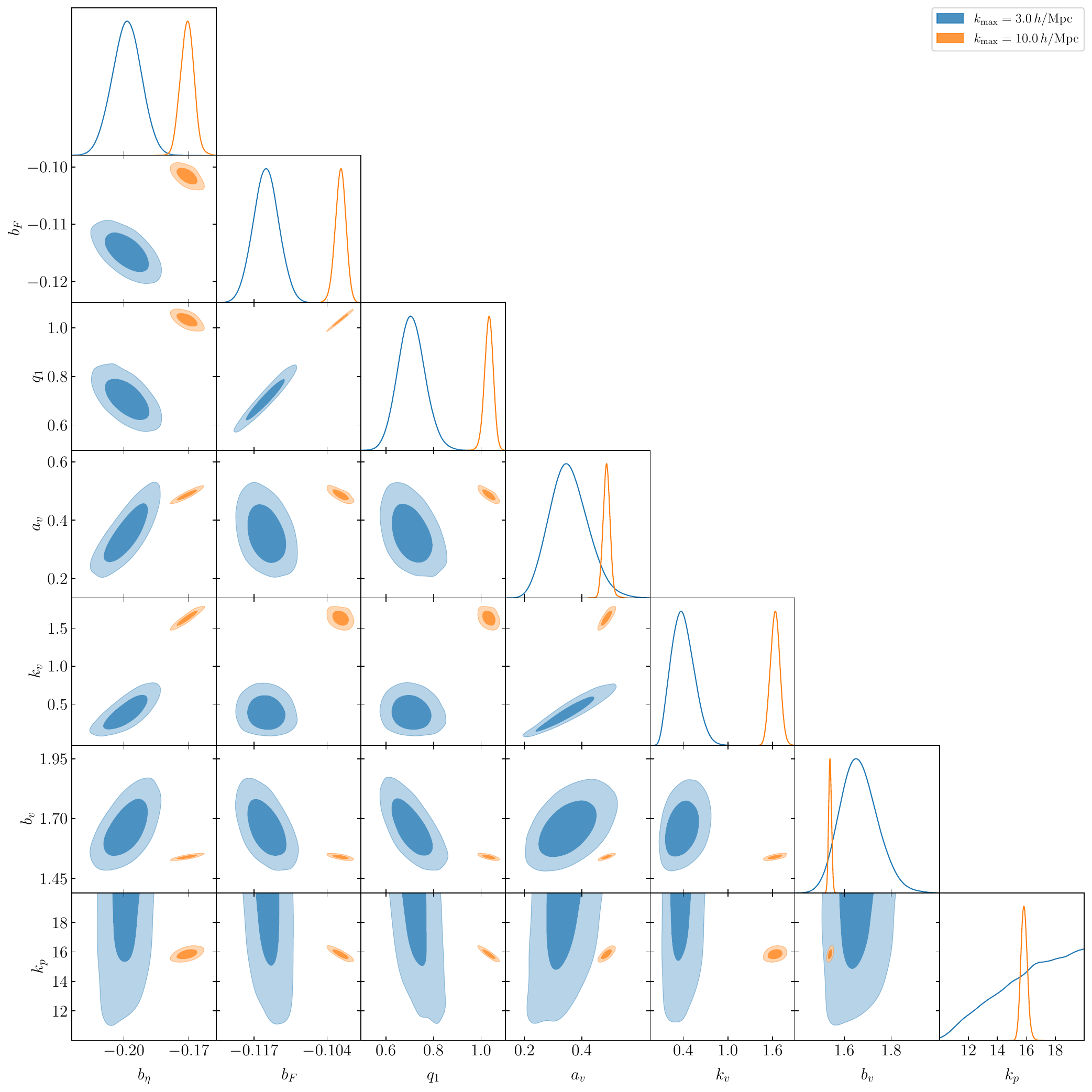}
\caption{\label{fig:triangle_kmax} Triangle plot for fitting $80\_1024$ Sherwood simulation at $z=2.4$. The orange corresponds to the posterior distributions from our fiducial AiP15 fitting procedure in which we fit up to $k_{\rm max}=10\,h\,\rm cMpc^{-1}$ while the blue indicates the case with $k_{\rm max}=3\,h\,\rm cMpc^{-1}$.}
\end{figure}

\begin{table}
\centering
\small
\caption{Best-fitting bias parameters of the AiP15 model for $80\_1024$ Sherwood simulation with different fitting procedures.}
\label{tab:parameters_testkmaxq2_bias}
\vspace{0.2cm}
\begin{tabular}{cccc}
\hline
Assumption & $z$ & $b_{\rm F}$ & $b_{\eta}$\\
\hline
$q_2\neq0$ 
& 2.4 & $-0.1194_{-0.0016}^{+0.0016}$ & $-0.1898_{-0.0031}^{+0.0031}$ \\
& 2.8 & $-0.1884_{-0.0025}^{+0.0025}$ & $-0.2637_{-0.0045}^{+0.0045}$ \\
& 3.2 & $-0.2752_{-0.0038}^{+0.0038}$ & $-0.2871_{-0.0063}^{+0.0064}$ \\
& 3.6 & $-0.3763_{-0.0054}^{+0.0054}$ & $-0.2731_{-0.0088}^{+0.0087}$ \\
& 4.2 & $-0.5510_{-0.0129}^{+0.0129}$ & $-0.2299_{-0.0085}^{+0.0084}$ \\
& 4.8 & $-0.8145_{-0.0135}^{+0.0134}$ & $-0.1961_{-0.0182}^{+0.0221}$ \\
\hline
$k_{\max}=3\,h\,\rm cMpc^{-1}$ 
& 2.4 & $-0.1154_{-0.0022}^{+0.0022}$ & $-0.1978_{-0.0064}^{+0.0070}$ \\
& 2.8 & $-0.1819_{-0.0035}^{+0.0035}$ & $-0.2699_{-0.0099}^{+0.0100}$ \\
& 3.2 & $-0.2604_{-0.0049}^{+0.0049}$ & $-0.3079_{-0.0139}^{+0.0139}$ \\
& 3.6 & $-0.3539_{-0.0066}^{+0.0066}$ & $-0.3155_{-0.0173}^{+0.0190}$ \\
& 4.2 & $-0.5028_{-0.0124}^{+0.0110}$ & $-0.3490_{-0.0300}^{+0.0305}$ \\
& 4.8 & $-0.7753_{-0.0193}^{+0.0176}$ & $-0.3716_{-0.0509}^{+0.0637}$ \\
\hline
\end{tabular}
\end{table}

\begin{table*}
\centering
\small
\setlength{\tabcolsep}{4pt}
\caption{Best-fitting non-linear $D_{\rm NL}$ parameters of the AiP15 model for $80\_1024$ Sherwood simulation with different fitting procedures.}
\label{tab:parameters_testkmaxq2_DNL}
\vspace{0.2cm}
\begin{tabular}{cccccccc}
\hline
Assumption & $z$ &
$q_1$ &
$q_2$ &
$k_{\mathrm v}$ &
$a_{\mathrm v}$ &
$b_{\mathrm v}$ &
$k_{\rm p}$ \\
&
&
&
&
[$h\,\mathrm{cMpc}^{-1}$]
&
&
&
[$h\,\mathrm{cMpc}^{-1}$] \\
\hline
$q_2\neq0$
&2.4&$0.290_{-0.053}^{+0.053}$&$0.438_{-0.031}^{+0.031}$&$0.200_{-0.072}^{+0.053}$&$0.233_{-0.020}^{+0.020}$&$1.551_{-0.006}^{+0.006}$&$12.32_{-0.22}^{+0.22}$\\
&2.8&$0.379_{-0.064}^{+0.064}$&$0.417_{-0.045}^{+0.045}$&$0.276_{-0.107}^{+0.069}$&$0.274_{-0.027}^{+0.024}$&$1.665_{-0.008}^{+0.008}$&$12.76_{-0.24}^{+0.21}$ \\
&3.2&$0.499_{-0.084}^{+0.075}$&$0.364_{-0.066}^{+0.071}$&$0.602_{-0.195}^{+0.141}$&$0.336_{-0.036}^{+0.032}$&$1.801_{-0.011}^{+0.011}$&$12.67_{-0.23}^{+0.20}$ \\
&3.6&$0.756_{-0.102}^{+0.101}$&$0.227_{-0.108}^{+0.098}$&$1.575_{-0.319}^{+0.315}$&$0.415_{-0.043}^{+0.043}$&$1.902_{-0.014}^{+0.014}$&$12.36_{-0.20}^{+0.20}$ \\
&4.2&$1.415_{-0.138}^{+0.135}$&$-0.222_{-0.175}^{+0.175}$&$4.467_{-0.474}^{+0.530}$&$0.480_{-0.055}^{+0.050}$&$1.951_{-0.018}^{+0.018}$&$11.91_{-0.19}^{+0.17}$ \\
&4.8&$2.733_{-0.182}^{+0.181}$&$-1.505_{-0.286}^{+0.287}$&$9.245_{-0.464}^{+0.612}$&$0.498_{-0.056}^{+0.058}$&$1.958_{-0.017}^{+0.018}$&$11.55_{-0.18}^{+0.16}$ \\
\hline
$k_{\max}=3\,h\,\rm cMpc^{-1}$
&2.4&$0.686_{-0.054}^{+0.054}$& $0$ &$0.400_{-0.167}^{+0.136}$&$0.370_{-0.072}^{+0.062}$&$1.667_{-0.086}^{+0.072}$&$111.32_{-51.40}^{+85.76}$\\
&2.8&$0.735_{-0.066}^{+0.066}$& $0$ &$0.611_{-0.225}^{+0.196}$&$0.430_{-0.099}^{+0.075}$&$1.607_{-0.100}^{+0.079}$&$107.42_{-62.86}^{+63.13}$ \\
&3.2&$0.894_{-0.083}^{+0.082}$& $0$ &$0.953_{-0.303}^{+0.301}$&$0.456_{-0.121}^{+0.091}$&$1.506_{-0.101}^{+0.082}$&$100.05_{-89.44}^{+95.87}$ \\
&3.6&$1.124_{-0.107}^{+0.094}$& $0$ &$1.447_{-0.362}^{+0.436}$&$0.478_{-0.146}^{+0.108}$&$1.464_{-0.103}^{+0.088}$&$93.68_{-85.89}^{+99.88}$ \\
&4.2&$2.069_{-0.220}^{+0.159}$& $0$ &$3.111_{-1.484}^{+1.340}$&$0.136_{-0.135}^{+0.093}$&$1.530_{-0.096}^{+0.082}$&$7.99_{-1.22}^{+0.97}$ \\
&4.8&$2.819_{-0.257}^{+0.207}$& $0$ &$6.129_{-2.276}^{+2.127}$&$0.151_{-0.151}^{+0.108}$&$1.682_{-0.105}^{+0.090}$&$8.62_{-1.58}^{+0.84}$ \\
\hline
\end{tabular}
\end{table*}

Throughout this work, we fit the AiP15 model to the Ly$\alpha$ forest 3D power spectrum over the range $k\leq10\,h\,\mathrm{cMpc}^{-1}$ and fix $q_2=0$. As discussed in Sec.~\ref{sec:P3D_analytic}, the inferred bias parameters are more sensitive to the adopted $k$-range than to fixing $q_2$. Here, we examine these tests in greater detail, focusing on their impact on the non-linear parameters and on the posterior degeneracies between the AiP15 parameters. We use the $80\_1024$ simulation and compare our fiducial fits with fits in which $q_2$ is allowed to vary freely and with fits restricted to $k_{\rm max}=3\,h\,\mathrm{cMpc}^{-1}$. The resulting evolution of the non-linear parameters is shown in Fig.~\ref{fig:DNL_zevo_test_kmax_q2}, while the best-fitting bias and non-linear parameters are listed in Tables~\ref{tab:parameters_testkmaxq2_bias} and \ref{tab:parameters_testkmaxq2_DNL}, respectively.

We first test the impact of fixing $q_2=0$ by repeating the $k_{\rm max}=10\,h\,\mathrm{cMpc}^{-1}$ fits while allowing $q_2$ to vary freely. As shown in Fig.~\ref{fig:DNL_zevo_test_kmax_q2}, the preferred value of $q_2$ evolves systematically with redshift, decreasing from positive values at low redshift towards zero and subsequently to negative values at the highest redshifts. This qualitative evolution is similar to that found by \cite{Chabanier_2024}, although our preferred values are systematically shifted towards higher $q_2$. Allowing $q_2$ to vary also changes the preferred values and redshift evolution of several of the remaining non-linear parameters, most notably $q_1$, $k_{\rm v}^{a_{\rm v}}$, and $k_{\rm p}$. These shifts demonstrate that $q_2$ is degenerate with the other parameters entering $D_{\rm NL}$ rather than providing an independent modification of the model.

The origin of these shifts is illustrated by the posterior distributions at $z=2.4$ and $z=4.2$ in Figs.~\ref{fig:triangle_q2_z24} and \ref{fig:triangle_q2_z42}, respectively. At $z=2.4$, when allowed to vary, $q_2$ exhibits strong correlations with several of the other non-linear parameters, and the additional degree of freedom consequently changes the preferred combination of AiP15 parameters. The bias parameters are also shifted through these degeneracies, whereas $b_{\rm v}$, which shows comparatively weak degeneracy with $q_2$, remains relatively stable. At $z=4.2$, the shifts between the two fitting procedures are substantially smaller and the preferred value of $q_2$ lies much closer to zero. This may reflect the weaker degree of non-linearity at higher redshift, whereas at $z=2.4$ the more evolved density field and larger contribution from shock-heated gas may make the decomposition of $D_{\rm NL}$ among its correlated parameters more sensitive to the inclusion of $q_2$. Nevertheless, as shown in Section~\ref{sec:P3D_analytic}, the overall redshift evolution of the inferred bias parameters remains qualitatively similar when $q_2$ is allowed to vary. In particular, allowing $q_2$ to vary does not remove the high-redshift upturn in $b_{\eta}$. The different behaviour of $b_{\eta}$ compared with \cite{Chabanier_2024} therefore cannot be attributed primarily to our choice of fixing $q_2=0$.

We next investigate the sensitivity of the inferred parameters to the range of scales included in the fit by restricting the analysis to $k_{\rm max}=3\,h\,\mathrm{cMpc}^{-1}$, compared with $k_{\rm max}=10\,h\,\mathrm{cMpc}^{-1}$ in our fiducial analysis. As discussed in Section~\ref{sec:P3D_analytic}, this choice has a substantially stronger impact on the inferred bias parameters, particularly $b_{\eta}$: restricting the fitting range removes the high-redshift upturn obtained in the fiducial fits and brings its evolution into better agreement with \cite{Arinyo-i-Prats_2015}, \cite{Chabanier_2024}, and the mEFT analysis of the same simulations in the accompanying work \cite{Autieri_2026}.

The non-linear parameters are also strongly affected by the restricted fitting range. Figure~\ref{fig:DNL_zevo_test_kmax_q2} shows that the preferred values of several parameters shift relative to the fiducial fits, while their uncertainties generally increase. This is particularly apparent for $k_{\rm p}$, which becomes poorly constrained at several redshifts when only modes with $k\leq3\,h\,\mathrm{cMpc}^{-1}$ are included, leading to the large variation in its best-fitting values seen in Table~\ref{tab:parameters_testkmaxq2_DNL}, particularly between $z=3.6$ and $4.2$. These shifts should therefore not be interpreted as a physical change in $k_{\rm p}$, but rather as a consequence of the loss of constraining power on this parameter. This behaviour is expected because $k_{\rm p}$ primarily controls the small-scale suppression of power and therefore cannot be robustly determined when those scales are excluded.

The posterior distributions at $z=2.4$, shown in Fig.~\ref{fig:triangle_kmax}, provide further insight into this dependence. Restricting the fitting range substantially broadens the constraints on several of the non-linear parameters and enhances degeneracies between the bias and non-linear parameters. The joint posterior distributions therefore demonstrate that changing the range of scales included in the fit changes the combination of AiP15 parameters required to describe $P_{\rm 3D,\alpha}$.

These tests therefore demonstrate two distinct effects. Allowing $q_2$ to vary redistributes the preferred values among the correlated parameters entering $D_{\rm NL}$, but does not qualitatively change the redshift evolution of the bias parameters. In contrast, restricting the fit to $k_{\rm max}=3\,h\,\mathrm{cMpc}^{-1}$ changes both the constraints on the non-linear parameters and, most importantly, the inferred evolution of $b_{\eta}$. This supports the interpretation presented in Section~\ref{sec:P3D_analytic} that the high-redshift upturn in $b_{\eta}$ in our fiducial fits is associated primarily with the inclusion of modes at $k\gtrsim3\,h\,\mathrm{cMpc}^{-1}$ and their coupling to the bias parameters through $D_{\rm NL}$. More generally, this scale dependence highlights that the parameters of the phenomenological AiP15 model can absorb information from both large and small scales, and should therefore be interpreted with care when used as physical bias parameters in full-shape analyses. Nevertheless, we retain $k_{\rm max}=10\,h\,\mathrm{cMpc}^{-1}$ and $q_2=0$ as our fiducial choices, since the primary aim of this work is to characterize $P_{\rm 3D,\alpha}$ and its numerical and astrophysical dependence over the full range of scales considered here.

\acknowledgments

TŠ and MV are grateful for the support by the Istituto Nazionale di Astrofisica Osservatorio Astronomico di Trieste (INAF-OATs) under the Theory grant `Cosmological Investigation of the Cosmic Web' (C93C23006820005). TŠ, GA and MV acknowledge the support by the Istituto Nazionale di Fisica Nucleare (INFN) INDARK grant. MV is also supported by IDEAS SISSA grant.
The simulations used in this work were performed using the Joliot-Curie supercomputer at the Très Grand Centre de Calcul (TGCC) and the Cambridge Service for Data Driven Discovery (CSD3), part of which is operated by the University of Cambridge Research Computing on behalf of the STFC DiRAC HPC Facility (www.dirac.ac.uk). We acknowledge the Partnership for Advanced Computing in Europe (PRACE) for awarding us time on Joliot Curie in the 16th call. The DiRAC component of CSD3 was funded by BEIS capital funding via STFC capital grants ST/P002307/1 and ST/R002452/1 and STFC operations grant ST/R00689X/1. This work also used the DiRAC@Durham facility managed by the Institute for Computational Cosmology on behalf of the STFC DiRAC HPC Facility. The equipment was funded by BEIS capital funding via STFC capital grants ST/P002293/1 and ST/R002371/1, Durham University and STFC operations grant ST/R000832/1. DiRAC is part of the National eInfrastructure. In addition, the authors are thankful for the computational resources provided by Istituto Nazionale di Astrofisica - Osservatorio Astronomico di Trieste (INAF-OATs) and Scuola Internazionale Superiore di Studi Avanzati (SISSA) with the Ulysses supercomputer which was used for postprocessing of the simulations.
The authors also acknowledge the developers of publicly available software which was used in this work including \textsc{CAMB} \citep{Lewis_2011_CAMB,Lewis_2013_CAMB}, \textsc{matplotlib} \citep{Hunter_2007}, \textsc{N-GenIC} \citep{Springel_2005,Angulo_2012}, \textsc{numpy} \citep{Harris_2020}, \textsc{NUTS} \citep{Hoffman_2011}, \textsc{scipy} \citep{Virtanen_2020} and \textsc{ZeNBu}. 


\bibliographystyle{utphys}
\bibliography{library}

\end{document}